\documentclass[nobibnotes,nofootinbib,prb,superscriptaddress,showpacs,papersize=a4paper,twocolumn]{revtex4-1}
\usepackage[pdftex]{graphicx}
\usepackage{bm}
\usepackage{etoolbox}
\usepackage[usenames,dvipsnames]{color}
\usepackage{mathtools}
\usepackage{amsmath}
\usepackage{amsfonts}
\usepackage{amssymb}
\usepackage{todonotes}
\usepackage{hyperref}

\DeclarePairedDelimiterX\braket[2]{\langle}{\rangle}{#1 \delimsize\vert #2}

\newcommand*{\dd}{\mathrm{d}}
\newcommand{\deriv}[2]{\dfrac{\partial #1}{\partial #2}}
\newcommand{\ADDED}[1]{ #1 }

\begin{document}

\title{Disorder and interaction in chiral chains: Majoranas vs complex fermions}

\author{J.\,F.~Karcher}
\thanks{These authors contributed equally to this article}
\affiliation{Institut f{\"u}r Nanotechnologie, Karlsruhe Institute of Technology, 76021 Karlsruhe, Germany}
\affiliation{Institut f{\"u}r Theorie der Kondensierten Materie, Karlsruhe Institute of Technology, 76128 Karlsruhe, Germany}

\author{M.~Sonner}
\thanks{These authors contributed equally to this article}
\affiliation{Institut f{\"u}r Nanotechnologie, Karlsruhe Institute of Technology, 76021 Karlsruhe, Germany}
\affiliation{Institut f{\"u}r Theorie der Kondensierten Materie, Karlsruhe Institute of Technology, 76128 Karlsruhe, Germany}
\author{A.\,D.~Mirlin}
\affiliation{Institut f{\"u}r Nanotechnologie, Karlsruhe Institute of Technology, 76021 Karlsruhe, Germany}
\affiliation{Institut f{\"u}r Theorie der Kondensierten Materie, Karlsruhe Institute of Technology, 76128 Karlsruhe, Germany}
\affiliation{L.\,D.~Landau Institute for Theoretical Physics RAS, 119334 Moscow, Russia}
\affiliation{Petersburg Nuclear Physics Institute,188300 St.\,Petersburg, Russia.}

\begin{abstract}
We study the low-energy physics of a chain of Majorana fermions  in the
presence of interaction and disorder, emphasizing the difference between Majoranas and conventional (complex) fermions.
While in the non-interacting limit both
models are equivalent (in particular, belong to the same symmetry class BDI and flow towards the same infinite-randomness critical fixed point), their behavior
differs drastically once interaction is added. Our density-matrix renormalization group calculations
show that the complex-fermion chain remains at the non-interacting fixed point. On the other hand,
the Majorana fermion chain experiences a spontaneous symmetry breaking and localizes for repulsive interaction.
To explain the instability of the critical Majorana chain with respect to a combined effect of interaction and disorder,
we consider interaction as perturbation to the
infinite-randomness fixed point and calculate numerically two-wavefunction correlation functions that enter interaction matrix elements.
The numerical results supported by analytical arguments exhibit a rich structure of critical eigenstate correlations.
This allows us to identify a relevant interaction operator that drives the Majorana
chain away from the infinite randomness fixed point. For the case of complex fermions, the interaction is irrelevant.
\end{abstract}
\maketitle

\section{Introduction}
\label{SectionIntro}

Topological states of matter represent one of the central directions of the
contemporary condensed matter physics \cite{haldane_nobel_2017}.  Systems with
topological order are usually characterized by a gap in the bulk and
``metallic'' states at the boundaries. These boundary states are robust against
disorder-induced Anderson localization as long as the disorder is not strong
enough to close the gap in the bulk\cite{nomura_topological_2007,
moore_birth_2010, ostrovsky_quantum_2007, ryu_$mathbbz_2$_2007}.

One-dimensional (1D) systems with topological phases are considered a potential
platform for quantum computing\cite{sarma_Majorana_2015,alicea_new_2012,
	leijnse_introduction_2012, beenakker_search_2013}, as the quantum state is
stored non-locally and cannot be destroyed by local, uncorrelated noise (as long
as the noise is not strong enough to close the bulk gap). For non-interacting
systems, the symmetry classification by Altland and Zirnbauer
\cite{altland_spectral_2001} combined with the analysis of topologies
\cite{kitaev_periodic_2009, schnyder_classification_2008, ryu_topological_2010,
	mirlin_anderson_2010}, extended also to various spatial symmetries
\cite{fu_topological_2007,fu_topological_2011}, has provided a systematic
picture of possible topological states. Despite the progress on extending this
classification to include weak interactions
\cite{fidkowski_topological_2011,fidkowski_effects_2010,morimoto_breakdown_2015},
it is still a formidable task to determine which topological phases are present
in a given interacting systems. While non-interacting topological phases are
robust against disorder-induced localization, this is not always the case for
topological states in interacting systems. In particular, in 2D superconductor
systems, the combined effect of disorder and interactions has been shown to
break entirely the topological protection
\cite{foster_interaction-mediated_2012, foster_topological_2014}. The underlying
mechanism is that disorder renders the interaction relevant in the
renormalization-group (RG) sense; see also
Refs.~\onlinecite{ostrovsky_interaction-induced_2010,
	burmistrov_enhancement_2012} for related physics.  The fact that the interplay
of interaction and disorder may crucially affect the physics has been known for
a while \cite{finkelstein_disordered_2010}; recent works show that it is
also of central importance for topological states of matter.

In this work, we explore the the effect of disorder and interaction on the low
energy physics of a chain of Majorana quasiparticles commonly called Kitaev
chain \cite{kitaev_unpaired_2001}.  Note that usually one studies the gapped
Kitaev chain, with zero-dimensional Majorana bound states at its ends. In this
paper, we will pay particular attention to the combined effect of disorder and
interaction on a gapless Majorana chain representing a one-dimensional wire with
counterpropagating Majoarana modes. The most local interaction one can have in
this system is a four-point Majorana interaction \cite{rahmani_phase_2015}.
Disorder is introduced by choosing the hopping parameters from a random
distribution.  This model could potentially be realized by vortex lattices
\cite{chiu_strongly_2015, pikulin_interaction-enabled_2015, chiu_proposed_2015}
in a thin film topological superconductor. In general, chains of parafermions
such as Majoranas can also be realized in superconductor-ferromagnet structures
along quantum spin Hall edges \cite{shivamoggi_Majorana_2010}. Further, the
(gapped) Kitaev chain Hamiltonian has been realized as an effective low energy
theory in InGaAs nanowires on top of a superconductor in a magnetic field
\cite{mourik_signatures_2012}. A gapless Majorana chain can be realized on the
edge of an array of such wires \cite{milsted_statistical_2015}. Other platforms
for generating Majorana chains include chains of magnetic atoms on top of a
superconductor \cite{nadj-perge_observation_2014}, as well as cold atoms in optical lattices
\cite{buhler_majorana_2014}. The phase diagram of a clean interacting Kitaev
chain was studied in Ref.~\onlinecite{rahmani_phase_2015}. 

We will compare the Majorana model to that of complex fermion hopping on a chain
with the chemical potential tuned to zero \cite{fisher_random_1994,
	balents_delocalization_1997}. In spin language, this model is equivalent to the
random bond XXZ model. In the absence of interaction, both Majorana and
complex-fermion models belong to the symmetry class BDI and are largely
equivalent. The only difference between them is that in the case of complex
fermions each pair of states related through chiral symmetry represent two
independent single body states, while in the case of the Majorana chain each
pair represents a single state. However, the situation changes dramatically once
one adds interaction. In the case of complex fermions, previous work based on
real-space RG analysis showed that weak interactions are irrelevant in the RG
sense \cite{fisher_random_1994,fisher_critical_1995} and thus do not change the
low energy properties of the system. This system flows into a peculiar critical
infinite-randomness fixed point. For the interacting disordered Majorana chain,
the behavior turns out to be very different.  We show that interaction drives
the system away from the infinite randomness fixed point, which leads to
localization in the case of (even weak) repulsive interaction. The localization
of a disordered Majorana chain with moderately strong repulsive interaction was
observed previously in Ref.~\onlinecite{milsted_statistical_2015}. We further
explain why the above two similar models behave so drastically different once
interaction is added.

The outline of the paper is as follows. We define the models and review previous
results  in Sec. \ref{SectionModels}. In Sec. \ref{SectionDMRG}, we present our
numerical results obtained with the density matrix renormalization
group\cite{white_density-matrix_1993} (DMRG) code OSMPS
\cite{jaschke_open_2018}. We consider first the clean interacting Majorana chain
that we drive out of criticality by staggering in order to explore emerging
topological phases. Then we turn to the DMRG study of combined effect of
disorder and interaction, both for complex fermions and for Majoranas. In the
case of complex fermions, we find that properties of a random chain are not
essentially influenced by interaction, in consistency with previous results. On
the other hand, we observe that the interacting disordered Majorana chain
localizes even for weak repulsive interaction. This localization is accompanied
by a spontaneous  breaking of symmetry between two topological phases that
manifests itself in correlation functions. To shed light on the physical origin
of these results, we employ in Sec. \ref{SectionClean} and \ref{sec:SectionIRFP} two
complementary approaches. Specifically, in Sec. \ref{SectionClean} we use
momentum-space RG methods to investigate the effect of weak disorder on the
interacting clean models. We show that disorder in both models is strongly
relevant rendering the clean fixed point unstable. We thus turn to the
complementary approach in Sec. \ref{sec:SectionIRFP}, where we start from an exact
treatment of disorder (which drives the system into the infinite-randomness
fixed point) and  consider interaction as perturbation. By combining the RG
treatment of interaction with a numerical study of wave-function correlations at
the infinite-randomness fixed point, we identify a relevant operator in the case
of the Majorana chain. No such operator exists in the case of the complex
fermionic chain in view of the cancellation between Hartree and Fock
contributions. This explains why the Majorana fermion chain is unstable with
respect to weak interaction, while the complex fermion chain is stable.

\section{Models}
\label{SectionModels}

In this Section we define two 1D models to be considered in this paper: that of
complex fermions, Sec.~\ref{smcomplex}, and of Majoranas,
Sec.~\ref{smMajoranas}. We also briefly review some previous results relevant to
this work.

\subsection{Complex Fermion chain}
\label{smcomplex}

We start with a spinless fermionic chain where the chemical potential is tuned
to zero,
\begin{align}
H&=\sum_j t_j (c_{j}^\dagger c_{j+1}+h.c.).
\end{align}
Every hopping term is between an even (e) and an odd (o) site. The Hamiltonian
possesses therefore a sublattice symmetry which is represented by the operator
$\mathcal{S} = \tau_z$, where $\tau_z$ is the Pauli matrix  operating on the even-odd subspace.
By using the  local $U(1)$ gauge freedom, we can
always choose the hopping matrix elements $t_j$  to be real. This implies a time reversal symmetry
represented by complex conjugation $\mathcal{T} = \mathcal{K}$ with $\mathcal{T}^2=1$.
Further, the system possesses in addition the particle hole symmetry $\mathcal{C}$ expressed by
$\mathcal{C}=\mathcal{K}\tau_z$, with $\mathcal{C}^2=1$. These symmetries place the model in the
BDI symmetry class.

\begin{figure}
	\centering
	\includegraphics[width=.95\columnwidth]{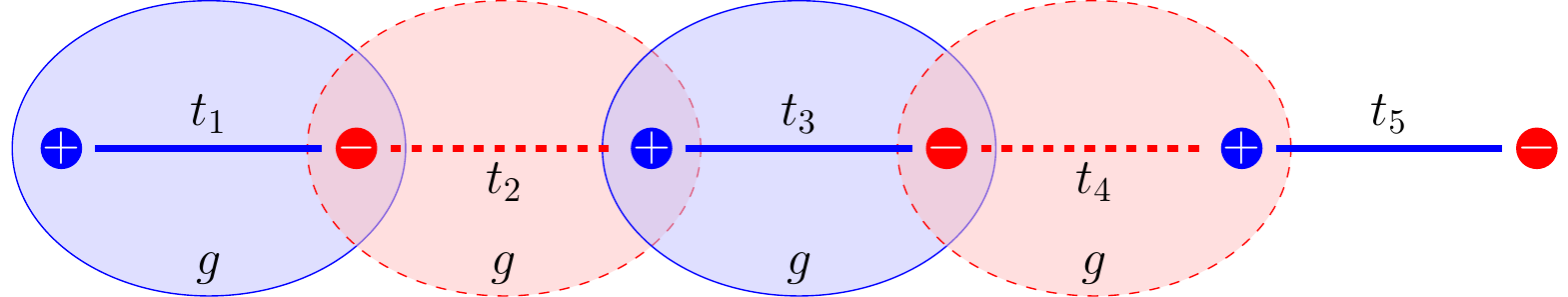}
	\caption{Sketch of the couplings of the complex-fermion chain with Hamiltonian
		\eqref{eq:ham_fer} and $r=1$. Couplings starting on odd sites are solid,
		those starting on even sites are dashed. Odd sites have blue color and are labeled by $+$, while even sites have red color and are labeled by $-$.
		The first few quartic interaction terms involving the sites $j$ and $j+1$ are indicated by blue (odd $j$) and red (even $j$)
		ellipses.}
	\label{fig:sketchFer}
\end{figure}

We introduce disorder by making the hopping matrix elements random. This
does not change the symmetry classification. The most local interaction that can
be added to this model is a two-point nearest-neighbor density-density
interaction. To keep the system at half filling, a chemical potential
proportional to the interaction strength has to be included. Since we will later
see that the sublattice structure of the  interaction is important, we
generalize the interaction to act on sites separated by a distance $r$:
\begin{align}
H &= \sum_j t_j (c_{j}^\dagger c_{j+1}+h.c.)  + g \sum_j p_j p_{j+r},
\label{eq:ham_fer}\\
p_j &= c_j^\dagger c_j - \frac{1}{2}.
\label{eq:pj}
\end{align}
The couplings of this model for $r=1$ are sketched in Fig. \ref{fig:sketchFer}.

\subsubsection{Spin representation}

Using the Jordan-Wigner transformation, one can map the model (\ref{eq:ham_fer}) onto a
random-bond, spin-$\frac12$ XXZ chain:
\begin{align}
H_{\mathrm{spin}} &= \sum_j t_j (\sigma_j^x\sigma_{j+1}^x+ \sigma_j^y\sigma_{j+1}^y) + g \sigma^z_j\sigma^z_{j+r}.
\label{eq:xxz_spin}
\end{align}
The $U(1)$ gauge freedom in the fermionic model corresponds to the spin-rotation
symmetry in the $XY$ plane. While the two models (\ref{eq:ham_fer}) and (\ref{eq:xxz_spin}) are equivalent, the Jordan-Wigner transformation is non-local, and so is the mapping between the correlation functions. The spin representation turns out to be particularly suitable for the DMRG analysis and will be used in this paper.

\subsubsection{Symmetries and topology}
\label{complex-sym-top}

To show that our interaction does not change the symmetry class, we consider the
many body generalizations of the above symmetries $\mathcal{T} = \hat{U}_T
\mathcal{K}, \mathcal{C} = \hat{U}_C \mathcal{K}, \mathcal{S} = \hat{U}_S$, see Ref.~\onlinecite{chiu_classification_2016}. They
can be obtained by defining the action of the  symmetry operators
on the creation and annihilation operators:
\begin{align}
\hat{T} c_j \hat{T}^{-1} &= (U_T)_{j,i} c_i = c_j,\\
\hat{C} c_j \hat{C}^{-1} &= (U_C)_{j,i} c_i^\dagger =  (-1)^j c_j^\dagger,\\
\hat{S} = \hat{T}\cdot\hat{C}.
\end{align}
This defines the action of  $\hat{T}, \hat{C}, \hat{S}$ on all operators and states in the Fock space.
In this many-body formulation, the time-reversal symmetry $\hat{T}$ and chiral symmetry $\hat{S}$
are represented by anti-unitary operators, while the particle hole symmetry $\hat{C}$ is
represented by a unitary  operator. In contrast to the single body
symmetry operators $\mathcal{C}$ and $\mathcal{S}$, the many body symmetry
operators $\hat{C},\hat{S},\hat{T}$ all commute with the Hamiltonian.

Let us now analyze the symmetries of the Hamiltonian (\ref{eq:ham_fer}).
First, all couplings are real, implying that $\hat{T}$ commutes with $H$. Second, the term $-1/2$ in Eq.~(\ref{eq:pj}), which corresponds to a proper choice of the chemical potential  ensures that the model is invariant under
$\hat{C}$. Further, the operators $\hat{T}$ and $\hat{C}$ square to unity. The interacting
model belongs therefore to the symmetry class BDI. It was shown that 1D interacting
systems of complex fermions belonging to this symmetry class (in absence of pairing terms) have a $\mathbb{Z}_4$ topological
invariant  \cite{morimoto_breakdown_2015}.

\subsubsection{Clean limit}

Let us briefly discuss the clean limit. If all matrix elements $t_j$ are equal,
$t_j = t$, and the interaction $g$  is not too strong, the low-energy theory of
the XXZ model (\ref{eq:xxz_spin}) is the Luttinger liquid. This is a conformal
field theory with central charge $c=1$. For the case of nearest-neighbor
interaction, $r=1$, the corresponding condition is\cite{luther_calculation_1975} $|g| <
t$. For $|g| > t$ the system is gapped.

One can drive the system away from the critical line by introducing a
staggering, $t_{2j} = t_{\rm e}$ and $t_{2j+1} = t_{\rm o}$, with $t_{\rm e} \ne
t_{\rm o}$. This will in general open a gap. More precisely, investigating the
RG relevance of the corresponding term in the bosonization language (see
analysis in Sec. \ref{SectionClean} below), we find that the staggering
immediately opens a gap  for $-1<g/t<0.7$, i.e., almost in the whole range of
$g/t$ corresponding to a critical theory. The gapped phases with $t_{\rm e} >
t_{\rm o}$ and $t_{\rm e} < t_{\rm o}$ are topologically distinct. This can be
easily seen by observing that in the limit $t_{\rm e} \rightarrow\infty$, the fermion
at the first site decouples from the rest of the chain, thus representing a
topological zero mode. This zero mode will persist for $t_{\rm e} > t_{\rm o}$
(although it will spread over a few sites). In the opposite case, $t_{\rm o}
\rightarrow \infty$, there is no zero mode. The $c=1$ critical theory (Luttinger
liquid) thus represents a boundary between two topologically distinct phases.

\subsubsection{Noninteracting limit}
\label{complex-non-int}

Consider now a non-interacting system ($g=0$) but in the presence of disorder,
i.e. with random hopping matrix elements $t_j$. This breaks translational
symmetry $j\rightarrow j+1$ for a given realization of disorder. However, if the
distributions of even $t_{2j}$ and odd $t_{2j+1}$ matrix elements are the same,
the system remains self-dual with respect to the transformation $j\rightarrow
j+1$. In spin language, the model corresponds to a disordered XY chain.
Analytically, the problem can be treated with a real space RG procedure
\cite{fisher_random_1994}. At the self-dual point, the system is critical
despite an RG flow towards strong disorder. This very peculiar fixed point is
termed infinite-randomness fixed point. By considering the scaling of the
disorder-averaged entanglement entropy, one can define an effective central
charge $c_\text{eff} =\ln 2$ characterizing this critical state
\ADDED{\cite{refael_entanglement_2004, refael_criticality_2009, laflorencie_scaling_2005}}.

\subsection{Majoranas}
\label{smMajoranas}

To introduce the second model---the one that is which is of the central interest
for this work---we start with a 1D chain of spinless fermions of length
$L$ with superconducting pairing matrix elements $\Delta_j$, hopping
$\tilde{t}_j$ and local chemical potential $\mu_j$. The pairing and
hopping are chosen to be real. The Hamiltonian reads
\begin{eqnarray}
H &=& \sum _{j=1}^{L} \mu_j c_j^\dagger c_j + \tilde{t}_j (c_j^\dagger c_{j+1}  + c_{j+1}^\dagger c_j) \nonumber \\
&+& \Delta_j (c_jc_{j+1} + c^\dagger_{j+1}c^\dagger_j).\label{eq:supercond_chain}
\end{eqnarray}
Now we rewrite each pair of fermionic creation and annihilation operators in
terms of two Hermitian Majorana operators $\gamma_j = \gamma_j^\dagger$:
\begin{align}
c_j=(\gamma _{2j} + i \gamma_{2j+1})/2;\qquad c_j ^\dagger = (\gamma_{2j} - i\gamma_{2j+1})/2
\label{eq:basis_change}.
\end{align}
The Majorana operators obey the commutation relations
\begin{align}
\{\gamma_i,\gamma_j\}= 2\delta_{ij};\qquad  \gamma_i^2=1.
\end{align}
Each Majorana operator represents half a degree of freedom.  The Hamiltonian becomes now
\begin{eqnarray}
H&=&\frac{i}{2}\sum_{j=1}^{L} [\mu_j \gamma_{2j}\gamma_{2j+1} +
(-\tilde{t}_j+\Delta_j) \gamma_{2j+1}\gamma_{2j+2} \nonumber \\
&+& (\tilde{t}_j+\Delta_j) \gamma_{2j}\gamma_{2j+3}].
\end{eqnarray}
If the hopping and pairing terms are chosen such that $\tilde{t}_j=-\Delta_j$, this simplifies to
\begin{align}
H=\sum_{j=1}^{2L} i t_j \gamma_{j} \gamma_{j+1}, \label{eq:kitaev_chain}
\end{align}
where we have introduced notations $t_{2j}=\mu_j/2$ and $t_{2j+1}=-\tilde{t_j}$.
This model is known as Kitaev chain\cite{kitaev_unpaired_2001}.

We now inspect the symmetries of the non-interacting Hamiltonian (\ref{eq:supercond_chain}).
The pairing terms in Hamiltonian (\ref{eq:supercond_chain}) break the global
$U(1)$ symmetry to the parity $\mathbb{Z}_2$. As usual for Bogolyubov-de Gennes models, the Hamiltonian has a particle hole
symmetry $\mathcal{C}= \mathcal{K}$. Since we chose all couplings
real, the system has time reversal symmetry $\mathcal{T} = \tau_z\mathcal{K}$. Both symmetry operators square to unity, thus the
model belongs to class BDI. The product of those two symmetries yields the
sublattice symmetry $\mathcal{S}=\tau_z$.

Now we include the interaction term.
Since $\gamma_n^2 = 1$,  the most
local interaction term couples four neighboring Majoranas
\cite{rahmani_phase_2015}:
\begin{align}
H&=\sum_{j=1}^{2L} i t_j\gamma_{j}\gamma_{j+1} + \sum_{j=1}^{2L} g_j \gamma_{j}\gamma_{j+1} \gamma_{j+2}\gamma_{j+3}. \label{eq:ham_disint}
\end{align}
Below we will allow for randomness in the hopping matrix elements $t_j$.  If the values of the interaction $g_j$ as well as the distribution of hopping matrix elements $t_j$ is the same for even and odd sites, the model is self-dual under translation by one side.

\begin{figure}
	\centering
	\includegraphics[width=.95\columnwidth]{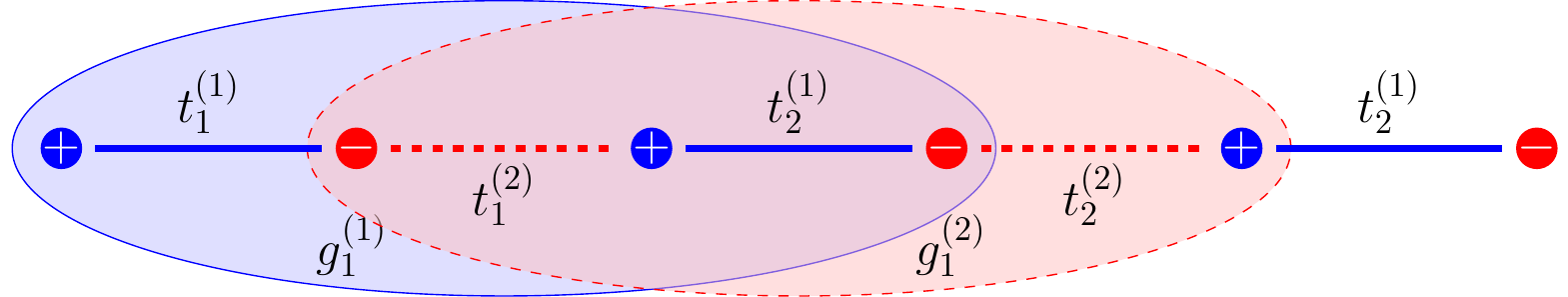}
	\caption{Sketch of the couplings in the Majorana Hamiltonians
		\eqref{eq:ham_disint},\eqref{eq:spins}. Couplings $t_{2j} $ are dashed, $t_{2j+1}$ solid. Odd sites have blue color and are labeled by $+$, while even sites have red color and are labeled by $-$. The first two quartic interaction
		terms with couplings $g_1^{(1)}$ and $g_1^{(2)}$ are indicated by  a blue and a red ellipse, respectively.
		Translation by one site swaps even and odd hopping and interaction terms.}
	\label{fig:sketch}
\end{figure}

\subsubsection{Symmetry and topology}
\label{sec:mod:bdi}

The symmetries $\mathcal{T}, \mathcal{C}$ and $\mathcal{S}$ can be extended to
the many-body setting in analogy with discussion in Sec.~\ref{complex-sym-top}
for the case of complex fermions. In terms of Majorana operators the symmetries
read
\begin{align}
\hat{T} \gamma_j \hat{T}^{-1} &= (-1)^j\gamma_j,\\
\hat{C} \gamma_j \hat{C}^{-1} &= \gamma_j,\\
\hat{S} = \hat{T}\cdot\hat{C}.
\end{align}
It is worth mentioning that for Bogolyubov-de Gennes Hamiltonians the
particle-hole symmetry is not a true many-body symmetry but rather a constraint
related to the Fermi statistics, see discussion in
Ref.~\onlinecite{kennedy_bott_2016}. This puts our model in interacting
symmetry-class BDI with $\mathbb{Z}_8$ topological classification, see
Ref.~\onlinecite{fidkowski_topological_2011}.

While the Hamiltonian (\ref{eq:ham_disint}) contains only nearest-neighbor
Majorana hopping $t_j$, any odd-range hopping  is in principle permitted by
symmetry. In particular, as we discuss below, the interaction generates third
nearest neighbor hopping on the mean-field level. An even-range hopping would
couple Majoranas from the same sublattice and break the chiral symmetry and the
time-reversal symmetry. Similarly, any interaction term containing an even
number of Majorana operators belonging to even sites (and thus an even number of
operators from odd sites), is consistent with the $\hat{T}$ and chiral
symmetries.

\subsubsection{Spin representation}

The interacting Kitaev chain (\ref{eq:ham_disint}) can be mapped onto a spin-$\frac12$-chain by means of Jordan-Wigner transformation:
\begin{eqnarray}
H &=& \sum_{j=1}^{L}  t^{(1)}_j\sigma^x_j - \sum_{j=1}^{L} t^{(2)}_j\sigma_j^z \sigma_{j+1}^z \nonumber \\
& - & \sum_{j=1}^{L} g^{(1)}_j \sigma^x_j\sigma^x_{j+1} - \sum_{j=1}^{L} g^{(2)}_j \sigma_j^z \sigma_{j+2}^z.
\label{eq:spins}
\end{eqnarray}
Here $t^{(1)}_j$ and $t^{(2)}_j$ correspond respectively to  odd ($t_{2j-1}$) and  even ($t_{2j}$) hopping matrix elements of Eq.~(\ref{eq:ham_disint}), and similarly for the interaction couplings $g$. The couplings of this model are sketched in Fig. \ref{fig:sketch}. 

\ADDED{
It is interesting to note that the odd couplings  $g_j^{(1)}$ and $t_j^{(1)}$ couple in the spin language to $x$ components, and the the odd couplings $g_j^{(2)}$ and $t_j^{(2)}$ to $z$ components. Translation by one site (even-odd transformation) exchanges $g_j^{(1)} \leftrightarrow g_j^{(2)}$ and  $t_j^{(1)} \leftrightarrow t_j^{(2)}$. Models related by this transformation are dual, although this duality is less obvious in the spin representation than in the Majorana representation. 
}

We will use the spin representation for the DMRG analysis below.

\subsubsection{Noninteracting limit}

In the non-interacting limit ($g=0$) the Hamiltonian \eqref{eq:spins} describes the transverse
Ising model. In the clean translational-invariant case (no staggering,  $t^{(1)}=t^{(2)}$) the system is critical
with a 1D Majorana low-energy theory and central charge $c=\frac12$. In the presence of random hopping,
the model is at the infinite-randomness fixed point \cite{fisher_critical_1995} as noted above in the context of complex fermions in Sec. \ref{complex-non-int}.
The difference between the two models in the absence of interaction is that two single-particle states of the complex-fermion model correspond to a single state of the Majorana model.
As a consequence, the effective central charge at the infinite-randomness fixed point  is halved, $c=(\ln 2)/2$.

\subsubsection{Clean limit}
\label{sec:clean-limit}

For the case of interacting model with homogenous couplings, $t_j = t$ and $g_j = g$, Rahmani et al.
\cite{rahmani_phase_2015} have determined the phase diagram:

\begin{itemize}

	\item \emph{Strong interaction.} The system is gapped for very strong interactions
	of both signs ($g>250$ or $g < - 2.86$). The translation symmetry gets spontaneously broken, and the transition between
	the topologically distinct phases is of first order type.

	\item \emph{Attractive interaction.} There is a critical phase up to very strong
	interactions $0< g < 250$. The low energy theory is a single Majorana mode with
	central charge $c = \frac12$. This phase is controlled by the same fixed point as the
	transverse Ising model and therefore dubbed Ising phase.

	\item \emph{Weak repulsive interaction.} For the case of repulsive interaction ($g<0$), the Ising phase is stable for sufficiently
	weak interactions, $g> - 0.28$.

	\item \emph{Intermediate repulsive interaction.} For repulsive interaction of intermediate strength, $-2.86 < g < -0.28$, a phase emerges with coexisting Luttinger-liquid and Majorana modes. Alternatively, one can say that a single Majorana mode of the non-interacting theory is promoted to three Majorana modes, which can be understood already by mean-field level treatment of the interaction. The central charges in this phase is  $c=\frac32$.

\end{itemize}

\section{DMRG results}
\label{SectionDMRG}

It is viable to calculate the groundstate properties of systems with length of
the order of a few hundred sites using methods based on matrix-product states (MPS). \ADDED{For these methods, spin models are most convenient. All DMRG calculations in this work have therefore been done on the spin representations, Eq. (\ref{eq:xxz_spin}) and Eq. (\ref{eq:spins}), using the software OSMPS \cite{jaschke_open_2018}. The maximum bond dimension was chosen to be 512, states with weight smaller than $10^{-8}$ were truncated.}

\subsection{Interacting Majorana chain with staggering}
\label{sec:majorana-clean-dmrg}

As we will later see, disorder drives an interacting Majorana chain into
different localized phases if the interaction is repulsive. To obtain an
overview over possible localized phases in the Majorana model, we first consider the clean model
and drive the system out of criticality by introducing staggering. We calculate the ground
state of the clean Majorana chain, Eq. (\ref{eq:spins}) with $t_i^{(1)} = t_1$, $t_i^{(2)} = t_2$, $g_i^{(1)} = g_1$, and $g_i^{(2)} = g_2$,
and $L = 96$ spin sites (which corresponds to $2L=192$ Majorana sites). We choose parameters in such a way that the relation $g_1/t_1 = g_2 /t_2$ is maintained; we use a
short-hand notation $g/t$ for this ratio. By using DMRG, we explore the range $-4<g/t<1$ of the interaction
strength, varying the staggering, $0<t_1/t_2=g_1/g_2<\infty$. For
the staggering region $0<t_1/t_2<1$, the hopping $t_1=1$ is fixed and $t_2$ is
varied, while for staggering above the self-dual line $1<t_1/t_2<\infty$,
$t_2=1$ is fixed and $t_1$ is varied.

The system with a given value of staggering $t_1/t_2$ is related to
the system with inverse staggering via duality transformation. In the Majorana
representation, this transformation corresponds simply to a translation by one lattice
site. On the other hand, in the spin language of Eq.~(\ref{eq:spins}) the duality transformation is much less trivial (and, in particular, non-local).

In the MPS representation the (von Neumann) entanglement entropy between two
subsystems split by a bond is readily available
\cite{schollwoeck_density-matrix_2011, jaschke_open_2018}. In a critical 1D system of length $L$ with open boundary conditions, the bond entropy
scales as a function of bond position $x$ as \cite{bazavov_estimating_2017}
\begin{align}
S(x) &= \frac{c}{6} \ln \left(\frac{2L}{\pi} \sin \frac{\pi x}{L} \right) +\gamma
\label{eq:scaling}
\end{align}
where $c$ is the central charge and $\gamma$ the topological entanglement
entropy. The slope of the dependence of the entanglement entropy on the scaling function  entering Eq.~(\ref{eq:scaling}) can thus
be used to extract the central charge of the system. In gapped systems, the
entanglement entropy saturates, i.e.,  $c=0$.

\begin{figure}
	\centering
	\includegraphics[width=.9\columnwidth]{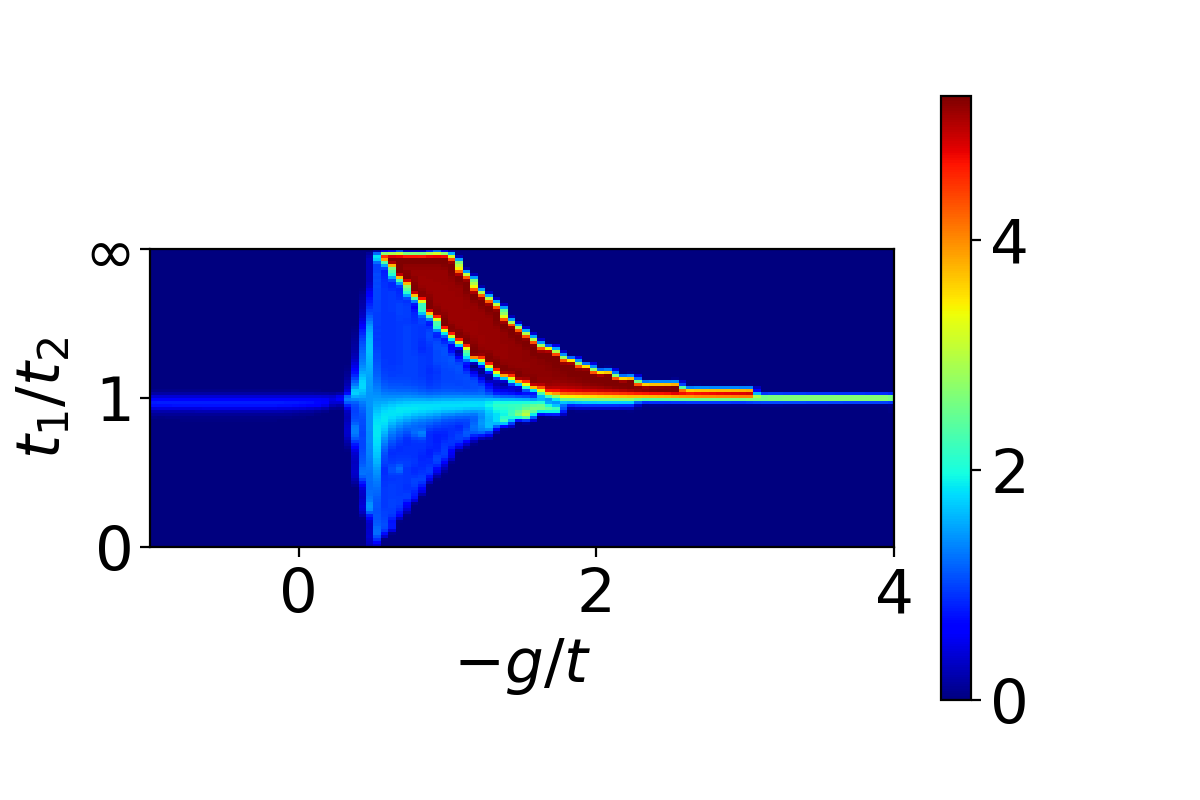}
	\caption{Central charge $c$ of the clean interacting Majorana chain vs interaction
		strength $g_1/t_1 = g_2/t_2 \equiv g/t$ and staggering $t_1/t_2=g_1/g_2$.
		On the self-dual line (no staggering, i.e.,  $t_1/t_2=1$), the results agree with  Ref.
		\onlinecite{rahmani_phase_2015}:
		the central charge is $c=\frac12$ for  $-g/t \lesssim 0.28$ and is then
		$c=\frac32$ until the system becomes gapped at strong repulsive interaction, $- g/t > 2.9$.
		In the Ising phase, the system is gapped everywhere apart from the critical line (i.e., by any staggering $t_1/t_2 \ne 1$).
		On the other hand, in the Ising+LL phase, adding staggering produces an extend critical region with $c=1$, see also a schematic phase diagram in Fig.~\ref{fig:phase_skizze}.
		The red patch is a peculiar  region where determination of $c$ by means of Eq.~(\ref{eq:scaling}) breaks down, see Appendix \ref{ap:red_patch} for more detail. In fact, this phase is gapped (as is also clear by inspecting its dual, $t_1/t_2 \to t_2/t_1$), i.e., the properly defined central charge is zero.}
	\label{fig:phase_cc}
\end{figure}

\begin{figure}
	\centering
	\includegraphics[width=.9\columnwidth]{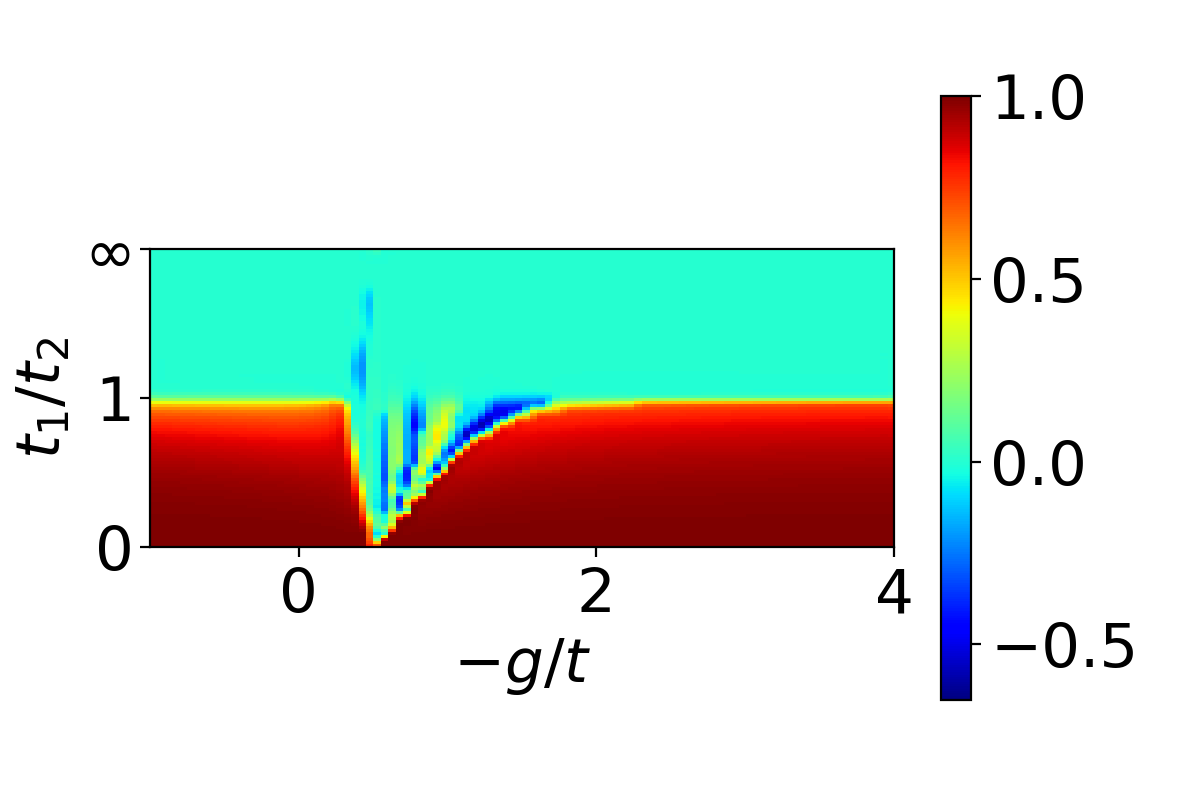}
	\caption{The $\langle \sigma^z_{24}\sigma^z_{48}\rangle$ correlator between
		spins on the sites $i=24$ and $i=48$ for the clean interacting Majorana chain
		as a function of interaction strength $g_1/t_1 = g_2/t_2 \equiv g/t$ and
		staggering $t_1/t_2=g_1/g_2$. In the gapped phases (cf. Figs.
		\ref{fig:phase_cc} and Fig.~\ref{fig:phase_skizze}) the correlator is equal to
		zero above the self-duality line and to unity below this line, thus helping to
		distinguish two topologically distinct phases.  In the critical region with
		$c=1$ around the Ising + LL phase  the correlator shows an oscillatory
		behavior, cf. Fig. \ref{fig:cleanweak}, right panels.}
		\label{fig:phase_sz}
\end{figure}

\begin{figure}
	\centering
	\includegraphics[width=1.00\columnwidth]{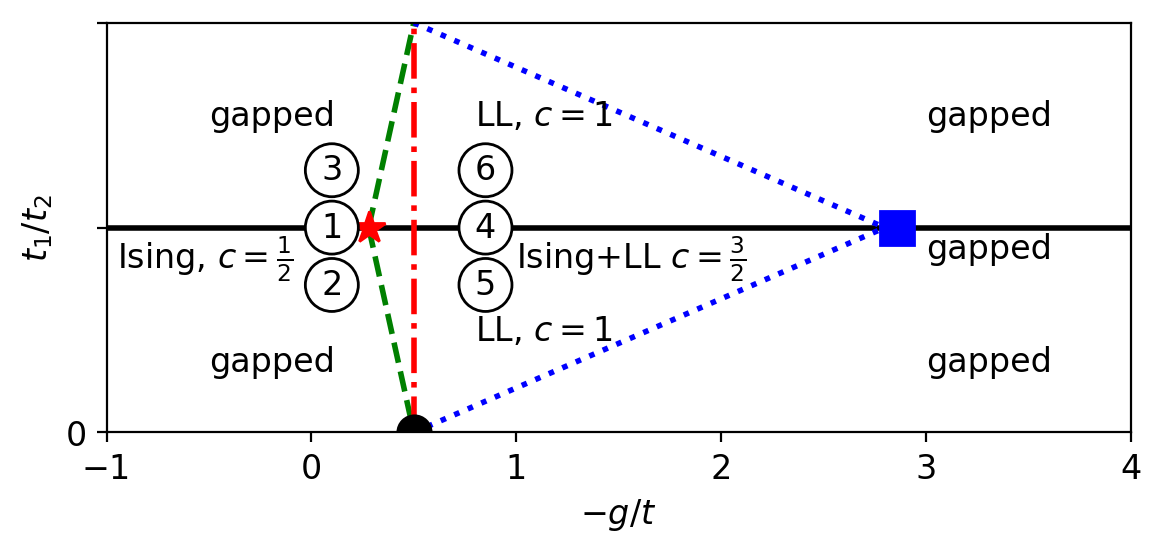}
	\caption{Schematic phase diagram of the clean interacting Majorana chain in
		the plane spanned by the interaction strength $g_1/t_1 = g_2/t_2 \equiv  g/t$
		and the staggering $t_1/t_2$. The labels from 1 to 6 correspond to the plots
		of the spin-spin correlator as a function of distance in Fig.
		\ref{fig:cleanweak} which are labeled in the same
		way. For $- g/t\lesssim 0.28$ the system on the self-dual line ($t_1/t_2 = 1$)
		is in the Ising phase with central charge $c=1/2$ (label 1). Introducing
		staggering yields two topologically distinct gapped phases (labels 2 and 3).
		At the point $- g/t\approx 0.28$ (marked by a red star) the system undergoes a
		Lifshitz transition into the Ising+LL phase with $c=3/2$ (label 4). This
		Ising+LL phase intersects our projection plane also in the vertical line at
		$-g/t=0.5$ (red dashed-dotted line). For intermediate interactions, a not too
		strong staggering leaves the system gapless but reduces its central charge
		down to $c=1$ (Luttinger liquid phases; bounded by green dashed and blue
		dotted lines, labels 5 and 6). These lines are drawn schematically, their
		exact form has not been determined. The black dot on the bottom of the diagram
		($g_1/g_2=t_1/t_2=0$ and $-g/t=0.5$) marks the first-order transition in the
		longitudinal Ising model. The blue square at $-g/t\approx 2.9$ on the
		self-duality line $t_1/t_2=1$ is the point of the transition to a gapped
		phase. The phase diagram is symmetric with respect to the duality
		transformation that  links points with the same value of $g/t$ and inverse
		values of $t_1/t_2$.
			}
	\label{fig:phase_skizze}
\end{figure}

In order to identify critical lines and regions, the central charge defined
according to Eq.~(\ref{eq:scaling}) is plotted in Fig. \ref{fig:phase_cc} via a
color map in the parameter plane spanned by the interaction strength $g/t$ and
the staggering $t_1/t_2$. Further, we show in a similar way in Fig.
\ref{fig:phase_sz} the long-range spin-spin correlation $\langle
\sigma^z_{L/4}\sigma^z_{3L/4}\rangle$. This plot helps  to differentiate between
topologically distinct  gapped regions. Figure \ref{fig:phase_skizze} provides
an overview over our results that are discussed in more detail below. In this
figure, numbers from 1 to 6 label different regions; the corresponding distance
dependences of spin correlations is shown (with the same labels) in Fig.
\ref{fig:cleanweak}. On the self-dual line, $t_1/t_2 = 1$, the range of
interaction strength $-4<g/t<1$ can be divided, in agreement with
Ref.~\onlinecite{rahmani_phase_2015}, into three intervals: the $c=\frac{1}{2}$
Ising phase for attractive and relatively weak repulsive interaction, $g/t >
-0.28$, the $c=\frac{3}{2}$ phase where the Ising sector coexists with a
Luttinger liquid sector for repulsive interaction in the interval
$-0.28>g/t>-2.9$, and a gapped phase for even stronger repulsive interaction,
$g/t < - 2.9$. This distinction remains useful also for understanding of phases
in the presence of staggering, as discussed below.

\subsubsection{Attractive and weak repulsive interaction}
\label{sec:clean-attractive}

In the absence of staggering, $t_1/t_2=1$, the system remains in the
non-interacting Ising phase for attractive interaction and for repulsive
interaction, $-g/t < 0.28$, as was found in
Ref.~\onlinecite{rahmani_phase_2015}. Indeed, we observe in Fig.
\ref{fig:phase_cc} that on the self-dual line the system is critical with a
central charge of $\frac12$ at this range of interactions. At finite staggering
the system is gapped, with two topologically distinct phases  (labeled 2 and 3
in Fig.~\ref{fig:phase_skizze}) that can be distinguished by the behavior of the
spin-spin correlator. For staggering $t_1/t_2=g_1/g_2>1$, which corresponds to
the topologically trivial phase in the fermionic picture, it decays quickly with
distance, see Fig. \ref{fig:phase_sz} and the top left panel of
Fig.~\ref{fig:cleanweak}. On the other hand,  in the symmetry-broken phase in
the spin language, $t_1/t_2=g_1/g_2<1$ (which is topologically non-trivial in
the fermion language), the correlator saturates at a constant value of order
unity at large distance, Fig. \ref{fig:phase_sz} and the bottom left panel of
Fig.~\ref{fig:cleanweak}. On the critical line $t_1/t_2=g_1/g_2=1$, the
correlator decays slowly (algebraically), as expected, see middle left panel of
Fig. \ref{fig:cleanweak}.

\subsubsection{Intermediate repulsive interaction}
\label{sec:clean-repulsive}

For stronger repulsive interaction $-0.28<g<-2.9$, the clean system without
staggering exhibits a Luttinger liquid sector in addition to the Ising sector as
has been already pointed out in Sec.~\ref{sec:clean-limit}. In this paper we
will call this phase ``Ising + LL'' phase, where ``LL'' stands for ``Luttinger liquid''. In
Ref.~\onlinecite{rahmani_phase_2015} this phase is called the ``floating''
phase, in analogy to a similar phase in the anisotropic next nearest neighbor
Ising model. It is characterized by a central charge of $c=\frac{3}{2}$. Our
numerical data in Fig. \ref{fig:phase_cc} confirm this behavior.

As Fig.~\ref{fig:phase_cc} demonstrates, the staggering
does not immediately lead to a gapped system in this interaction range. Instead, there is an extended region of
finite staggering with a central charge of $c=1$ around the no-staggering line. This can be understood as a result of the
Luttinger-liquid sector being stable to weak staggering, with the Ising sector becoming gapped. An argument based on RG analysis is given in Sec. \ref{SectionClean}. 
More precisely, there are two such phases with $c=1$, labeled 5 and 6  in Fig.~\ref{fig:phase_skizze}, which are separated by the line with $c=3/2$ (label 4).

\begin{figure}
	\centering
	\hspace*{-0.3cm}
	\includegraphics[width=.45\columnwidth]{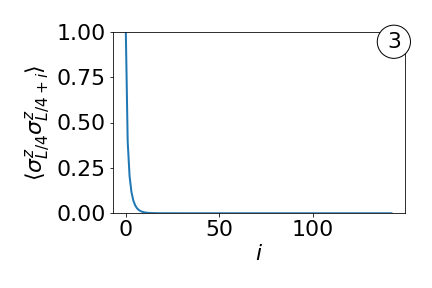}
	\includegraphics[width=.45\columnwidth]{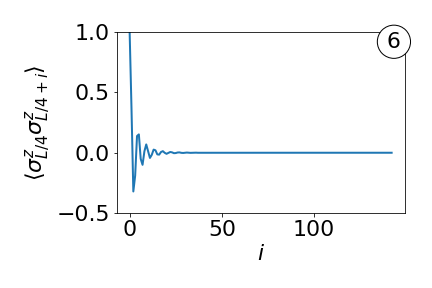}\\
	\includegraphics[width=.45\columnwidth]{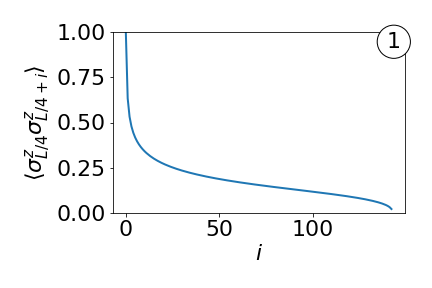}
	\includegraphics[width=.45\columnwidth]{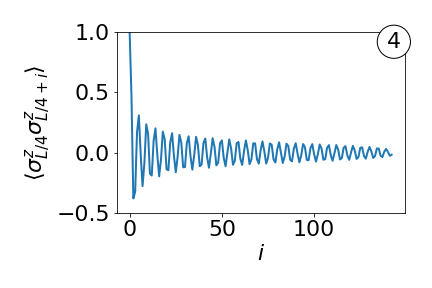}\\
	\includegraphics[width=.45\columnwidth]{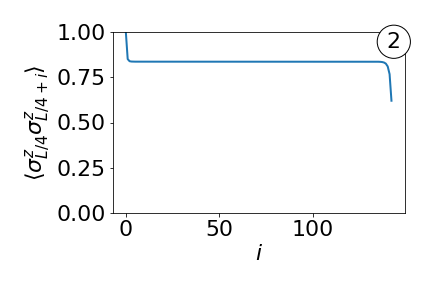}
	\includegraphics[width=.45\columnwidth]{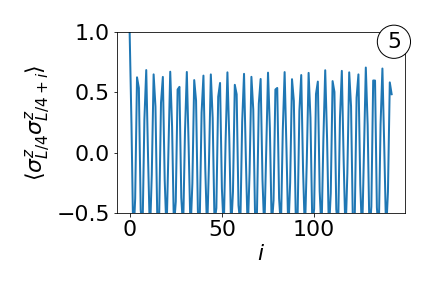}
	\caption{Spin-spin correlator $\langle\sigma^z_{L/4} \sigma^z_{L/4+i} \rangle$
	for the clean Majorana chain in spin formulation, Eq. \eqref{eq:spins}, at
	weak repulsive interaction $g/t= -0.10$ (left) and medium repulsive
	interaction $g/t=-0.85$ with no staggering, $t_1/t_2=1$ (middle), and
	staggering $t_1/t_2=1.39$ (top) and $t_1/t_2=0.72$ (bottom). The labels from 1
	to 6 correspond to those in Fig. \ref{fig:phase_skizze}. The system size is
	$L=190$, \ADDED{in the indices $L/4$ denotes the integer part $[190/4] = 47$.} 
	In the case of weak repulsive interaction, the correlator is
	strictly positive, while in the  case of medium repulsive interaction, the
	correlator oscillates as a function of distance and can take on negative
	values. On the self-dual line (middle), both correlators decay slowly
	(presumably algebraically) to zero. Above the self-dual line the correlators
	decay in both regimes quickly (presumably exponentially) to zero. Below the
  self-dual line, the correlator becomes constant for weak repulsive interaction
  and oscillates with a constant amplitude for medium repulsive interaction. 
  \ADDED{The drop of the correlator in the bottom left panel (with label 2)  near $i=3L/4$ (i.e., at the right end of the curve) is a boundary effect.}
  }
	\label{fig:cleanweak}
\end{figure}

In these extended critical regions, the spin-spin correlator is an oscillating
function of distance, as detailed in Fig. \ref{fig:cleanweak}. The oscillation
decay above the no-staggering line (label 6, top right panel), while their amplitude
remains constant below this line (label 5, bottom right panel).  On the line without
staggering, the oscillations decay very slowly (label 4, middle right panel). The
non-decaying oscillation in the extended critical region below the self-dual
line are also visible  in Fig. \ref{fig:phase_sz}.

At extreme staggering $t_1/t_2=0$, the model reduces to the longitudinal Ising
model. This model exhibits a first order transition at the point $g/t=0.5$.
The critical region with central charge $c=1$ is separated from the gapped
region of the Ising phase by a line connecting this point ($g/t= 0.5$ and
$t_1/t_2=g_1/g_2=0$; marked by a black dot in Fig. \ref{fig:phase_skizze}) with
the point of the Lifshitz transition on the critical line ($g/t\approx -0.28$
and $t_1/t_2=g_1/g_2=1$; marked by a red star in Fig. \ref{fig:phase_skizze}).
Additionally, there is a vertical critical line (red) connecting the black dot to its dual.
This line is also clearly visible in the picture of the central charge, Fig.
\ref{fig:phase_cc}, as it has a central charge of $c=\frac32$. 

\ADDED{At variance with the horizontal $c=\frac32$ line that is determined by the condition of no staggering, the vertical $c=\frac32$ line is not fixed by any simple symmetry. We have thus performed additional checks to verify its position. First, in order to exclude finite-size effects, we have considered twice larger systems ($L=192$) in this part of the phase diagram. The results demonstrated that neither the obtained value $c=\frac32$ nor the position of the line change with $L$. This implies that the vertical  $c=\frac32$ line is indeed a property of the system in the thermodynamic limit. Second, we have looked more carefully at the precise location of the line and found that it is not exactly at $-g/t=0.5$, although very close to it. As an example,  we find that the $c=\frac32$ line crosses the horizontal line $t_1/t_2=0.72$ at $-g/t \approx 0.45$. This indicates that the ``vertical'' $c=\frac32$ line is not exactly straight but rather shows a small deviation from the line  $-g/t = 0.5$. }

 Analogous to the horizontal (no-staggering) critical line, the value $c=\frac32$ can be understood as a superposition of a Luttinger liquid ($c=1$) and a Majorana mode ($c=\frac12$) due to a topological phase boundary. 
\ADDED{
To shed light on the reason for the emergence of the vertical $c=\frac32$ line, we have performed a mean-field analysis by generalizing that of Ref.~\onlinecite{sen_critical_1991} to our problem. In this way, we have approximately mapped an interacting fermionic Hamiltonian to a non-interacting (mean-field) one and obtained the condition for gap closing. This condition yields a two-dimensional surface in the whole (three-dimensional) space of parameters ($t_2/t_1$, $g_1/t_1$, and $g_2/t_1$). The surface can be computed numerically. We observe numerically that this two-dimensional surface intersects the two-dimensional surface determined by the condition $t_1/t_2=g_1/g_2$ (that is used in our DMRG numerics) on two lines -- the horizontal and the vertical ones.  The numerically obtained position of the vertical line is close to $-g/t=0.5$.  With the superimposed extended Luttinger liquid phase, we have $c=\frac32$ on these lines. In analogy with the horizontal line, the vertical line corresponds to the gap closing in the Ising sector, which corresponds to a phase boundary between topologically distinct phases. }

Another interesting point is the red patch appearing in the upper plane
seemingly violating the duality of the model. This is more than a numerical
artifact and has to do with corrections to the scaling form of the entanglement
entropy \eqref{eq:scaling} in gapped phases. We refer to Appendix
\ref{ap:red_patch} for a more detailed discussion.

\subsubsection{Strong repulsive interaction}

With increasing strength of repulsive interaction $-g/t$, the extended critical region around the no-staggering line gradually shrinks, see Fig. \ref{fig:phase_cc}.
For sufficiently strong interaction $-g/t > 2.9$ this region vanishes and, moreover, the line of no-staggering becomes gapped.

\subsection{Interacting Majorana chain with disorder}
\label{sec:majorana-int-dis-dmrg}

We now introduce disorder in the interacting Majorana chain model by choosing
hopping $t_j$ as random independent variables, with a homogeneous distribution over the interval $[0.5,1.5]$.
All hopping matrix elements have now the same distribution, so that there is no staggering.

\ADDED{In general, critical lines can move in phase space as function of disorder strength \cite{Gergs_Topological_2016,mcginley_robustness_2017}. 
However, the critical line at no staggering is pinned by self-duality. Therefore,  it should remain critical in the presence of both disorder and interaction unless spontaneous symmetry breaking takes place, see a more detailed discussion in Sec.~\ref{majorana-disorder-weak-repulsive-interaction} below. 
}

Since the average value of the hopping matrix elements is unity, the value of the interaction $g$ has now the same meaning as $g/t$ in the analysis of the clean system.
We consider three different ranges of interaction strength: (i) attractive interaction
$0<g<250$, (ii) weak repulsive interaction $0>g>-0.28$ and (iii) medium repulsive
interaction $-0.28>g>2.86$. We calculate the effective central charge in these regions of interaction  by analyzing the disorder-averaged
entanglement entropy via Eq. (\ref{eq:scaling}).

\subsubsection{Attractive interaction}

\begin{figure}
	\centering
	\includegraphics[width=.45\columnwidth]{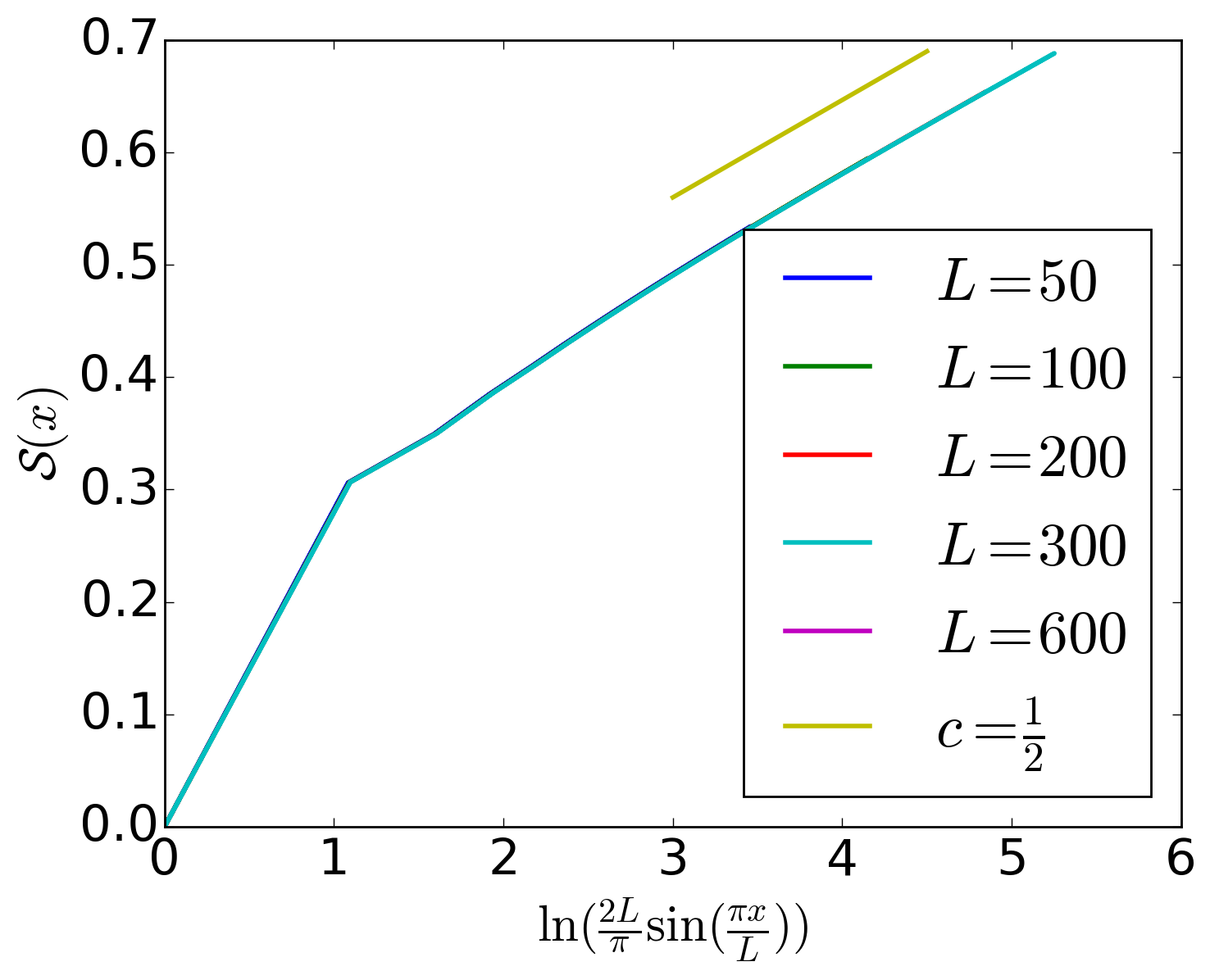}
	\includegraphics[width=.45\columnwidth]{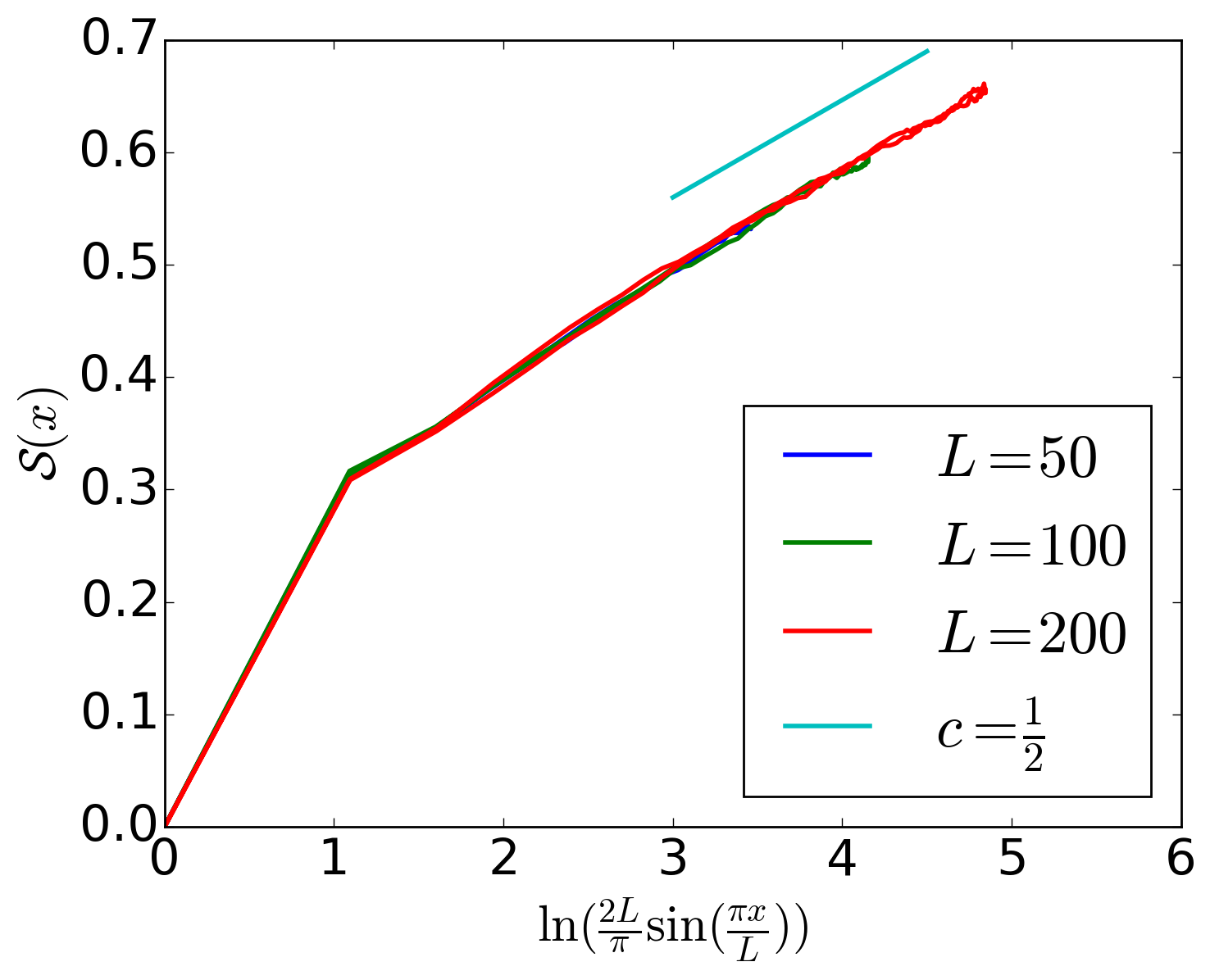}
	\caption{Entanglement entropy of the clean (left) and disordered (right) Majorana
		chain with attractive interaction $g=1$ vs the scaling function Eq.
		(\ref{eq:scaling}) for different system sizes. For the clean system, the central
		charge is $c=\frac{1}{2}$ in agreement with Ref.~\onlinecite{rahmani_phase_2015}. For the
		disordered system, the effective central charge is also found to be  $c=\frac{1}{2}$.}
	\label{fig:att_ent}
\end{figure}

For attractive interaction $0<g<250$, the clean system is in the Ising
phase\cite{rahmani_phase_2015} with a central charge of $\frac12$, see Sec.~\ref{sec:clean-attractive} and left panel of Fig.
\ref{fig:att_ent}. On the other hand, the disordered non-interacting system has an effective
central charge of $c_\text{eff}=\frac{\ln 2}{2}\approx 0.35$ as was found
in Ref.~\onlinecite{refael_criticality_2009}. Our numerics confirms this value.

Remarkably, in the presence of both disorder and interaction, the central
charge returns to the value of the clean system $c_\text{eff}=\frac{1}{2}$, see
Fig. \ref{fig:att_ent} (right panel).  For higher attractive interaction, the disorder
averaging requires less samples in order to give a smooth function of the
entanglement entropy vs scaling function than for lower interaction. This
serves as an additional indication that disorder does not play an important role for the Majorana chain with
attractive interaction.

\subsubsection{Weak repulsive interaction}
\label{majorana-disorder-weak-repulsive-interaction}

\begin{figure}
	\centering	\includegraphics[width=.45\columnwidth]{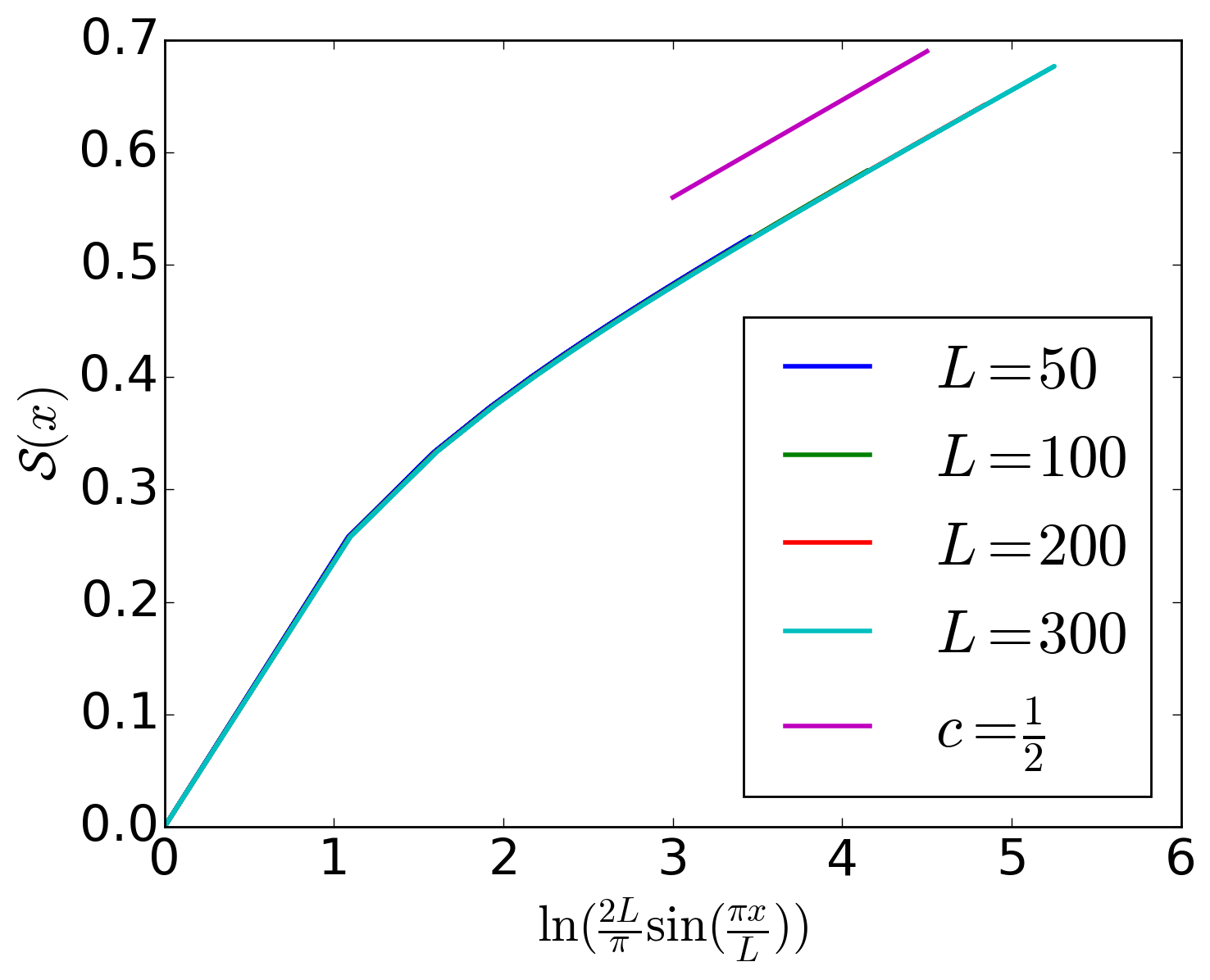}
	\includegraphics[width=.45\columnwidth]{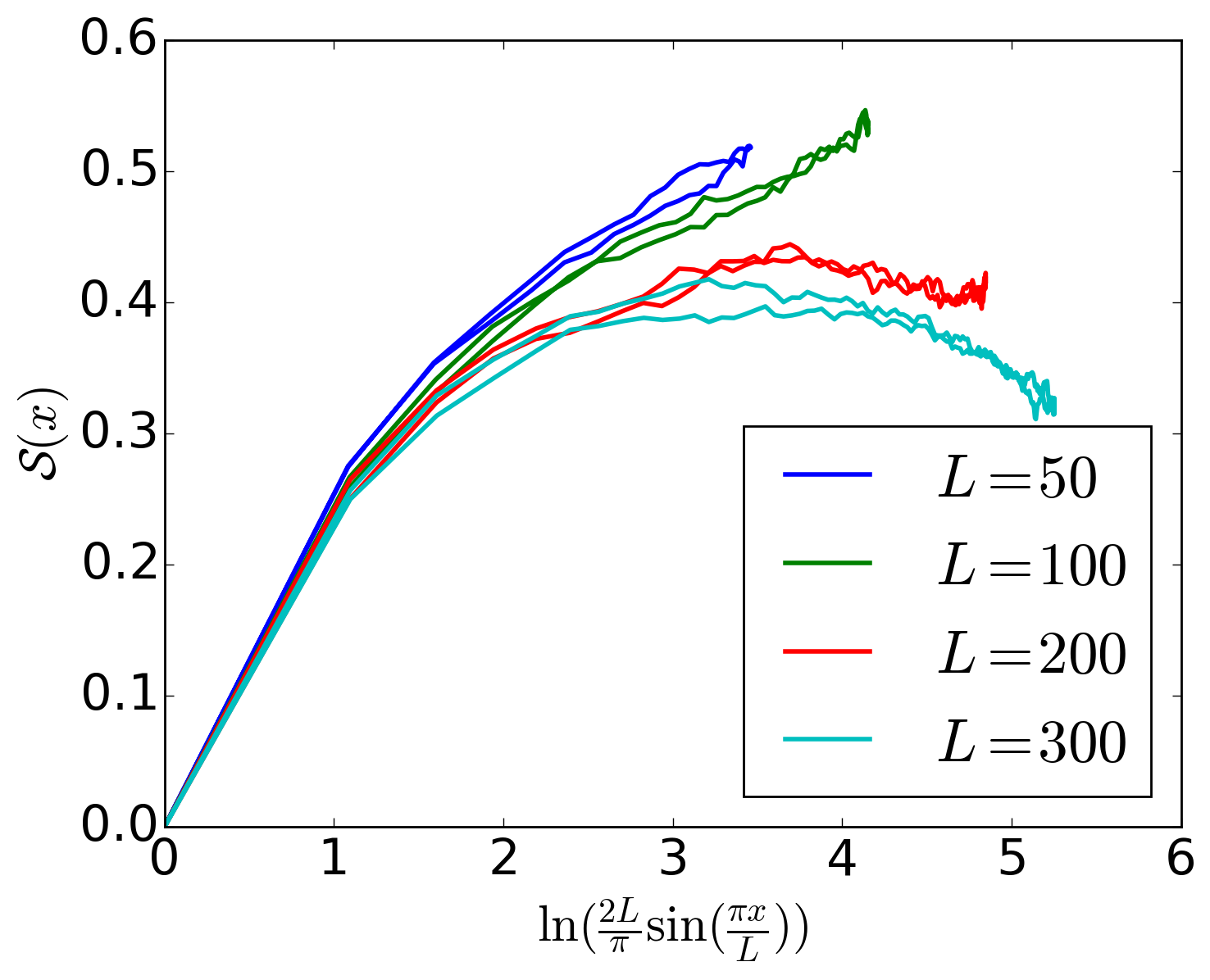}
	\caption{Entanglement entropy of the clean (left) and disordered (right) Majorana
		chain with weak repulsive interaction $g=-0.1$ vs the scaling function Eq.
		(\ref{eq:scaling}) for different system sizes. In the clean system, the central
		charge stays at $c=\frac{1}{2}$, while in the disordered case the entanglement entropy
		saturates indicating localized behavior.}
	\label{fig:weak_ent}
\end{figure}

The clean system stays critical with $c=\frac{1}{2}$ for weak repulsive
interaction \cite{rahmani_phase_2015}, $-0.28 < g <0$, see
Sec.~\ref{sec:clean-attractive} and the left panel of Fig. \ref{fig:weak_ent}.
We find that adding disorder leads to localization, see right panel of Fig.
\ref{fig:weak_ent}. This appears to happen for arbitrarily weak repulsive
interaction and arbitrarily weak disorder. Due to duality, the critical lines
have to be mirror symmetric around the self-dual line with respect to
staggering. This holds also when the system is disordered. For this reason, the
critical line cannot simply bend away from the self dual line. If the system
localizes on the self-dual line, there are therefore two possibilities: (i) the
critical line splits up into two lines with equal central charge, leaving a
gapped region around the self-dual line, or (ii) the critical line terminates,
and the transition between the region above and below the self-dual line becomes
first order. It is shown in Appendix \ref{sec:majorana-disorder-mean-field} by treating the
interaction at the mean-field level that the criticality  is pinned
to the self-dual line for all interaction values and disorder strengths. This
excludes the option (i), thus implying that the possibility (ii) is realized.

\begin{figure}
	\centering
	\includegraphics[width=.45\columnwidth]{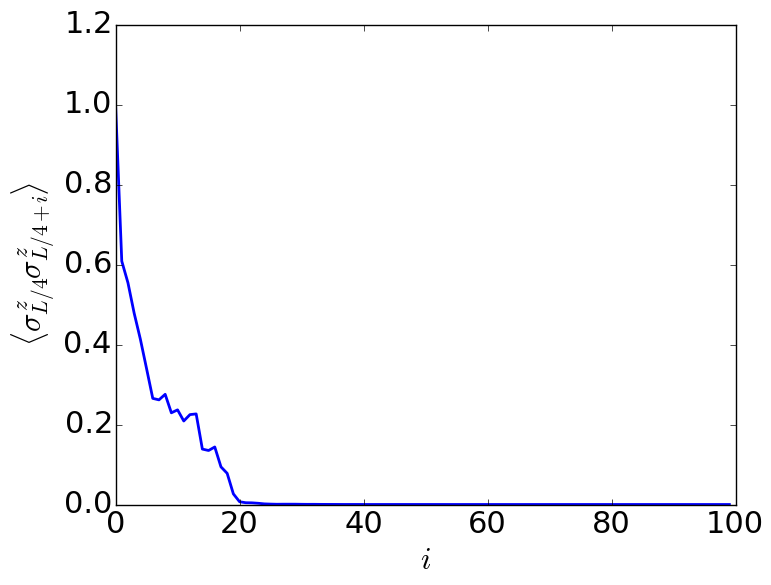}
	\includegraphics[width=.45\columnwidth]{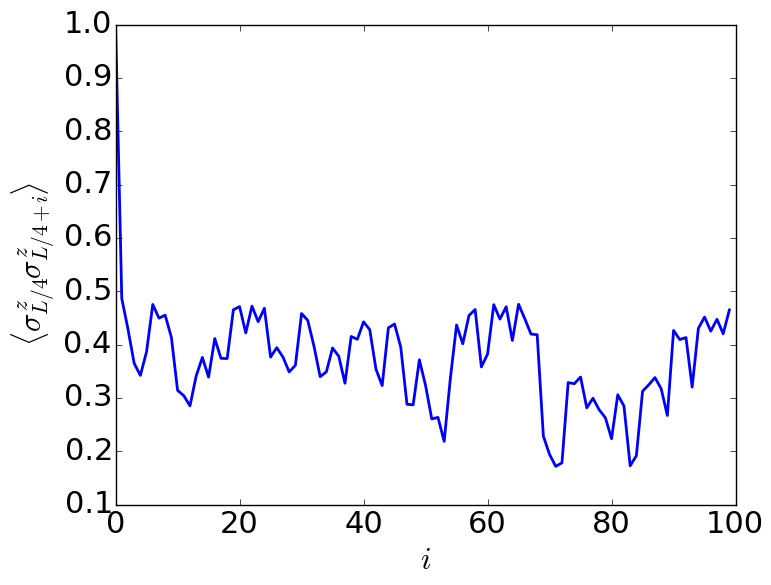}
	\caption{Spin-spin correlator $\langle\sigma^z_{L/4} \sigma^z_{L/4+i} \rangle$
	of the Majorana chain with weak repulsive interaction $g=-0.1$ at length
	$L=200$. The two panels represent two different disorder configurations. In
	the left panel, the correlator decays quickly to zero, which is analogous to
	the behavior in the presence of staggering $g_1/g_2=t_1/t_2>1$, see top right
	panel of Fig. \ref{fig:cleanweak}. In the right panel, the correlation
	function fluctuates, staying of order $O(1)$. This is similar to the region
	with staggering $g_1/g_2=t_1/t_2<1$, see bottom right panel of Fig.
	\ref{fig:cleanweak}. This behavior reflects the fact that disorder breaks
	spontaneously the symmetry with respect to duality transformation, placing the
	system in one of two topological phases.}
	\label{fig:disweak}
\end{figure}

We thus conclude that, for a disordered system with repulsive interaction, the
symmetry gets spontaneously broken, and the system undergoes a first-order
transition on the self-dual line. This is also reflected in the distance
dependence of the spin correlation function. Specifically, we find that,
depending on the disorder configuration, this correlation function shows one of
two types of behavior: it either very quickly decays to zero or fluctuates
around a value of order unity. This is illustrated in Fig. \ref{fig:disweak}
where the results for two disorder configurations are shown. These two types of
behavior correspond to two topologically distinct phases, as is clear from the
comparison of two panels of Fig. \ref{fig:disweak} with the top left and bottom
left panels of Fig. \ref{fig:cleanweak}. In the latter figure, the topologically
distinct phases were induced by staggering (in a clean model) breaking
explicitly the symmetry with respect to the duality transformation. We now see
that adding disorder breaks spontaneously the symmetry of the system on the
no-staggering line, placing it into one of the two  topologically distinct
phases. The transition between these two topological phases becomes thus first
order.

\subsubsection{Medium repulsive interaction}

\begin{figure}
	\centering
	\includegraphics[width=.45\columnwidth]{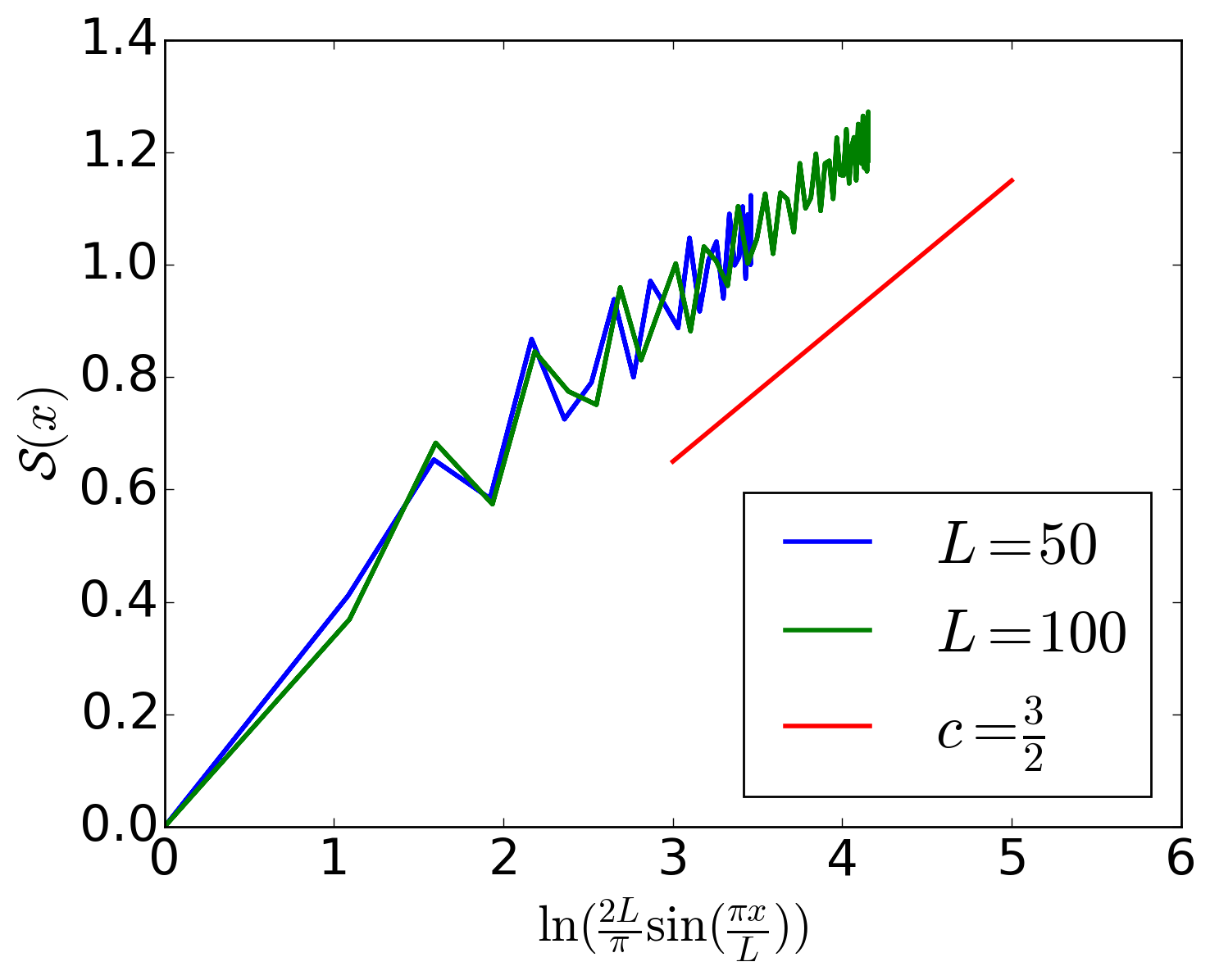}
	\includegraphics[width=.45\columnwidth]{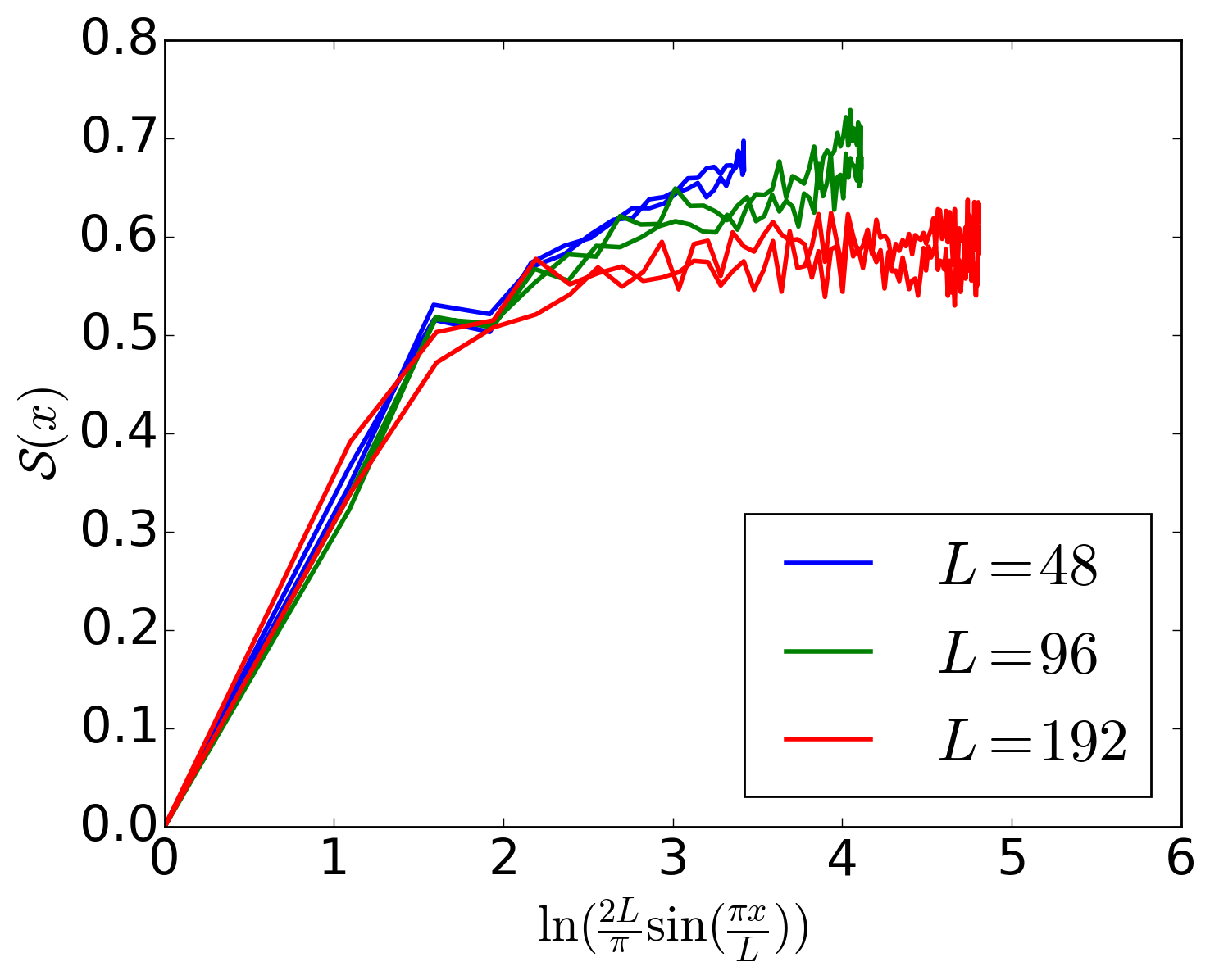}
	\caption{Entanglement entropy of the clean (left) and disordered (right) Majorana
		chain with medium repulsive interaction $g= - 0.5$ vs the scaling function Eq.
		(\ref{eq:scaling}). The central charge of the clean system is $\frac{3}{2}$ as predicted
		\cite{rahmani_phase_2015}. On the other hand,  the entanglement entropy saturates for
		the disordered case, implying localization. }
	\label{fig:med_ent}
\end{figure}

If the repulsive interaction is in the interval $-2.86<g<-0.28$ the clean system
is in the Ising+LL\cite{rahmani_phase_2015}
phase which is characterized by a
central charge of $\frac{3}{2}$, see Sec.~\ref{sec:clean-repulsive} and left panel of Fig. \ref{fig:med_ent}.
Our results show that, similar to
the case of weak repulsive interaction, disorder leads to localized behavior also in this range of interaction,
see right panel of Fig. \ref{fig:med_ent}. This was also found in Ref.
\onlinecite{milsted_statistical_2015}.

\begin{figure}
	\centering
	\includegraphics[width=.45\columnwidth]{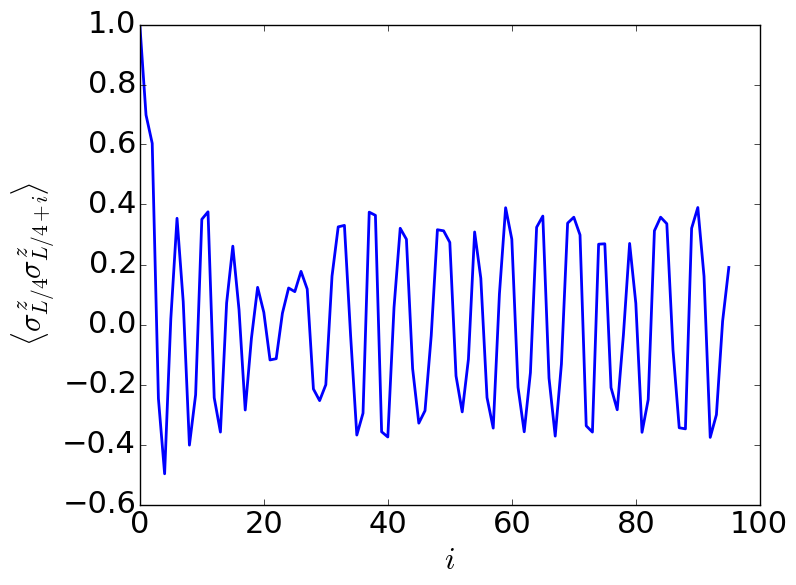}
	\includegraphics[width=.45\columnwidth]{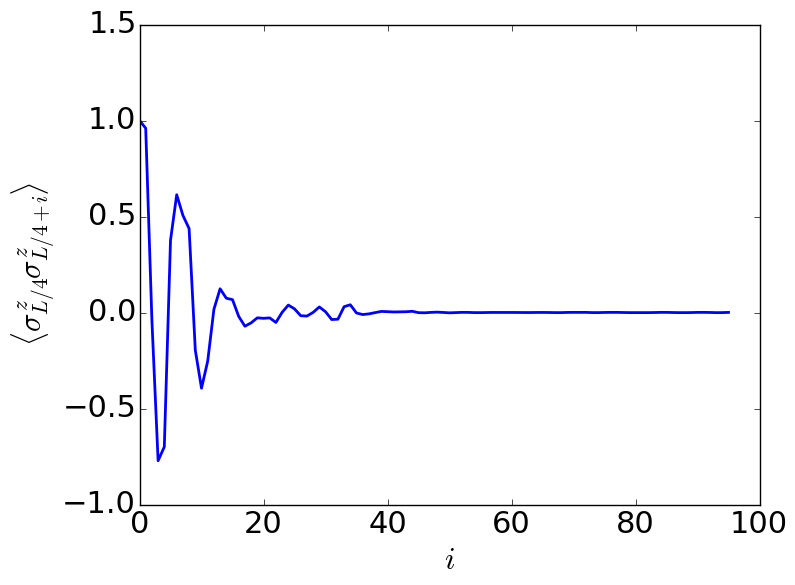}
	\caption{Spin-spin correlators $\langle\sigma^z_{L/4} \sigma^z_{L/4+i}
	\rangle$ of the Majorana chain for medium repulsive interaction $g=-0.5$ with
	length $L=200$. Two panels show results for two different disorder
	configurations that lead to vastly different behavior. In the plot for the
	first disorder configuration, the spin correlator oscillates around zero with
	an amplitude essentially independent of distance. This behavior is analogous
	to the one induced by staggering in the region below the self-dual line, see
	bottom right panel of Fig. \ref{fig:cleanweak}. For the other disorder
	configuration, the spin-spin correlator oscillates and quickly drops to zero.
	This behavior corresponds to the one induced by staggering in the region above
	the no-staggering line, see top right panel of Fig. \ref{fig:cleanweak}. The
	disorder thus breaks spontaneously the symmetry between the two topologically
	distinct phases. In both phases, the correlator takes negative values for some
	distances,   at variance with the case of weak repulsive  interaction,
	Fig.~\ref{fig:disweak}.}
	\label{fig:disfloating}
\end{figure}

As in the case of weak repulsive interaction, the spontaneous symmetry breaking
by disorder can be visualized by inspecting the spin-spin correlation function
for individual realizations of disorder. We find again two distinct types of
behavior that are illustrated in Fig. \ref{fig:disfloating}: oscillations
without decay or with a quick decay. The behavior shown in the left panel of
Fig. \ref{fig:disfloating} corresponds to that in the clean model in the
Ising+LL phase with staggering $g_1/g_2=t_1/t_2<1$, see bottom right panel of
Fig. \ref{fig:cleanweak}, while the behavior shown in the right panel of
Fig. \ref{fig:disfloating} corresponds to that in the clean model with
staggering $g_1/g_2=t_1/t_2>1$, see top right panel of Fig.
\ref{fig:cleanweak}. Thus, the symmetry between the two topological phases
gets broken spontaneously by disorder in full analogy with the
weak-repulsion regime. A comparison of Figs. \ref{fig:disweak} and
\ref{fig:disfloating} reveals an interesting difference between the
weak-repulsion and intermediate-repulsion topological phases. Specifically, in
the latter case the correlator shows oscillations around zero, thus changing
sign.

\subsection{Disordered Fermionic chain}

\begin{figure}
	\includegraphics[width=.45\columnwidth]{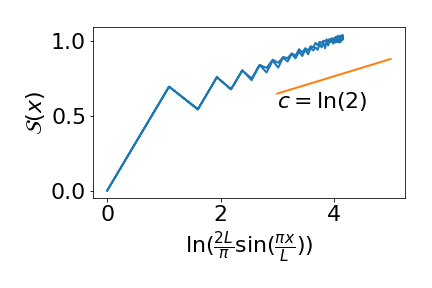}
	\includegraphics[width=.45\columnwidth]{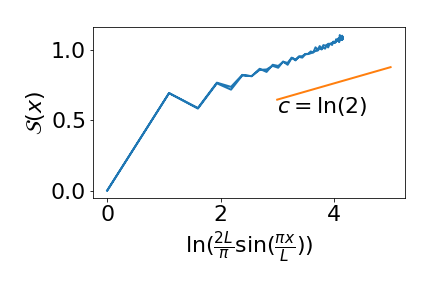}\\
	\includegraphics[width=.45\columnwidth]{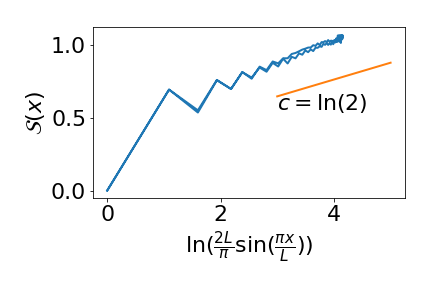}
	\includegraphics[width=.45\columnwidth]{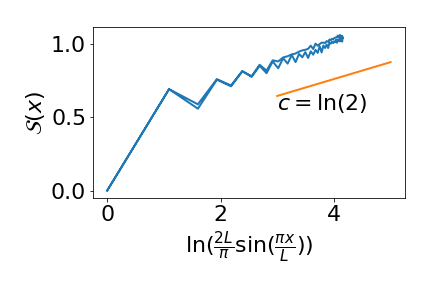}
	\caption{Disorder-averaged entanglement entropy vs scaling function for an
	interacting fermionic system with Hamiltonian Eq. (\ref{eq:ham_fer}) and
	parameters $L=100$, random hoppings $t_j$ drawn from the uniform distribution
	over $[0.5,1.5]$, as calculated by DMRG. Left panels: attractive  interaction
	$g=0.1$, Right panels: repulsive interaction $g=-0.1$.
	The interaction
	distance is $r=1$ (top panels) and $r=2$ (bottom panels). The scaling of the
	entanglement entropy corresponds to the value of the central charge $c=\ln 2$,
	as for a non-interacting system. This indicates that the interaction term is
	irrelevant in the RG sense.
  }
	\label{fig:fermi_dmrg}
\end{figure}

We turn now to the DMRG results for a disordered interacting fermionic chain,
Eq. (\ref{eq:ham_fer}). The results for the entanglement entropy are shown in
Fig. \ref{fig:fermi_dmrg} for the cases of odd ($r=1$) and even ($r=2$)
interaction distances. We observe that the \ADDED{(sufficiently weak)} interaction does not modify the
behavior of the disordered system: both for $r=1$ and $r=2$ the interacting
system remains critical and has the central charge $c=\ln 2$ characteristic for
the infinite-randomness fixed point. This implies the RG-irrelevance of the
interaction.

\section{Renormalization group around the clean fixed point}
\label{SectionClean}

Numerical results of Sec.~\ref{sec:majorana-int-dis-dmrg} for a disordered interacting Majorana chain indicate that in the presence of disorder an interaction
of either sign becomes relevant. To get the corresponding analytical insight, one has to consider a model with both interaction and disorder, which is an extremely challenging problem. In this Section we approach this problem by starting from a clean interacting Majorana chain and exploring the effect of weak disorder.

The stability of the clean fixed points of the interacting fermionic and Majorana models
can be probed by a weak-disorder momentum-space RG analysis.
For this purpose, we consider the low-energy theory in the continuum limit. In the case
of the complex fermionic chain, this is a Luttinger liquid (LL) theory. In the
Majorana case, it is either a Majorana theory ($c=\frac12$, Ising phase) or a
Majorana theory with an additional LL sector ($c=\frac32$, Ising +LL phase),
depending on the interaction strength. The density-density parts of the
interaction are quadratic in Luttinger theory and renormalize the Luttinger
parameter $K$.

In these continuum theories, disorder appears as a random-mass term. Choosing nonzero
average of the mass or a constant non-vanishing mass corresponds to
staggering.  By including such terms, one can draw conclusions
about the stability with respect to staggering, which is another goal of the present section.
This should help understanding the appearance of
extended gapless phases that were found by DMRG numerical analysis in Sec. \ref{sec:majorana-clean-dmrg}.

We will show below that at any of the  fixed points of the clean Majorana chain (Ising or Ising + LL), the disorder becomes relevant and
flows to the strong-coupling regime. This happens also for the complex-fermion fixed point (Luttinger liquid) if the interaction is not too strong. This will lead us to the complementary analysis in Sec. \ref{sec:SectionIRFP}, where we treat disorder exactly and the interaction as a perturbation.

\subsection{Majorana: $c=1/2$ fixed point}
\label{majorana-12-fixed-point}

The continuum decomposition in slow modes $\gamma_{R/L}$ of the lattice Majorana
operators $\gamma_j$ is
\begin{align}
\gamma_j &= \gamma_R + (-1)^j \gamma_L.
\end{align}
For a Majorana low energy theory disorder  corresponds to a
random-mass term of the form:
\begin{align}
S_m^{\mathrm{maj}} &= \int \dd \tau \dd x \; m(x) \gamma_R(\tau,x)\gamma_L(\tau,x).
\end{align}
A constant mass $m(x) = m_0$ corresponds to a staggering; it directly opens a gap of size $m_0$.

The disorder is assumed to be Gaussian white noise with  $\langle m(x) m(y) \rangle = D\delta(x-y)$; one can also include a staggering by introducing a non-zero mean $\langle m(x) \rangle = m_0$. Treating the disorder by using the replica trick, one straightforwardly finds that the term
generated by disorder has (upon disorder averaging)  the scaling dimension $1$ and is therefore relevant in the RG sense. This term
drives the system away from the clean fixed point. However, this does not necessarily
mean that the system becomes gapped. For example, in the non-interacting case (and in the absence of staggering) the system flows to the critical infinite-randomness fixed point \cite{fisher_critical_1995}. It means, however, that an analysis based on RG around the clean fixed point is insufficient to understand the infrared physics of the problem and suggests a complementary approach as implemented in Sec.~\ref{sec:SectionIRFP}.

Finally it should be noted that no relevant
interaction term can be written down in a Majorana low-energy theory. Indeed, the interaction should involve at least four Majorana
operators with scaling dimension $\frac{1}{2}$ each and two derivatives with
dimension $-1$. The most relevant term thus has scaling dimension $-2$ and is
strongly irrelevant.

\subsection{Complex fermions: Luttinger liquid ($c=1$) fixed point}
\label{luttinger-fixed-point} 

Lattice operators $c_j$ are related to their continuum versions $\psi_{R/L}$ as
follows
\begin{align}
c_j &= i^j\psi_R  + (-i)^j \psi_L.
\end{align}
In the presence of interaction $g\neq 0$, bosonization has to be employed. Here
the following conventions relating the fermionic fields $\psi_{R/L}$ to the
bosonic fields $\phi, \theta$ are used:
\begin{align}
\psi_{R/L} &= U_{R/L} \exp \left( \phi \pm \theta \right).
\end{align}
The Klein factors $U_{R/L}$ are not important in any of the following
considerations.

The exact dependence of the Luttinger parameter $K$ on the parameters of the
lattice model is known\cite{luther_calculation_1975}:
\begin{align}
g/t = -\cos \left( \pi/ 2 K \right).
\end{align}
Disorder and staggering introduce a mass term of the form:
\begin{align}
S_m^{\mathrm{LL}} &= \int \dd \tau \dd x \; m(x) (\psi_R^\dagger(\tau,x)\psi_L(\tau,x) + h.c.).
\end{align}
The scaling dimension of a constant mass term is $2-K$. This means that  it is
relevant for $K<2$, which corresponds, in terms of the microscopic parameters, to the interval $-1<
g/t < 0.7 $ covering almost the whole range of critical theories, $|g/t| < 1 $.

\begin{figure}
	\includegraphics[width=.45\columnwidth]{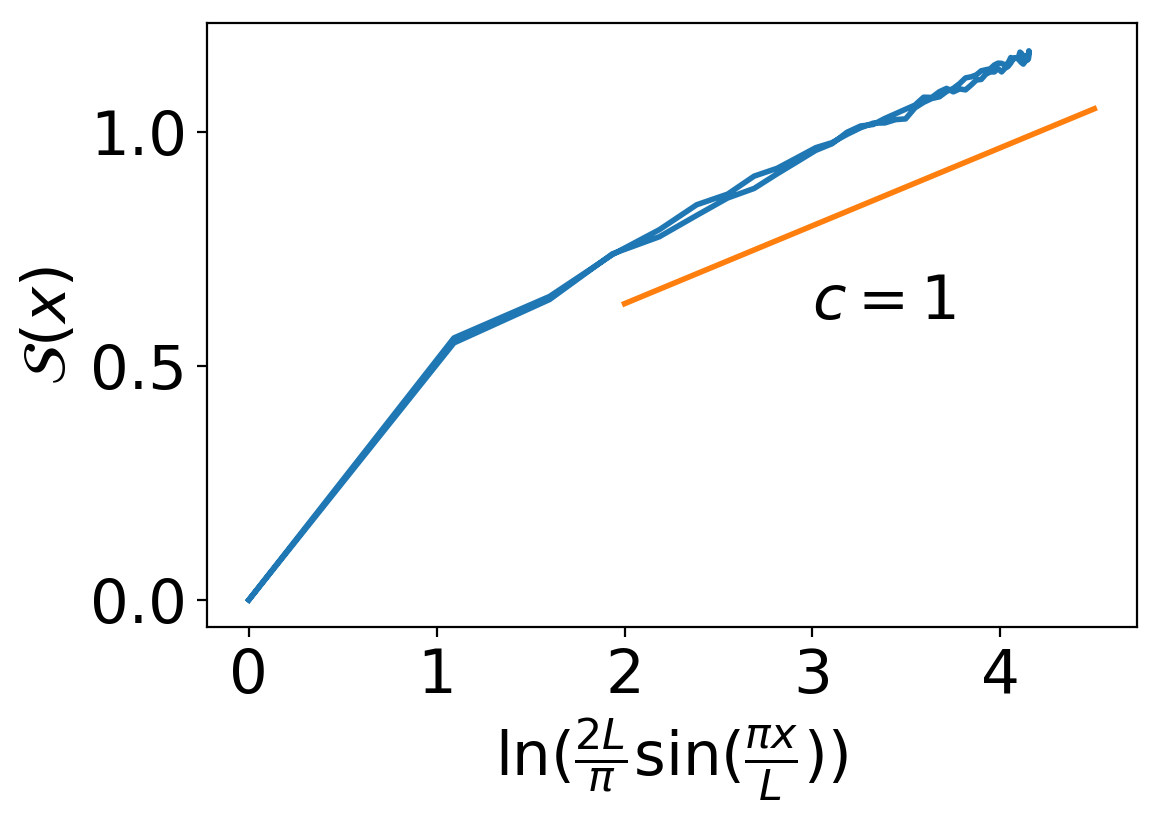}
	\caption{\ADDED{DMRG results for disorder-averaged entanglement entropy vs scaling function for an
			interacting fermionic system with Hamiltonian Eq. (\ref{eq:ham_fer}) and
			parameters $L=100$, random hoppings $t_j$ drawn from the uniform distribution
			over $[0.5,1.5]$. The interaction is attractive and strong, $g=0.8$, which distinguishes this figure from Fig.~\ref{fig:fermi_dmrg}.
			The scaling of the
			entanglement entropy corresponds to the value of the central charge $c=1$,
			as for a clean system.
			The system is at the Luttinger-liquid fixed point with $K>\frac32$ that is stable towards weak disorder, see discussion in Sec. \ref{luttinger-fixed-point}.}
	}
	\label{fig:lutt_dmrg}
\end{figure}

The scaling dimension of the quartic term generated by disorder, as  obtained by
the replica field-theory approach, is $3-2K$. It depends thus on the Luttinger parameter $K$
whether the disorder is relevant or not. Specifically, for $ g/t < 0.5 $ the disorder is relevant, while for $0.5< g/t <
1 $ the model remains at the clean fixed point in the presence of weak disorder. \ADDED{We have checked the latter prediction by DMRG, see Fig. \ref{fig:lutt_dmrg} for the scaling of the entanglement entropy at strong attractive interaction, $g=0.8$, and sufficiently weak disorder.  We find $c=1$, as expected for the system at the Luttinger-liquid fixed point. }
Around the non-interacting limit, i.e. for $K$ sufficiently close to unity, the disorder is strongly relevant, as expected.

We also briefly discuss allowed interaction terms as perturbations to the Luttinger liquid fixed point. They are of three types. First, the density-density interaction is marginal and simply modifies the value of $K$. Second, terms that are of higher order in $\psi$ or contain gradients are strongly irrelevant. Finally, the
staggering yields sine and cosine terms that are relevant in a range of $K$ (in particular, around the weak-interaction point $K=1$). On the self-dual line, these latter terms are absent.

\subsection{Majorana chain: Ising+Luttinger liquid ($c=3/2$) fixed point}
\label{sec:ising-luttinger-weak-disorder-rg}

We turn now to the  $c=\frac{3}{2}$ fixed point of the clean Majorana chain that emerges in a range of medium-strength repulsive interactions, as discussed above.
It was suggested in Ref. \onlinecite{rahmani_phase_2015} that, at this
fixed point, the low-energy theory consist of Majorana and Luttinger-liquid
sectors, see also Sec.~\ref{sec:clean-limit} and \ref{sec:clean-repulsive}.
This can be understood by considering the quadratic form of the action including the
third-nearest-neighbor hopping which is generated by mean-field treatment of the interaction (or, alternatively, under RG flow):
\begin{eqnarray}
H&=&i\sum_j \left[t_j \gamma_j\gamma_{j+1} + t' \gamma_j\gamma_{j+3} \right].
\label{eq:rg:H0}
\end{eqnarray}
The third-nearest-neighbor hopping term modifies the dispersion such that there are now three Majorana modes, or, equivalently,
a fermionic mode emerge in addition to the Majorana mode.
The lattice Majorana operator $\gamma_j$ then has the following low-energy decomposition \cite{rahmani_phase_2015}:
\begin{align}
\gamma_j &= 2\gamma_{L}+2(-1)^{j}\gamma_{R}\nonumber\\
&+\exp(-i k_0 j)\Psi^{\dagger}_{L}+\exp(+i (k_0+\pi) j)\Psi^{\dagger}_{R} +h.c. \label{eq:lowdec}\;,
\end{align}
where $k_0$ is the effective Fermi momentum. The interaction $g\gamma_j\gamma_{j+1}\gamma_{j+2}\gamma_{j+3}$ generates now the density-density interaction of the fermions $\Psi_{R}$, $\Psi_{L}$. To treat this interaction exactly, we employ the  bosonization approach. Another interaction term couples the resulting Luttinger liquid to the Majoranas with strength $g'$, see Eq.~\eqref{eq:rg:int}.

Next, let us discuss the stability with respect to staggering. The kinetic term $\gamma_j\gamma_{j+1}$ has oscillatory components with
wave vectors $k_i =0$, $k_0$, $k_0+\pi$, $2k_0$, $2k_0+\pi$, and $\pi$.
A constant mass term $m(x)=m$ describing staggering couples to the $\pi$-component of the kinetic term:
\begin{align}
S_m &= \int\mathrm{d}^2r \left[-8m \gamma_L\gamma_R+4m\cos k_0 \cos 2\theta \right].
\end{align}
The Majoranas are then immediately gapped out. On the other hand, the cosine term in the Luttinger-liquid sector is relevant only for $K>\frac12$. There is therefore a region of the interaction strength where the Luttinger liquid is stable towards staggering. This explains the existence of the extended gapless phase with $c=1$ observed numerically, see Fig. \ref{fig:phase_cc} and the schematic phase diagram in Fig.~\ref{fig:phase_skizze}.

Now we analyze the effect of disorder that is treated as a weak perturbation. Combining the oscillatory components of the kinetic term $\gamma_j\gamma_{j+1}$ (with the six wave vectors listed above) with the corresponding Fourier components of the random mass yields non-oscillatory contributions. We get therefore six independent disorder couplings
$D_{k_i}$ that coincide at the beginning of the RG flow but renormalize differently. Details on implementation of the RG procedure are presented in Appendix \ref{AppendixRG}.  In Eq.~\eqref{eq:rg:terms}, the disorder-induced terms in the action (with the replica formalism used to average over disorder) are presented.  While the forward scattering $D_0$ cannot be gauged away straightforwardly, a more detailed calculation shows that it does not change the results presented here.

\begin{table}
	\centering
	\caption{The RG scaling dimension and relevance range of couplings in the low-energy theory of the  Ising+LL phase. Forward scattering is gauged away, see Appendix \ref{AppendixRG}. The five remaining (dimensionless) coupling constants
		corresponding to disorder are labeled $y_{k_i}$, where $k_i$ refers to the
		momentum component. The dimensionless interaction strength is denoted by $y'=g' a u^{-1}$, where $a$ is the lattice spacing and $u$ the LL velocity.  The clean Ising+LL phase of the Majorana chain is characterized by $K<1$ and remains stable with respect to coupling between the Ising and LL sectors as long as \cite{rahmani_phase_2015} $\frac{1}{4}<K<1$.}
	\centering
	\label{tab:couplings}
	\begin{tabular}{ccc}
		coupling & dimension & relevant in\\
		\hline
		$y_{k_0}$ & $2-\frac12(K+K^{-1})$ & $0.27 < K < 3.8$\\
		$y_{k_0+\pi}$ & $2-\frac12(K+K^{-1})$ & $0.27 < K < 3.8$\\
		$y_{2k_0}$ & $3-2(K+K^{-1})$ & $0 < K < 2$\\
		$y_{2k_0+\pi}$ & $3-2K$ & $K<1.5$\\
		$y_{\pi}$ & $3 - 2K^{-1}$ & $0.67 < K$\\
		$y'$ & $1-K^{-1}$ & $1<K$
	\end{tabular}
\end{table}

In Table \ref{tab:couplings}, we list the scaling dimensions of the disorder couplings resulting from the corresponding RG equations. They determine the range of $K$ in which the disorder-induced terms are RG-relevant. We observe that at least one of the couplings is relevant for any value of $K$, i.e. the disorder always drives the system aways from the clean fixed point.
In analogy with the conventional Giamarchi-Schulz RG \cite{giamarchi_anderson_1988}, the  RG equations for the disorder-induced couplings are complemented by the flow
equation for the Luttinger constant $K$:
\begin{align}
\deriv{K}{\ell} &= -\dfrac{1}{2}\left[K^2-\dfrac{(1+K^2)(3-2K)}{2}\right]y_{2k_0+\pi} \nonumber \\
&+\dfrac{1}{2}\left[1-\dfrac{(1+K^2)(3-2/K)}{2}\right]y_{\pi}
\label{eq:rg:end0}.
\end{align}
Here $y_{2k_0+\pi} = \pi^{-1}D_{2k_0+\pi}au^{-2}$ and $y_{\pi} = 16\cos ^2k_0D_{\pi}au^{-2}$ are dimensionless coupling constants for the disorder-induced terms with momentum component $k_i$ in terms of lattice spacing $a$ and Luttinger-liquid velocity $u$.
In Eq.~(\ref{eq:rg:end0}),  we have kept only the contribution of the couplings $y_{2k_0+\pi}$ and $y_{\pi}$ to the renormalization of $K$.
In principle, the other couplings $y_{k_i}$ also contribute to this renormalization; however, they are less relevant for $K$ around unity, so that we have neglected  their contributions.

A brief summary of main conclusions that we draw from this RG is as follows.  First, the Ising+LL clean fixed point is stable towards interaction.
Indeed, this phase is characterized by a repulsive interaction, hence $K<1$, so that the $y'$ coupling is irrelevant. In fact,  a higher order coupling between the LL and Majorana sectors becomes relevant for very strong interaction \cite{rahmani_phase_2015}, $K<1/4$, so that the range of stability in the absence of disorder is $1/4<K<1$.
Second, over an extended parameter regime, the staggering is irrelevant in agreement with the numerical results of Sec.~\ref{sec:clean-repulsive}, see Fig. \ref{fig:phase_cc}. Third, and most importantly, the disorder at the Ising+LL fixed point always runs to strong coupling. In other words, this fixed point is unstable with respect to disorder.

The results obtained in Sec.~\ref{SectionClean} demonstrate that the weak-disorder analysis is not sufficient for Majorana chain, both in the $c=1/2$ and $c=3/2$ phases of the clean system. The RG relevance of disorder is also supported by the analysis in Appendix \ref{sec:majorana-disorder-mean-field} where the exact treatment of disorder is combined with mean-field treatment of the interaction.  The disorder is also RG relevant for the complex-fermion chain if the interaction is not too strong.
These results motivate us to perform in Sec.~\ref{sec:SectionIRFP} a complementary analysis. We will start there from an exact treatment of disorder and will include interaction as a weak perturbation.

\section{Strong randomness fixed point:  Eigenfunction statistics and effect of interactions}
\label{sec:SectionIRFP}

In Sec. \ref{SectionClean}, we have seen that the combined effect of interaction
and disorder cannot be understood as a perturbation around the clean interacting
fixed point. Specifically, we have established that disorder is strongly
relevant at the clean fixed point, thus quickly increasing under  RG. We know
that, in the absence of interaction, this RG flow leads to the critical
infinite-randomness fixed point. It is thus a natural question whether this
fixed point is stable or not with respect to interaction. This question is
addressed in the present section. Our analysis has much in common with the
investigation of stability of 2D surface states of disordered topological
superconductors with respect to interaction
\cite{foster_interaction-mediated_2012, foster_topological_2014}. A closely
related physics controls the enhancement of superconducting and ferromagnetic
instabilities by disorder in 2D systems \cite{burmistrov_enhancement_2012,
burmistrov_superconductor-insulator_2015}.
Further, there are close connections with the analysis of the anomalous scaling
dimension of interaction in context of the study of decoherence and the
dynamical critical exponent at the quantum-Hall transition with short-range
interaction\cite{lee_effects_1996,wang_short-range_2000,burmistrov_wave_2011}.

In the clean system, the relevance or irrelevance of an operator can be often established by a relatively straightforward power
counting. As an example, this was done in Sec. \ref{SectionClean} to show that interactions are
RG-irrelevant at the clean fixed point of the Majorana chain. In the presence of disorder, the situation is much more complex,
since the multifractal nature of wavefunctions as well as a non-trivial scaling of the density of
states have to be taken into account. Formally, this disorder-induced renormalization of the interaction $U$ can be expressed by an
RG equation of the form\ADDED{\cite{foster_interaction-mediated_2012,foster_topological_2014}
\begin{align}
\frac{\mathrm{d}\ln U}{\mathrm{d}\ln L} &= x_1 - x_2^{(U)}. 
\label{eq:ir_rg}
\end{align}
Here $x_1$ is the scaling dimension of the density of states of a non-interacting system, with $x_1 >  0$ and $x_1 < 0$ corresponding to the cases of vanishing and diverging density of states, respectively.
Further, $x_2^{(U)}$ is the scaling dimension of the four-fermion interaction operator with respect to the non-interacting theory. For a detailed derivation of Eq.~(\ref{eq:ir_rg}) we refer the reader to Appendix C of Ref. \onlinecite{foster_topological_2014}. If the right-hand side of Eq. (\ref{eq:ir_rg}) is positive, the interaction is relevant at the non-interacting fixed point; otherwise it is irrelevant. 

For a short-range interaction, and in the case when cancellations of the Hartree-Fock type (see below) are not operative, the scaling dimension $x_2^{(U)}$ is equal to the dimension $x_2$ of the squared density of states (which is also a local four-fermion operator).  For the clean system $x_2$ is simply equal to $2x_1$ but for a disordered system one has in general $x_2 < x_1$ in view of multifractality (characterizing strong fluctuations of the density of states) \cite{evers_anderson_2008,foster_topological_2014, gruzberg_classification_2013}. 
Specifically, 
\begin{equation}
\label{x2}
x_2 = \Delta_2 + 2x_1,
\end{equation}
where $\Delta_2 < 0$ is the anomalous dimension of the fourth moment of the eigenfunction ($\langle U_{i\alpha}^4\rangle$ in the notations used below). 
In this situation of the maximally relevant interaction (no suppression due to Hartree-Fock cancellation or other reasons), Eq.~(\ref{eq:ir_rg}) takes the form
\begin{align}
\frac{\mathrm{d}\ln U}{\mathrm{d}\ln L} &= - x_1 - \Delta_2. 
\label{eq:x1D2}
\end{align}
The sum of two exponents $-x_1$ and $-\Delta_2$ in the r.h.s. of Eq.~(\ref{eq:x1D2}) determines the scaling with $L$ of the product $\nu C_H$ of the density of states $\nu$  and the Hartree-type correlation function $C_H$  [defined in Eq.~(\ref{corr-wave-func-H}) below] for $r=0$.

In general, $x_2^{(U)} \ge x_2$ since the effect of the interaction can be suppressed due to Hartree-Fock-type cancellation. In this generic situation, we have, in analogy with Eq.~(\ref{x2}), 
\begin{equation}
\label{x2U}
x_2^{(U)} = \Delta_2^{(U)} + 2x_1,
\end{equation}
where $\Delta_2^{(U)}$ is the anomalous dimension of the eigenstate correlation function $C_{HF}$ corresponding to the matrix element of the interaction (and thus taking into account possible Hartree-Fock-type cancellations; see, e.g., Eq.~(\ref{correlator-complex-fermions}) for the case of complex fermions below). Substituting Eq.~(\ref{x2U}) into 
Eq.~(\ref{eq:ir_rg}), we get 
\begin{align}
\frac{\mathrm{d}\ln U}{\mathrm{d}\ln L} &= - x_1 - \Delta_2^{(U)}. 
\label{eq:x1D2U}
\end{align}
The sum of the exponents $-x_1$ and $-\Delta_2^{(U)}$ in the r.h.s. of Eq.~(\ref{eq:x1D2U}) corresponds to the scaling with $L$ of the product $\nu C_{HF}$ of the density of states and the correlation function $C_{HF}$. Below we determine the explicit form of this correlation function by inspecting the expectation value of the interaction operator and analyze the scaling of the product $\nu C_{HF}$ with $L$ for the models of complex fermions (Sec.~\ref{sec:complex-fermion-correlations}) and for the Majorana model (Sec.~\ref{sec:majorana-correlations}). 

If $- x_1 - \Delta_2^{(U)} < 0$, the interaction is RG-irrelevant, i.e., the non-interacting fixed point is stable with respect to inclusion of not too strong interaction. In the opposite case, $- x_1 - \Delta_2^{(U)} > 0$, the interaction is RG-relevant and drives the system away from the non-interacting fixed point. It was found in previous works on the effect of interaction at critical points of higher spatial dimensionality ($d>1$, with a particular focus on 2D systems)\cite{foster_interaction-mediated_2012, foster_topological_2014,ostrovsky_interaction-induced_2010,burmistrov_enhancement_2012,lee_effects_1996,wang_short-range_2000,burmistrov_wave_2011} that both these scenarios can be realized. Whether the interaction is relevant or irrelevant depends on the specific non-interacting critical theory considered (i.e., spatial dimensionality as well as symmetry and topology class). As we show below, both scenarios are also realized in the context of the present work (1D critical systems of class BDI): the interaction is irrelevant in the case of complex fermions and relevant in the Majorana model.

The present problem has much in common with $d>1$ Anderson-localization critical points studied in previous works where the multifractality induces strong correlations between eigenstates at different spatial points and different energies (often referred to as Chalker scaling). In fact, critical singularities are particularly strong in the present case. In the more conventional situation, both the density of states $\nu$ and the eigenstate correlation function $C_{HF}$ (and, correspondingly, their product) exhibit a power-law scaling with $L$, so that the indices $x_1$ and $\Delta_2^{(U)}$ are constant (i.e., independent on $L$). On the other hand, we will see below that in the present problem   $\nu$ and (in the complex-fermion case) $C_{HF}$ scale exponentially with $\sqrt{L}$, which means that $x_1$ and $\Delta_2^{(U)}$ are $L$-dependent and increase (by absolute value) at large $L$ as $\sqrt{L} / \ln L$.  This is a manifestation of the fact that the 1D critical point studied here is characterized by very strong multifractality.  What we are interested in is the sign of $- x_1 - \Delta_2^{(U)}$ at large $L$ which controls the behavior (increase or decrease) of $\nu C_{HF}$ in the limit $L \to \infty$. 
}

\begin{figure}
	\centering
	\includegraphics[width=.49\columnwidth]{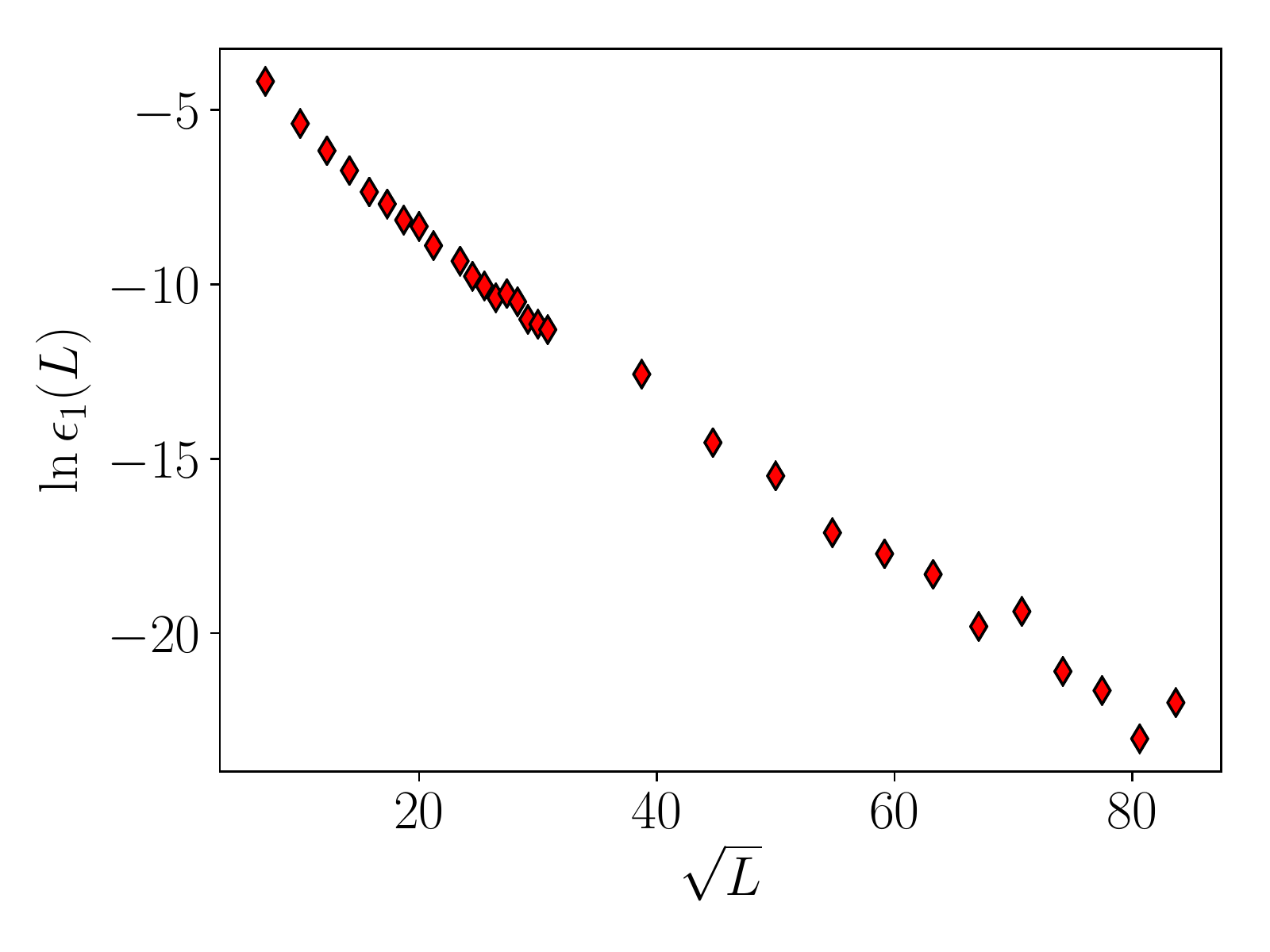}
	\includegraphics[width=.49\columnwidth]{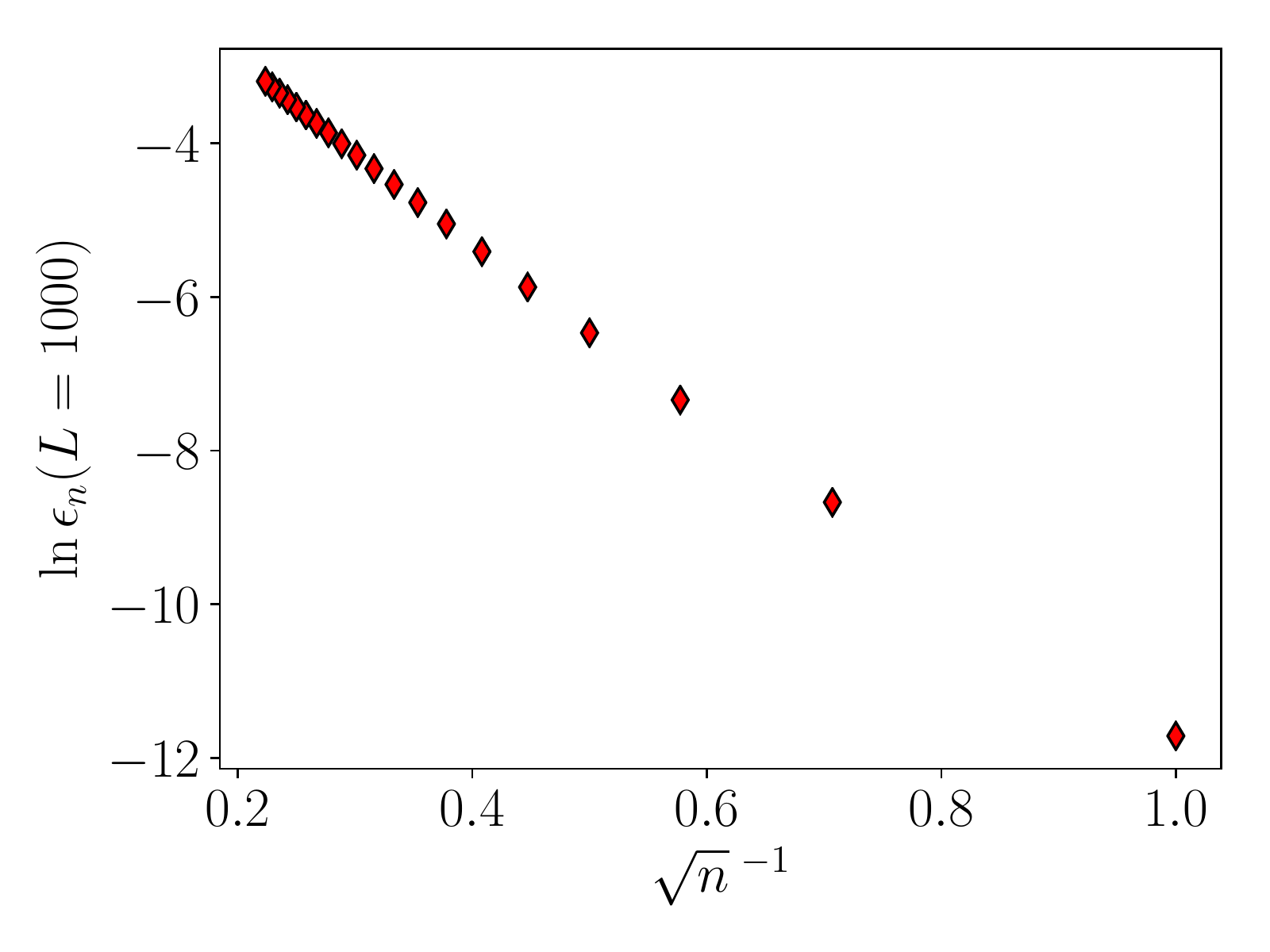}
	\caption{ Numerical verification of Eq.~(\ref{eq:epsilon_scale}) for the scaling of energies of the low-lying single-particle states.
		{\it Left:}  Average energy of the lowest eigenstate $\epsilon_1$ as a function
		of the square root of the system size, confirming the scaling
		$ - \ln \epsilon_1\propto \sqrt{L}$. {\it Right:} average energy $\epsilon_n$ of the $n$'th eigenstate vs $1/\sqrt{n}$ in a system of size
		$L=1000$, confirming the scaling $- \ln \epsilon_n \propto n^{-1/2}$ for sufficiently low energies.
		Combination of the scaling behavior observed in both panels confirms Eq. (\ref{eq:epsilon_scale}).
	}
	\label{fig:energ}
\end{figure}

For systems of the symmetry class BDI in
one dimension with an odd number of channels, the density of states at
low energies $\epsilon$ exhibits the well known Dyson singularity\cite{dyson_dynamics_1953,balents_delocalization_1997, mckenzie_exact_1996,Titov_Fokker_2001}:
\begin{align}
\nu(\epsilon) &\sim \frac{1}{\epsilon |\ln \epsilon|^3 }.
\label{nu-epsilon}
\end{align}
We can use this result to calculate the position of the n-th level in a system of the length $L$:
\begin{equation}
\int_{0} ^{\epsilon_n} \nu(\epsilon) \mathrm{d} \epsilon =  \frac{n}{L},
\end{equation}
which yields
\begin{align}
\epsilon_n \sim \exp\left(-c\sqrt{\frac{L}{n}}\right),&& c = O(1).
\label{eq:epsilon_scale}
\end{align}
We have verified the scaling (\ref{eq:epsilon_scale}) numerically for the model with the nearest-neighbor hopping matrix elements uniformly distributed over the
interval $t_j\in [0,1]$. The numerical data shown in Fig.~\ref{fig:energ} fully confirm the analytical prediction, with the coefficient $c \approx 0.5$.
Thus, we can write down the density of states around the lowest energy state $\epsilon_1$ as a function
of the length $L$:
\begin{align}
\nu(0,L) &\sim \frac{\exp(c\sqrt{L})}{L^{\frac{3}{2}}}.
\label{nu-L}
\end{align}
This behavior is not of power-law type, i.e., it is not characterized by a critical exponent in the usual sense.
We can define, however, an $L$-dependent scaling exponent \ADDED{$x_1 (L) = - \partial \ln \nu / \partial \ln L$}, with the result
\begin{align}
\ADDED{ - x_1(L)} &= c\frac{\sqrt{L}}{\ln (L)}-\frac{3}{2}. 
\label{Delta1}
\end{align}

The result (\ref{Delta1}) for the scaling dimension of the density of states is
valid both for the Majorana and complex fermions, since these models are
equivalent in the absence of interaction. (The only difference is that the
number of states is halved in the case of Majoranas.) On the other hand, we will show
that the scaling dimension $\Delta^{(U)}_2$ of the interaction is completely
different in these two models. We will explore the scaling of interaction by a
numerical approach, \ADDED{supporting the results by analytical arguments.}

\subsection{Scaling of interaction}
\label{scaling-interaction}

In order to determine the scaling of the interaction operators, we express the interaction matrix elements in terms of linear combinations of products of single-particle eigenfunctions. These expression in terms of the eigenfunctions are then numerically averaged over the disorder. The numerical results will be also supported by analytical considerations (Appendix \ref{app:wave-func-corr-analytics}).

We start by writing the most general
non-interacting Hamiltonian of a 1D system of size $L$ of symmetry
class BDI \cite{evers_anderson_2008}:
\begin{align}
H=\frac{1}{2} \left(\begin{array}{c c}\mathbf{c}_A^\dagger & \mathbf{c}_B^\dagger\end{array}\right)
\left(\begin{array}{c c}0&\underline{h}\\\underline{h}&0\end{array}\right)
\left(\begin{array}{c}\mathbf{c}_A\\ \mathbf{c}_B\end{array}\right),
\label{eq:bdimat}
\end{align}
where $\underline{h}$ is a real matrix and $c_{A,B}$,   $c_{A,B}^\dagger$ are onsite
operators acting on the two sublattices. In the case of the complex fermionic
chain, these are fermionic creation and annihilation operators, in the
case of the Majorana chain we have $c_A= \gamma_A = c_A^\dagger$ and $c_B= i\gamma_B = -c_B^\dagger$, where $\gamma_{A,B}$
are the real Majorana operators in Eq. (\ref{eq:kitaev_chain}).
Diagonalizing the $L\times L$ matrix in  Eq.~(\ref{eq:bdimat}), one can rewrite  the Hamiltonian
in the basis of operators which correspond to the single particle excitations of the system,
\begin{align}
H&=\frac{1}{2} \left(\begin{array}{c c}\mathbf{d}_+^\dagger & \mathbf{d}_-^\dagger\end{array}\right)
\left(\begin{array}{c c}\underline{\epsilon}&0\\0&-\underline{\epsilon}\end{array}\right)
\left(\begin{array}{c}\mathbf{d}_+\\ \mathbf{d}_-\end{array}\right),\\
c_i &= \sum_\alpha U_{i,\alpha} d_\alpha.
\label{eq:oprel}
\end{align}
Here $ \underline{\epsilon}$ is a diagonal matrix with eigenvalues $0 < \epsilon_1 < \epsilon_2 < \ldots < \epsilon_{L/2}$. 
In the case of complex fermions, the eigenvectors $U_{i\alpha}$ are
just the conventional single-particle wavefunction $\Psi_\alpha(i)$. The
ground state $|\Omega\rangle$ of the Hamiltonian can be written in terms of the
operators $d$ and the zero-particle state $|0\rangle$:
\begin{align}
|\Omega \rangle &= \prod _{\alpha,\epsilon_\alpha<0} d^\dagger_\alpha | 0 \rangle \label{eq:opgs}.
\end{align}
This immediately yields the action of the $d$ operators on the ground state:
\begin{align}
d_\alpha |\Omega\rangle &= 0 &\text{for }\epsilon_\alpha>0,\\
d^\dagger_\alpha |\Omega\rangle &= 0 &\text{for } \epsilon_\alpha<0.
\end{align}
A general $q$-body interaction operator can be expressed as sum of products of
annihilation and creation operators of the following type:
\begin{eqnarray}
\hat{O} &=& \prod _{i=1}^q c^\dagger_{a_i} \prod_{j=q+1}^{2q} c_{a_j} \nonumber \\
&=& \sum_{\{\alpha_i,\alpha_j\}}\prod _{i=1}^q U_{a_i,\alpha_i}
d^\dagger _{\alpha_i} \prod_{j=q+1}^{2q} U_{a_j,\alpha_j} d_{\alpha_j}.
\label{eq:opexp}
\end{eqnarray}
The
expectation value of the operator $\hat{O}$ over any eigenstate of a non-interacting system can now be calculated by
substituting Eq. (\ref{eq:oprel}) into Eq.(\ref{eq:opexp}):
\begin{align}
\langle \hat{O} \rangle &= \sum_{\{\alpha_i, \alpha_j\}}\prod_{i,j} U_{a_i,
	\alpha_i}  U_{a_j, \alpha_j}  \left\langle \prod _{i=1}^q d^\dagger_{\alpha_i} \prod_{j=q+1}^{2q} d_{\alpha_j}
\right\rangle.
\label{eq:opsum}
\end{align}
The expectation value that stands as a last factor on the right-hand side of
Eq.~(\ref{eq:opsum}) is non-zero only if the states $\alpha_i$ and $\alpha_j$
are pairwise identical; in this case, it is equal to $+1$ or $-1$, depending on
parity of the permutation of indices.  The right-hand side of
Eq.~(\ref{eq:opsum}) thus represents an algebraic sum of products of
single-particle eigenfunctions.

The terms in Eq. (\ref{eq:opsum}) are therefore the matrix elements of the
interaction operator expressed as products of the eigenvector amplitudes
$U_{i\alpha}$.
For the conventional case of two-body interaction, $q=2$, on which we focus
below, Eq.~(\ref{eq:opsum}) reduces, in accordance with the Wick theorem,to a sum over pairs of states $\alpha_1$,
$\alpha_2$.  For a given choice of sites $a_1, \ldots a_4$ and eigenstates
$\alpha_1$, $\alpha_2$,  there will be two different terms in
Eq.~(\ref{eq:opsum}) (plus analogous terms obtained by an interchange $\alpha_1
\leftrightarrow \alpha_2$), that have a meaning of Hartree and Fock terms. These
two terms correspond to the order of subscripts $\alpha_1\alpha_2 \alpha_1
\alpha_2$ and  $\alpha_1\alpha_2 \alpha_2 \alpha_1$ of $d$ operators in
Eq.~(\ref{eq:opsum}).
As usual, the Fock term will enter with a relative minus sign due to Fermi
statistics. We will see below that, in close analogy with Refs.
\onlinecite{lee_effects_1996,wang_short-range_2000,burmistrov_wave_2011}, a
major cancellation between the Hartree and Fock terms will play a crucial role
for the RG-irrelevance of the interaction in the case of complex fermions.  In
the case of Majorana system, there is a third term, originating from the
following order of indices $\alpha_1\alpha_1 \alpha_2 \alpha_2$, as discussed in detail in Sec.~\ref{sec:majorana-correlations}.  It has a
meaning of the Cooper term, and its emergence it is not surprising since
Majorana excitations are characteristic for superconducting systems. As we show
below, the presence of this term spoils the cancellation, making the total
interaction matrix element relevant in the RG sense.

 In general, the disorder averaged value of matrix elements under consideration
is a function of the system size and of the energies of the $q=2$ eigenvectors
involved. To obtain the scaling of these functions numerically, matrices of the
form Eq. (\ref{eq:bdimat}) for different system sizes were generated and the
lowest $20$ eigenvectors calculated. Then for each pair of eigenvectors the corresponding matrix elements entering Eq. (\ref{eq:opsum}) were calculated.
This procedure yields pairs of energies and the associated matrix
elements, which then have to be averaged over disorder configurations. This is
done by making a histogram and averaging the matrix elements over each energy bin. It is worth emphasizing that for the cases of logarithmic dependence of the matrix elements on energy, the correct choice of averaging procedure is crucial. In these cases, the bin sizes are chosen such that the number of data points is the same in every
bin. 

Even though we deal here with eigenstates of a non-interacting problem, the corresponding numerical analysis is a rather challenging endeavour.
This is particularly true in the regime of strong  Hartree-Fock cancellations that plays a central role below. 
In this situation, the default double precision that provides approximately 15 decimal digits is by far insufficient.
As will be shown below, the Hartree and Fock terms can be the same within hundreds of digits for large systems. The calculations have therefore been performed with at least 500 decimal digit floating point arithmetics. 

Since for large ($L\gg 1$) systems full diagonalization becomes slow (typically $\mathcal{O}(L^3)$)) and memory intensive (at least $\mathcal{O}(L^2)$)), a transfer matrix approach is chosen to compute the first few eigenvectors $U_{i,\epsilon_i}$. 
The characteristic polynomial $\lambda(\epsilon)$ is evaluated by $L$ column expansions in $\mathcal{O}(L^2)$.
The first 20 eigenenergies $\epsilon_i$ closest to zero are obtained as roots of $\lambda(\epsilon)$. 
The $\epsilon_i$ are plugged into the transfer matrix equation \eqref{eq:transfer} to find $U_{i,\epsilon_i}$.

For all following calculations, the hopping parameters are  chosen to be uniformly distributed over the
interval $t_j\in [0,1]$.

\subsection{Complex fermion chain}
\label{sec:complex-fermion-correlations}

We start with the model of the complex fermionic chain described by Hamiltonian
Eq. (\ref{eq:ham_fer}). Due
to chiral symmetry, each state with positive energy has a partner state with
negative energy. For zero chemical potential, in the non-interacting ground state all
states of negative energy are occupied and all of positive energy are free. The relevance of the interaction in the infrared limit is controlled by its matrix elements evaluated on low-lying eigenstates. To obtain the appropriate eigenstate correlation function, we inspect the expectation value of the interaction,  Eq.
(\ref{eq:opsum}). For each pair of sites $i$, $j$, we have a contribution
\begin{eqnarray}
\langle c_i^\dagger c_{i+r}^\dagger c_{i}c_{i+r}\rangle &=&
\sum_{\alpha\beta\gamma\delta}
U_{i\alpha}U_{i\beta}U_{i+r,\gamma}U_{i+r,\delta} \langle d^\dagger_\alpha
d^\dagger_\beta d_\gamma d_\delta \rangle \nonumber \\
&=& \sum_{\{\alpha\beta\}}
\left(U_{i\alpha}U_{i\alpha}U_{i+r,\beta}U_{i+r,\beta}\right.\nonumber\\
&- &\left. U_{i\alpha}U_{i\beta}U_{i+r,\alpha}U_{i+r,\beta}\right),
\label{correlator-complex-fermions}
\end{eqnarray}
with the summation in the last expression going over pairs of filled states.
The two terms in brackets  after the last equality sign in Eq.~(\ref{correlator-complex-fermions}) correspond to the conventional Hartree and Fock diagrams. 
We define the corresponding correlation functions of two single-particle eigenfunctions as functions of
energies, distance, and system size:
\begin{align}
C_{\rm H} (\epsilon_\alpha,\epsilon_\beta, r, L) &= \langle U_{i\alpha}U_{i\alpha}U_{i+r,\beta}U_{i+r,\beta} \rangle _{\rm dis}, \label{corr-wave-func-H} \\
C_{\rm F} (\epsilon_\alpha,\epsilon_\beta, r, L) &= \langle U_{i\alpha}U_{i\beta}U_{i+r,\alpha}U_{i+r,\beta} \rangle _{\rm dis},  \label{corr-wave-func-F}  \\
C_{\rm HF}(\epsilon_\alpha,\epsilon_\beta, r, L) &= \langle U_{i\alpha}U_{i\alpha}U_{i+r,\beta}U_{i+r,\beta}\nonumber\\
&-U_{i\alpha}U_{i\beta}U_{i+r,\alpha}U_{i+r,\beta} \rangle _{\rm dis},  \label{corr-wave-func-HF}
\end{align}
where $\langle \ldots \rangle_{\rm dis}$ denotes the disorder averaging. Below,
we analyze the scaling of the full correlation function $C_{\rm HF} = C_{\rm H}
- C_{\rm F}$ in order to determine the scaling exponent $\Delta^{(U)} _2$ of the
interaction. It was verified in Refs.
\onlinecite{lee_effects_1996,wang_short-range_2000,burmistrov_wave_2011} that
this scaling dimension also controls the scaling of interaction matrix elements
also in the second order of the perturbation theory. We thus expect that that
the analysis of the scaling of the correlation function
(\ref{corr-wave-func-HF}) with energy and the distance is sufficient for
establishing the relevance or irrelevance of the interaction near the
non-interacting fixed point.

\ADDED{The following comment concerning the $r$ dependence is in order here. Our DMRG results above dealt with short range interaction $r\sim 1$ only. At the same time, one may be also interested in effects of long-range interaction, in which case one needs to know the scaling of correlations functions of the type (\ref{corr-wave-func-HF}) with $r$. Furthermore, the analysis of correlations of eigenstates at the infinite-randomness fixed point constitutes by itself a very interesting problem (with $r$ dependence being an important ingredient), as it represents a remarkable example of strong-coupling Anderson-localization critical point (see also a discussion in Sec.~\ref{summary}). Since the $r$ dependence of the correlation functions (\ref{corr-wave-func-H}), (\ref{corr-wave-func-F}), (\ref{corr-wave-func-HF})  can be tackled by the same approach, we analyze below the correlation functions not only for $r \sim 1$ but also for arbitrary $r$.  In the end, when we study the RG relevance of the short-range interaction, we focus on the correlations at $r \sim 1$. This comment applies also to the Majorana chain, Sec.~\ref{sec:majorana-correlations}. }

\subsubsection{Single-wavefunction correlations}
\label{sec:single-wafefunc-corr}

Terms where the two wavefunctions are identical, i.e. $\alpha=\beta$, do not
contribute to the interaction matrix element $C_{\rm HF}$ as the Hartree and Fock terms cancel
each other exactly. Nevertheless, it is useful to start our analysis by considering the single-wavefunction correlations for two reasons.
First,  they can be particularly well understood analytically and can
serve as a benchmark to our numerical calculations. Second, we will see below that some of properties of the  single-wavefunction correlations translate to correlations of two eigenstates that are important for the interacting models.
We define the two-point, single-wavefunction correlation function $C_2$:
\begin{align}
C_2(\epsilon_\alpha,r,L) &= \langle U_{i\alpha}U_{i\alpha}U_{i+r,\alpha}U_{i+r,\alpha} \rangle _{\rm dis}.
\end{align}

For zero energy, the wavefunction $U_{r}$ can be expressed exactly in terms of a given realization of disorder \cite{balents_delocalization_1997}. The zero-energy wavefunctions belong entirely to one of the two sublattices (i.e., vanish on the other sublattice). If one
looks at the wave function moments at a  single point, their scaling is similar to that of a fully localized
wavefunction\cite{balents_delocalization_1997,evers_anderson_2008}:
\begin{align}
\langle U_{r}^{2q}\rangle\sim \frac{1}{L},
\end{align}
for $q>0$. At the same time, the spatial decay of the correlation function $C_2$ at zero energy is only algebraic, which is a property of a critical system
\cite{balents_delocalization_1997}:
\begin{align}
C_2(0,r,L) & \sim \begin{cases}
r^{-\frac{3}{2}}L^{-1}, & \text{r even};\\
0 & \text{r odd}.
\end{cases}
\label{eq:htself_balents}
\end{align}
For finite energy, this formula for even-$r$ correlations is expected to hold as long as the distance $r$ is smaller than  the localization length, $r\lesssim \xi_\epsilon$. The latter was predicted
\cite{balents_delocalization_1997} to scale with energy as
\begin{align}
\xi_\epsilon \propto |\ln \epsilon|^2.
\label{xi-epsilon}
\end{align}
Using Eq.~(\ref{eq:epsilon_scale}) with $n = 1$, we see that $\xi_\epsilon \sim L$ for the lowest eigenstate.

As to odd-distance correlations, they are not exactly zero for a non-zero energy $\epsilon$.
Indeed, the absence of odd-distance correlations,
Eq.~(\ref{eq:htself_balents}), is a consequence of the chiral symmetry which is exact at $\epsilon=0$ but is violated at non-zero energy and
gets progressively more strongly broken when the energy increases.
Thus, the odd-$r$ correlations should be strongly  suppressed relative to even-$r$ correlations at low energies, with the suppression becoming stronger with lowering energy.
As shown in Appendix~\ref{AppendixMF}, the corresponding suppression factor is $\sim \epsilon^2$ for odd $r\sim 1$.

\begin{figure}
	\centering
	\includegraphics[width=.9\columnwidth]{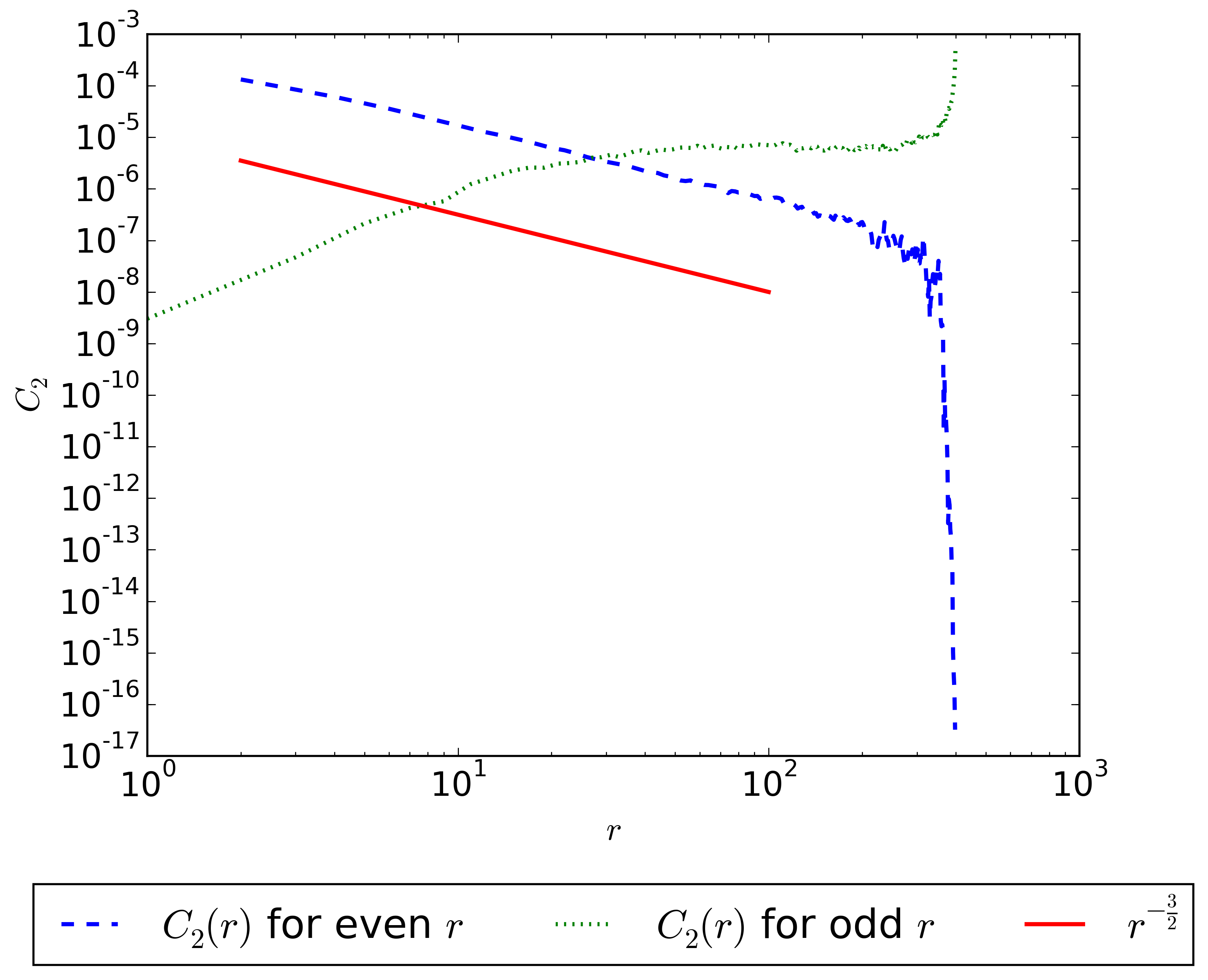}
	\caption{Single-wavefunction correlation function
		$C_2(\epsilon_1,r,L=400)$ vs distance $r$ for the lowest-energy
		state in a disordered complex-fermion chain of size $L=400$, for
		even and odd distances $r$. For small even distances, $C_2$ scales as $r^{-3/2}$ in agreement with Eq.
		(\ref{eq:htself_balents}). At distances $r$ approaching $L$ even correlations are strongly
		suppressed. Odd correlations are strongly suppressed for small distances in consistency with Eq.
		(\ref{eq:htself_balents}) and with the result of Appendix \ref{AppendixMF} but become large for $r$ comparable to $L$.}
	\label{fig:htself_r}
\end{figure}

We confront now the analytical predictions with numerical simulations.
In Fig. \ref{fig:htself_r}  we plot there the $r$ dependence of the correlation function $C_2(\epsilon_1,r,L=400)$, separately for even and odd $r$. For even $r$, we observe the $r^{-3/2}$ scaling, in agreement with Eq.~(\ref{eq:htself_balents}). This scaling holds with a good accuracy up to $r\approx L/2$.  As to the odd-distance correlations, they are strongly suppressed for small $r$ in comparison to even-distance ones, again in consistency with theoretical expectations.  Curiously, when $r$ approaches the system size $L$, the odd correlations become much stronger that the even correlations. This behavior will, however, play no role for our analysis, since we consider a finite-range interaction, i.e., $r \sim 1$.

\begin{figure}
	\centering
		\includegraphics[width=.9\columnwidth]{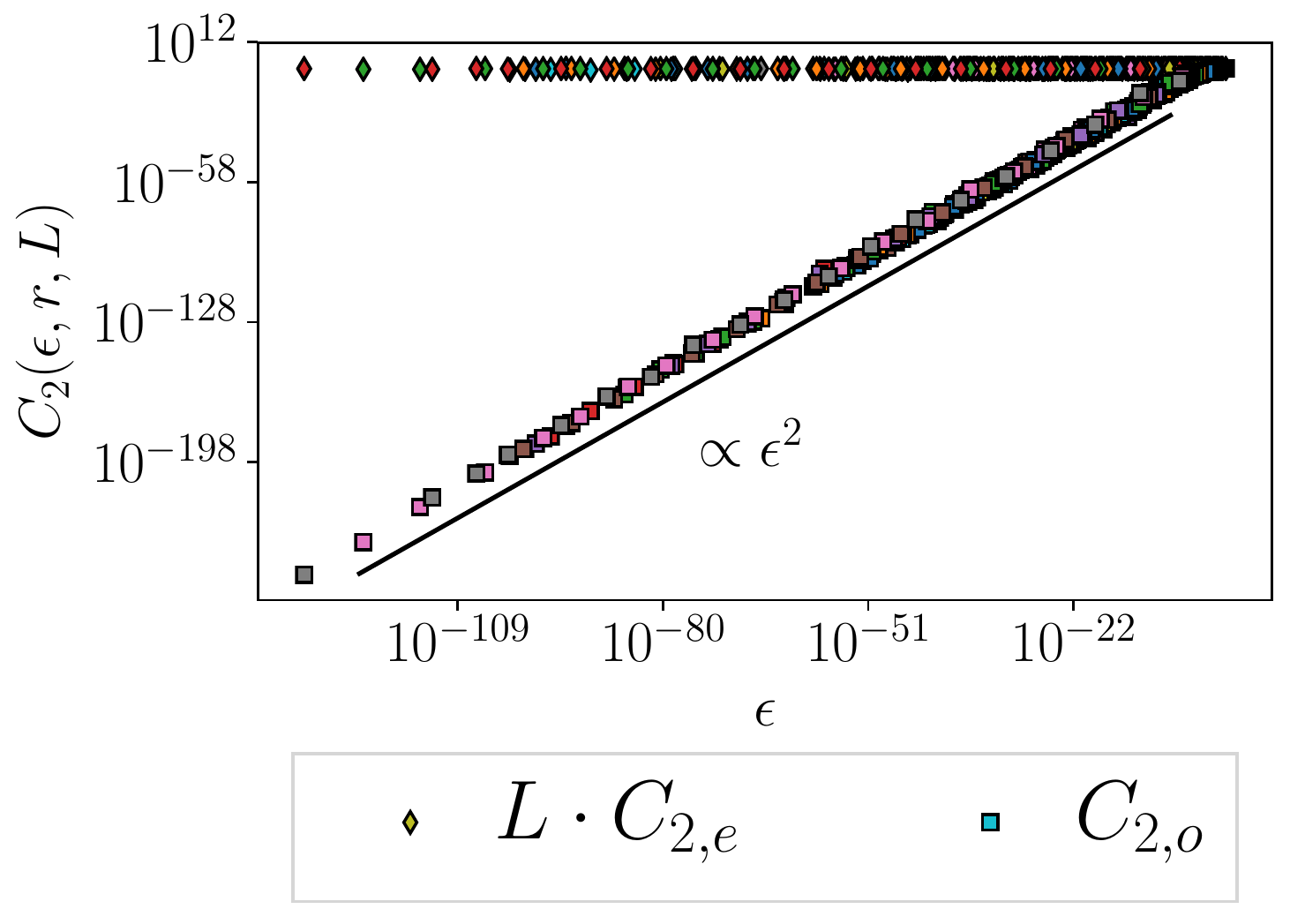}
		\caption{Single-wavefunction correlation function for short even distance,
			$L C_2(\epsilon ,r=2,L)$, and short odd distance, $C_2(\epsilon ,r=1,L)$,
			vs energy $\epsilon$ in systems with size $L$ from 100 to 10000 (distinct colors). For $r=2$
			the correlation function is independent on energy, while for $r=1$
	   it scales as $\epsilon^2$ (and thus is strongly suppressed at low energy), as predicted analytically.}
		\label{fig:htself_e}
\end{figure}

In Fig. \ref{fig:htself_e} we show the numerically obtained energy dependence of
the correlation function $C_2$ for fixed $L=1200$ and fixed small separation
$r$. Specifically, we choose $r=2$ for the even case and $r=1$ for the odd case.
It is seen that the even-distance correlations are essentially independent of
$\epsilon$. This is the expected behavior: indeed, for $r \sim 1$, the condition
$r \ll \xi_\epsilon$ is fulfilled as long as  $|\ln \epsilon| \gg 1$, i.e.,
essentially in the whole range of $\ln \epsilon$. On the other hand, the
odd-distance correlations strongly increase with energy. Specifically, the data unambiguously demonstrate the $\epsilon^2$ behavior of $C_2(\epsilon,r,L)$ for small odd $r$ discussed above and derived analytically in Appendix~\ref{AppendixMF}. It is worth emphasizing the enormously broad range of variation of the energy $\epsilon$ and the correlation function $C_2$ (odd $r$) in Fig. \ref{fig:htself_e}:  about 130 and 260 orders of magnitude, respectively!

\begin{figure}
	\centering
	\includegraphics[width=.49\columnwidth]{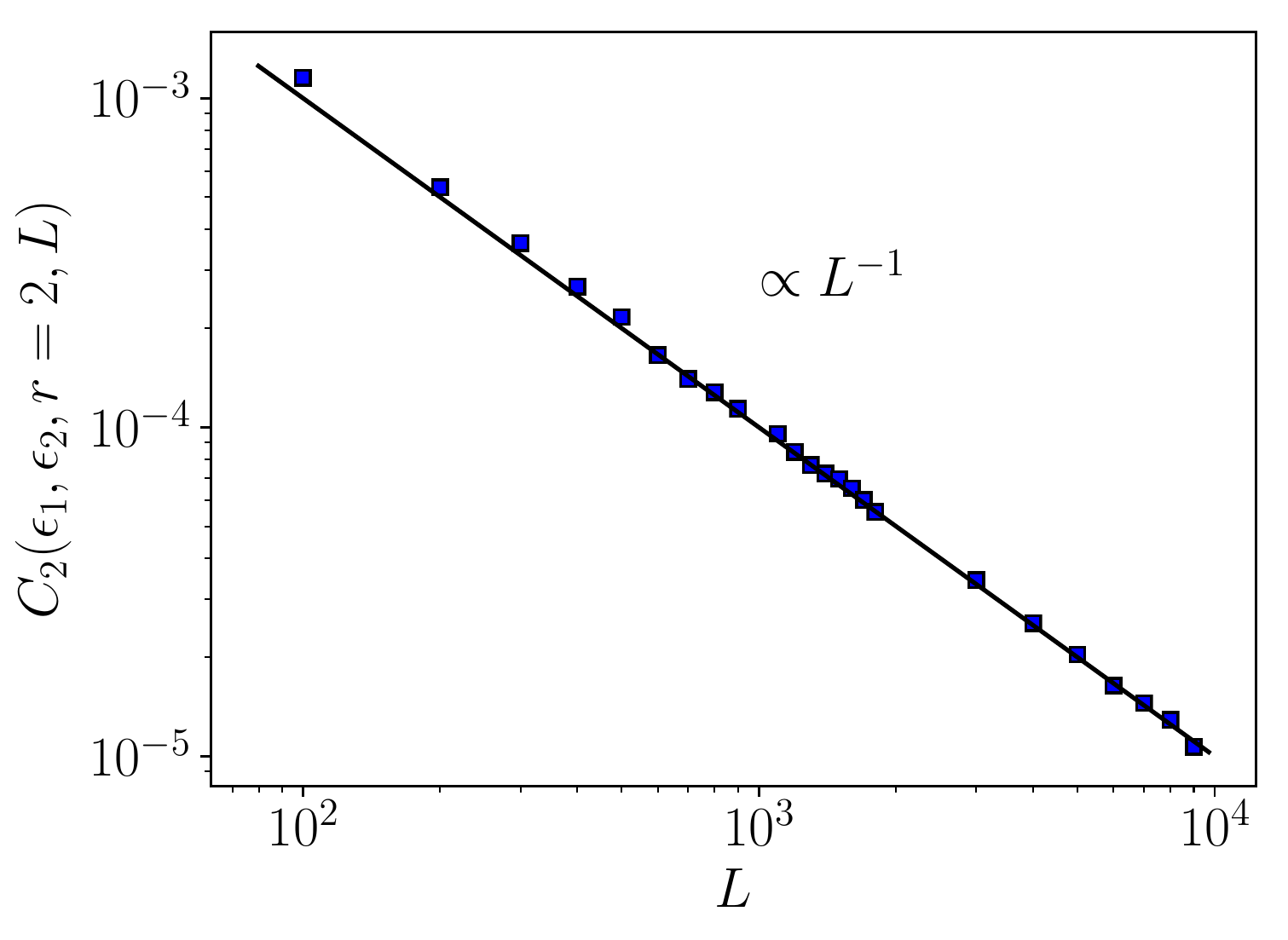}
	\includegraphics[width=.49\columnwidth]{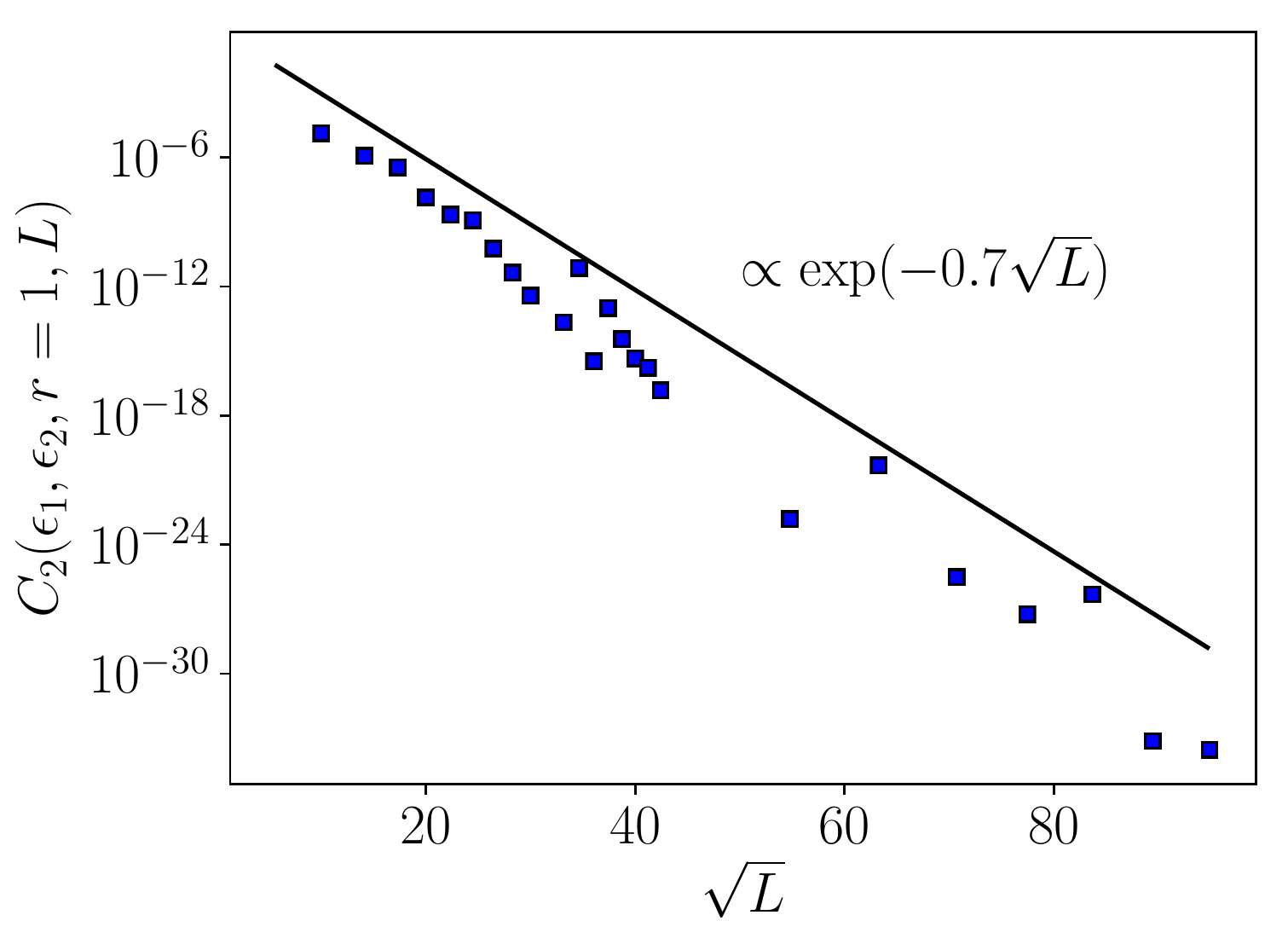}
	\caption{Single-wavefunction correlation function for short even distances
	$C_2(\epsilon_1,r=2,L)$ (left panel) and short odd distances
	$C_2(\epsilon_1,r=1,L)$ (right panel) vs system size $L$. For $r=2$ the data
	confirm the analytically predicted scaling, $C_2 \sim L^{-1}$,   see first
	line of Eq. (\ref{eq:htself_balents}). For odd distance, the correlation
	function decreases quickly with $L$ since the lowest energy $\epsilon_1$
	approaches zero exponentially fast, $\epsilon_1 \sim \exp\{-c\sqrt{L}\}$ and
	in view of $C_2 \propto \epsilon_1^2$, see Appendix~\ref{AppendixMF}.}
	\label{fig:htself_L}
\end{figure}

Finally, in Fig.~\ref{fig:htself_L} we show dependence of the correlation
function $C_2(\epsilon_1,r,L)$ on the system size $L$ for even ($r=2$) and odd
($r=1$) distance. In the even case, the correlation function does not depend on
energy for small $r$, so that the fact that $\epsilon_1$ is different from zero
and varies with $L$ is of no importance. The expected result is given by the
first line of Eq.~(\ref{eq:htself_balents}). The numerical data in the right
panel of Fig.~\ref{fig:htself_L} confirm the predicted $L^{-1}$ scaling. For odd
$r$ the decay of $C_2(\epsilon_1,r,L)$ with $L$ should be exponentially fast  due to $C_2(\epsilon,r,L) \sim \epsilon^2$ and the fact that the
energy $\epsilon_1$ approaches zero exponentially with increasing $L$, see Eq.(\ref{eq:epsilon_scale}). This yields the analytical expectation
$C_2(\epsilon_1,r,L) \sim \exp(-2c\sqrt{L})$, in full agreement with the data in the right panel of Fig.~\ref{fig:htself_L}.

\subsubsection{Two wavefunction correlations}
\label{sec:two-wavefunc-corr}

\begin{figure}
	\centering
	\includegraphics[width=.48\columnwidth]{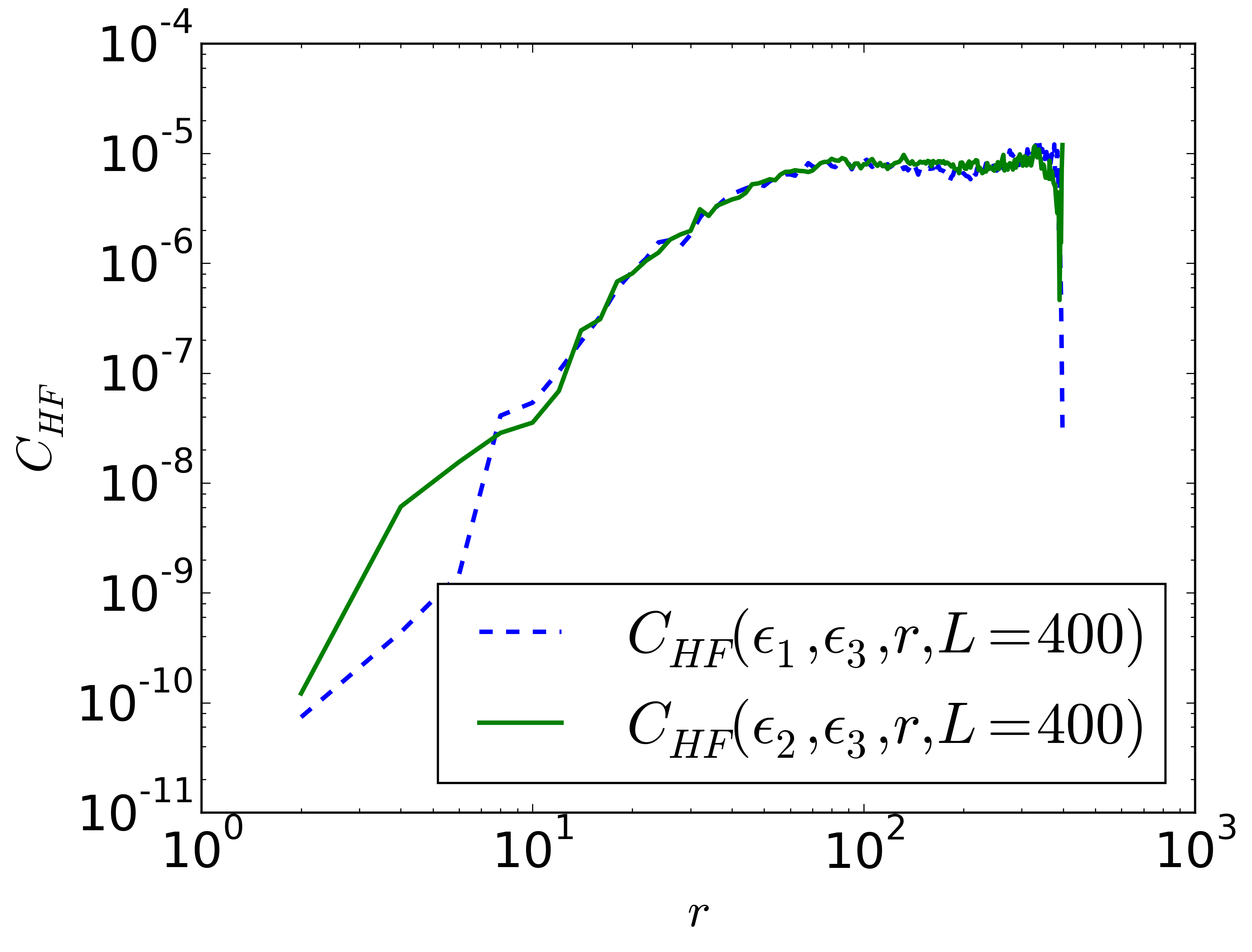}
	\includegraphics[width=.48\columnwidth]{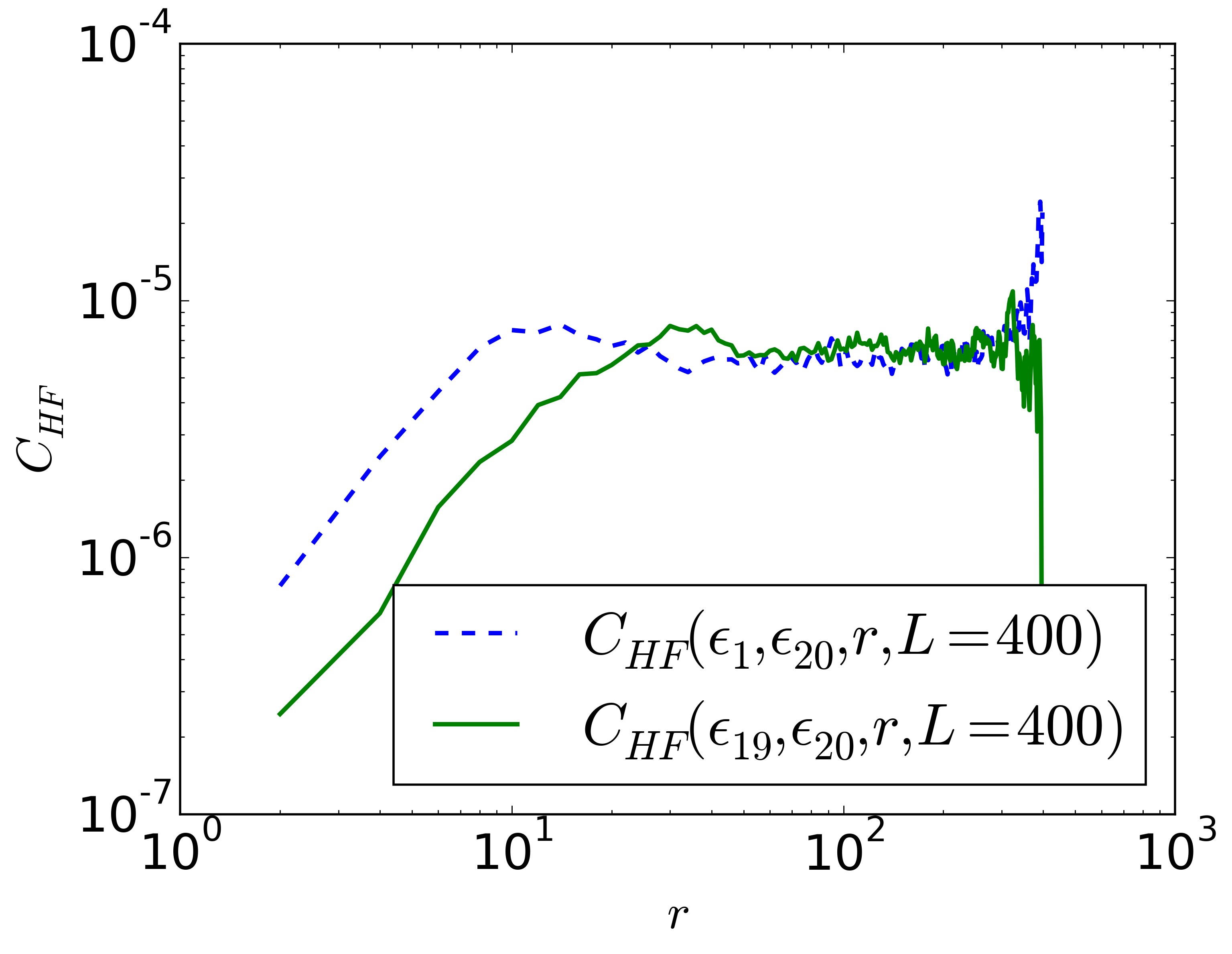}
	\caption{{\it Left:} $C_{HF}(\epsilon_1,\epsilon_3,r,L=400)$ and
		$C_{HF}(\epsilon_2,\epsilon_3,r,L=400)$ as functions of the even distance $r$.
		Two curves are essentially identical, which confirms insensitivity of the correlation function to the lower energy, as long it is much smaller than the larger one.
		{\it Right:}  $C_{HF}(\epsilon_1,\epsilon_{20},r,L=400)$ and
		$C_{HF}(\epsilon_{19},\epsilon_{20},r,L=400)$ as functions of the even distance $r$. The two curves show again a similar behavior but now there is a difference in a factor of order unity ($\approx 3$) between them at small $r$. This is because in this plot we consider higher-energy state, and, in particular, $\epsilon_{19}$ is close to $\epsilon_{20}$.}
	\label{fig:htfo_gs_check}
\end{figure}

Matrix elements for two-wavefunction correlations, Eqs.~(\ref{corr-wave-func-H})--(\ref{corr-wave-func-HF}),
are calculated using two eigenstates with different energies for a given disorder configuration, and then averaging over disorder.
The energy levels are on average distributed as $\epsilon_n \sim \exp(-c\sqrt{L/n})$,
which means that, for $L \gg 1$ and $n \sim 1$, one of the energies will almost always be much larger than the other one.
Since the energy breaks the chiral symmetry, it is expected that the matrix elements will essentially depend only on the larger of the two energies and only weakly on the lower one.  To verify numerically this expectation,  we compare in the left panel of Fig.
\ref{fig:htfo_gs_check} the Hartree-Fock correlation functions [see Eq.~(\ref{corr-wave-func-HF})]
$C_{HF}(\epsilon_1,\epsilon_3,r,L=400)$ of the lowest and third lowest energy
levels with $C_{HF}(\epsilon_2,\epsilon_3,r,L=400)$ of the second lowest and the
third lowest energy levels, for even $r$. As we will show below, this correlation function exhibits, for low energies, very strong dependence on the higher of
the two energies.  At the same time, the two curves in the left panel of Fig.
\ref{fig:htfo_gs_check} are nearly identical (within statistical fluctuations), which confirms the essential  insensitivity to the value of the lower energy.
In the right panel of Fig.~\ref{fig:htfo_gs_check}, we show analogous data by choosing now a higher excited state $\epsilon_{20}$ and varying the state with lower energy from $\epsilon_1$ to $\epsilon_{19}$. One still expects to see the same scaling behavior for the two correlation functions; however, since $\epsilon_{19}$ and $\epsilon_{20}$ are nearly equal for this value of $L$, a difference in a numerical factor of order unity is expected. This is exactly what is observed in the right panel of Fig.~\ref{fig:htfo_gs_check}.
In the numerical analysis below, we will choose the state with the lowest energy ($\epsilon_1$) as one of the two states for which the correlation function is calculated. This energy is always much smaller than another eigenstate energy (that will be denoted as $\epsilon$), which simplifies the scaling analysis at criticality.

The correlation functions at criticality depend thus on the energy $\epsilon$, the length $L$ and the distance $r$. As for the single-eigenstate correlation function, Sec.~\ref{sec:single-wafefunc-corr}, the behavior for even and odd distances $r$ is very different.
At low energy $\epsilon$, and short even distances, it is natural to expect that $C_H$ behaves,  in similarity with
with $C_2$, as a power-law in $r$ and $L$. Such a power-law behavior is also analogous to that of eigenfunction correlation functions at critical points of localization-delocalization transitions in systems of higher dimensionality, see Ref.~\onlinecite{evers_anderson_2008}.
As to the expected for of the energy dependence, we recall that, at the critical point that we study,  the logarithm of the energy scales as a power law of the length,
see Eq.~(\ref{eq:epsilon_scale}). Therefore, it is natural to expect a power-law scaling of $C_H$ with respect to $\ln \epsilon$. Therefore, for short even distances $r$ and low energy $\epsilon$, the correlation function $C_{H}$ is expected to have the scaling form (see also
\cite{foster_notes_2018}):
\begin{equation}
C_H(0,\epsilon,r,L) \sim \frac{|\ln \epsilon|^\alpha}{r^\beta L^\gamma}, \qquad r \text{ -- even}.
\label{scaling-alpha-beta-gamma}
\end{equation}
This equation should hold at criticality, so that the necessary condition is $r \lesssim \xi_\epsilon$.
We determine now the exponents $\alpha$, $\beta$, and $\gamma$ by a numerical analysis. We will also support the numerical results by analytical considerations
(details of which are presented in Appendix  \ref{app:wave-func-corr-analytics}) yielding the values of the exponents $\alpha$ and $\gamma$.

\begin{figure}
	\centering
	\includegraphics[width=.49\columnwidth]{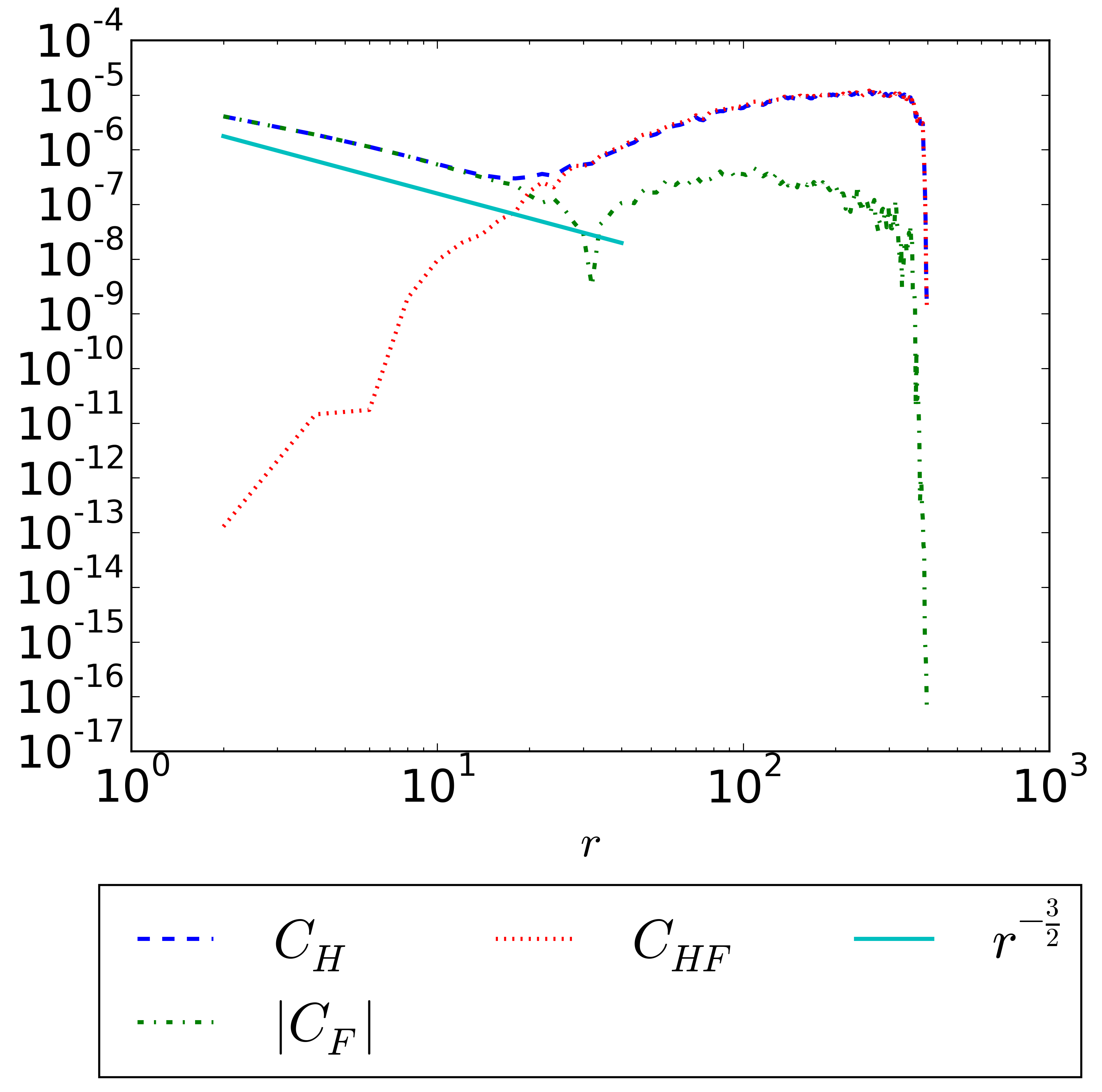}
	\includegraphics[width=.49\columnwidth]{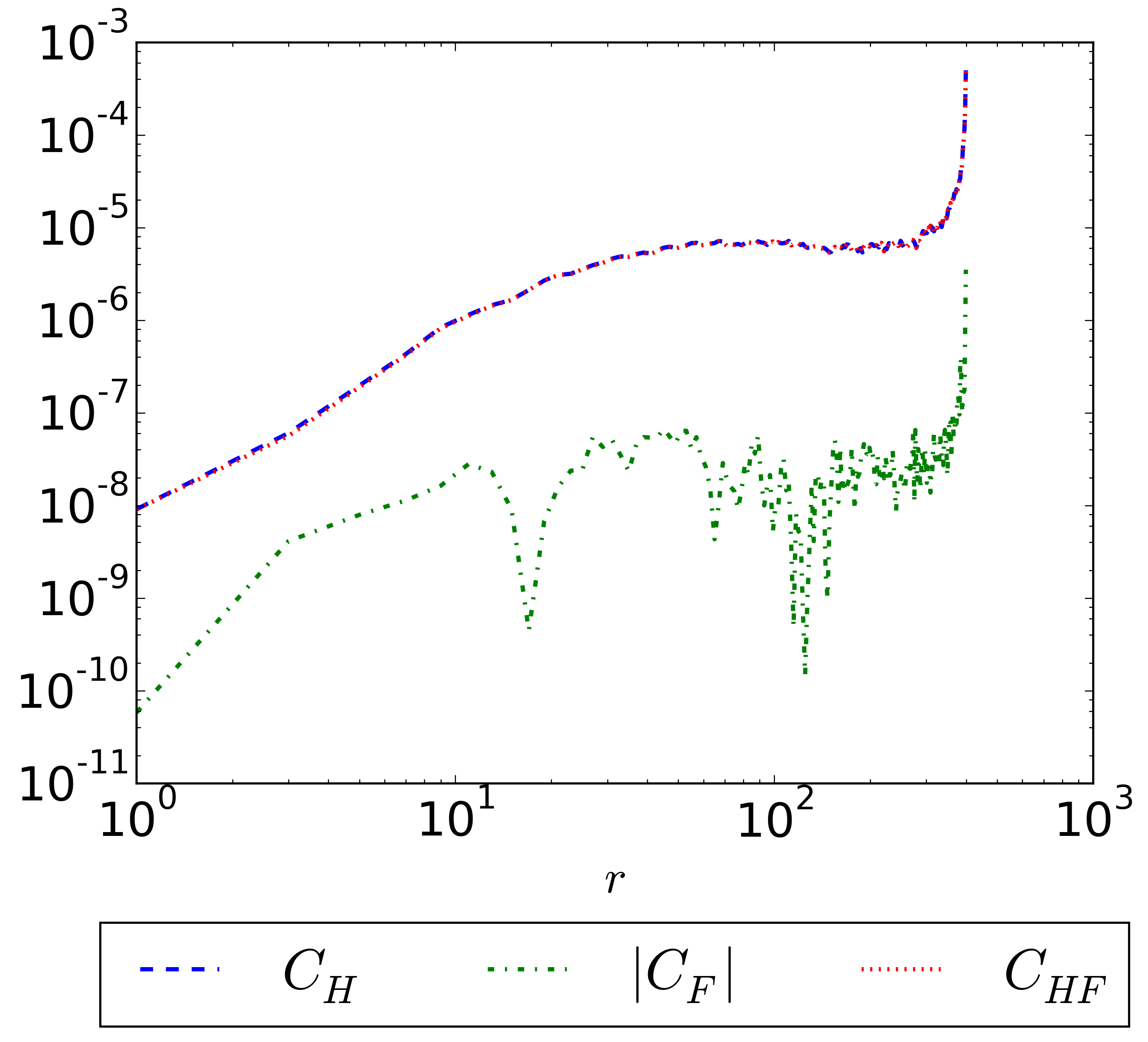}
	\caption{Hartree, $C_H(\epsilon_1,\epsilon_2,r,L=400)$, and Fock,
		$C_F(\epsilon_1,\epsilon_2,r,L=400)$, correlation functions and their difference
		$C_{HF}(\epsilon_1,\epsilon_2,r,L=400)$ plotted as functions of the distance $r$.
		{\it Left:} Even $r$. The scaling of $C_H$ at criticality (distance $r$ much smaller than the correlation length $\xi_\epsilon$), is of the $r^{-\frac{3}{2}}$ form, implying that the index $\beta$ in Eq.~(\ref{scaling-alpha-beta-gamma}) is $\beta=3/2$. The Fock term is nearly equal to the Hartree term in this critical regime, so that $C_{HF}$ is very strongly suppressed at small $r$. At $r\approx 40$, the Fock term becomes much smaller than the Hartree one and changes sign.
		 {\it Right:} Odd $r$. In the critical regime (small $r$) the Hartree term is strongly suppressed. The Fock term is still smaller, so that $C_{HF} \simeq C_H$.}
	\label{fig:htfo}
\end{figure}

\begin{figure}
	\centering
	\includegraphics[width=.49\columnwidth]{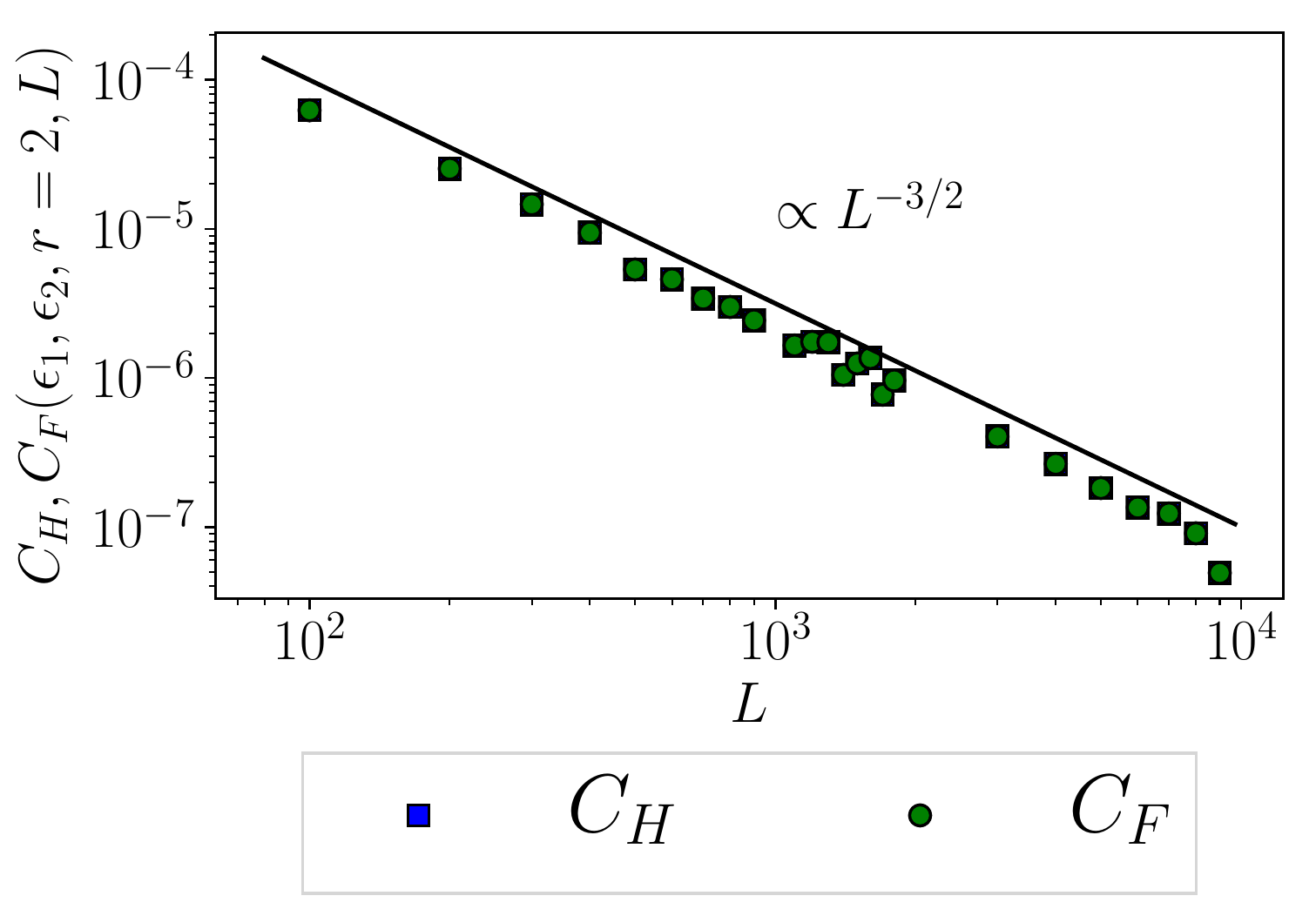}
	\includegraphics[width=.49\columnwidth]{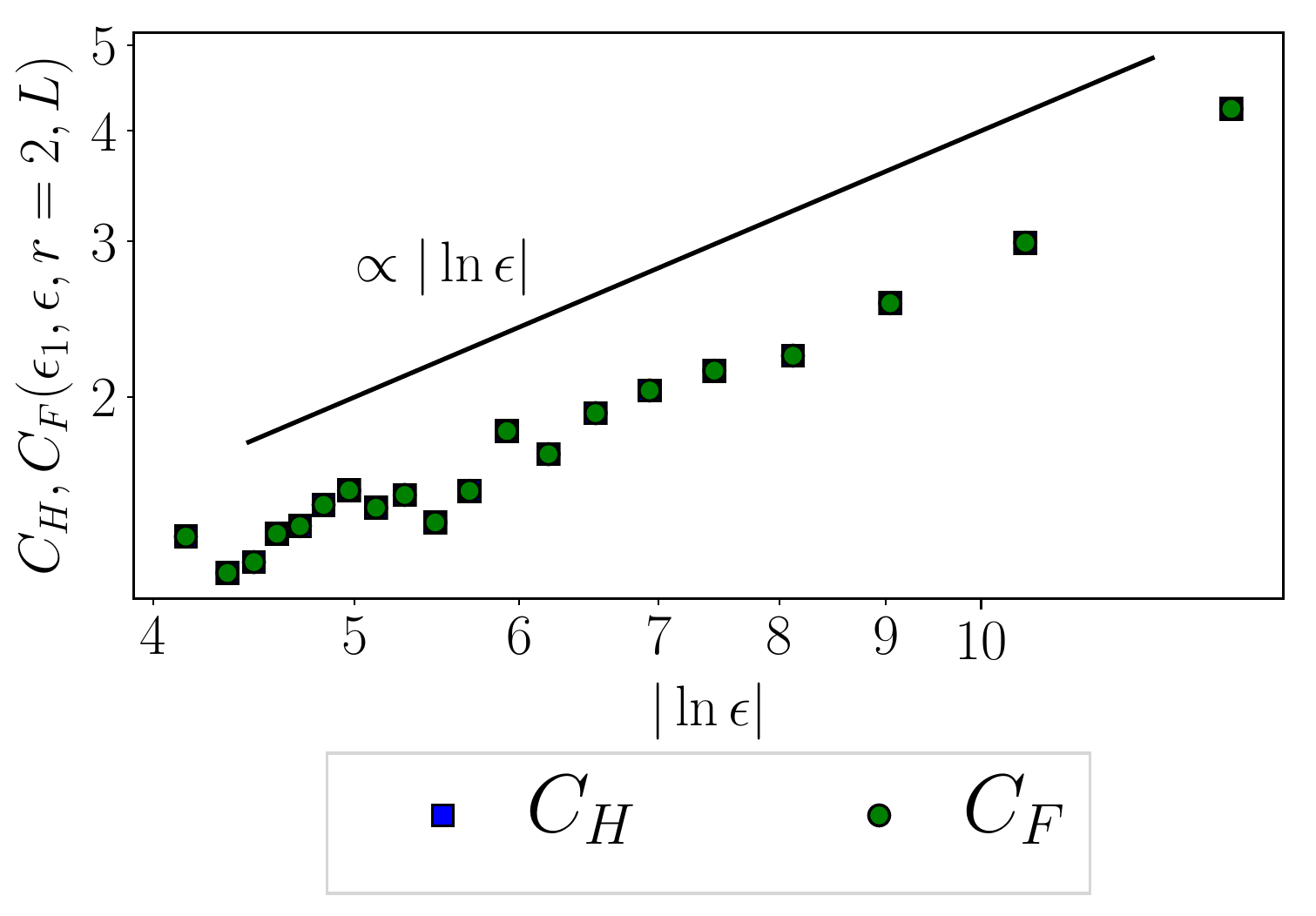}
		\\
	\includegraphics[width=.49\columnwidth]{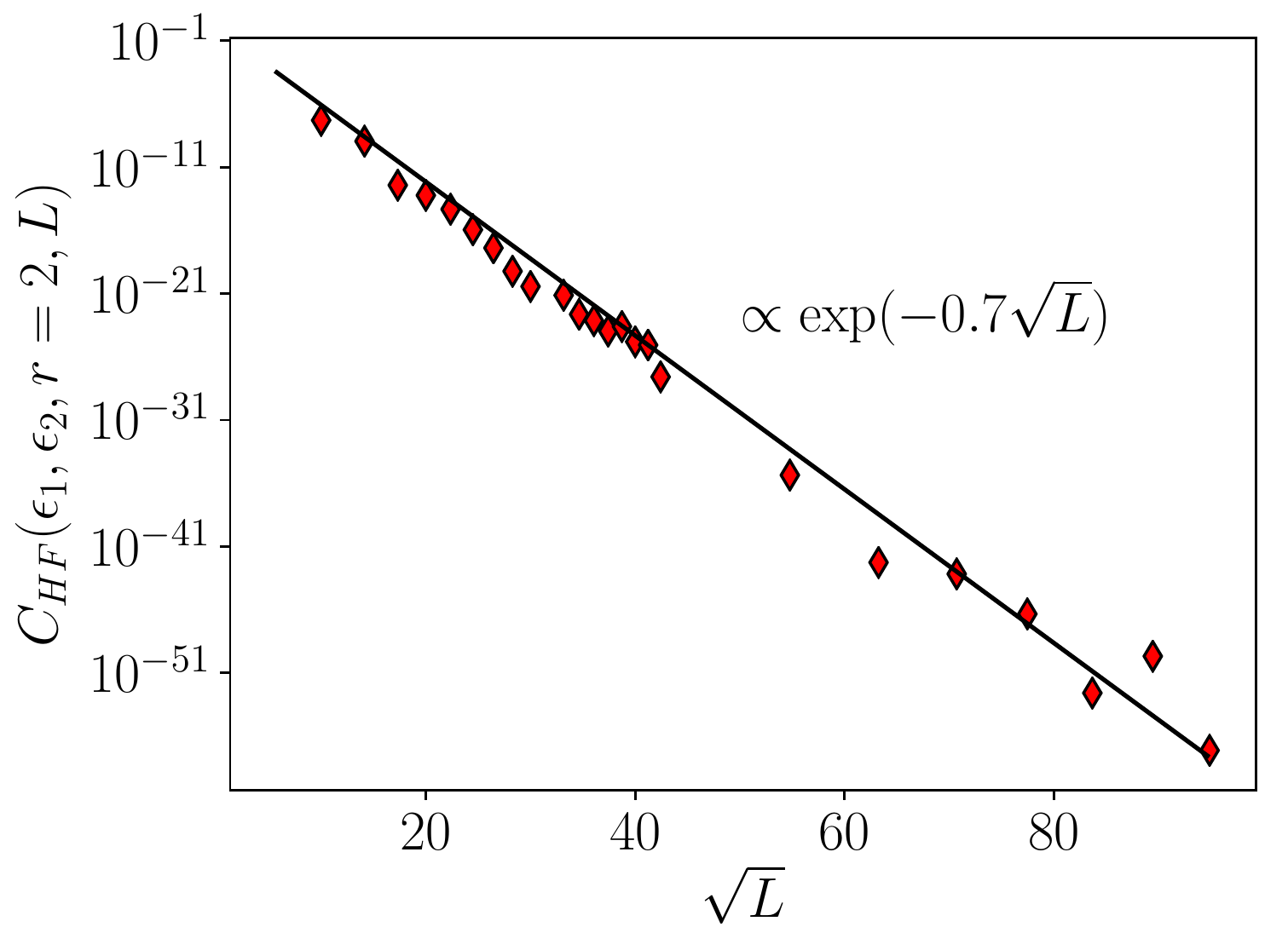}
	\includegraphics[width=.49\columnwidth]{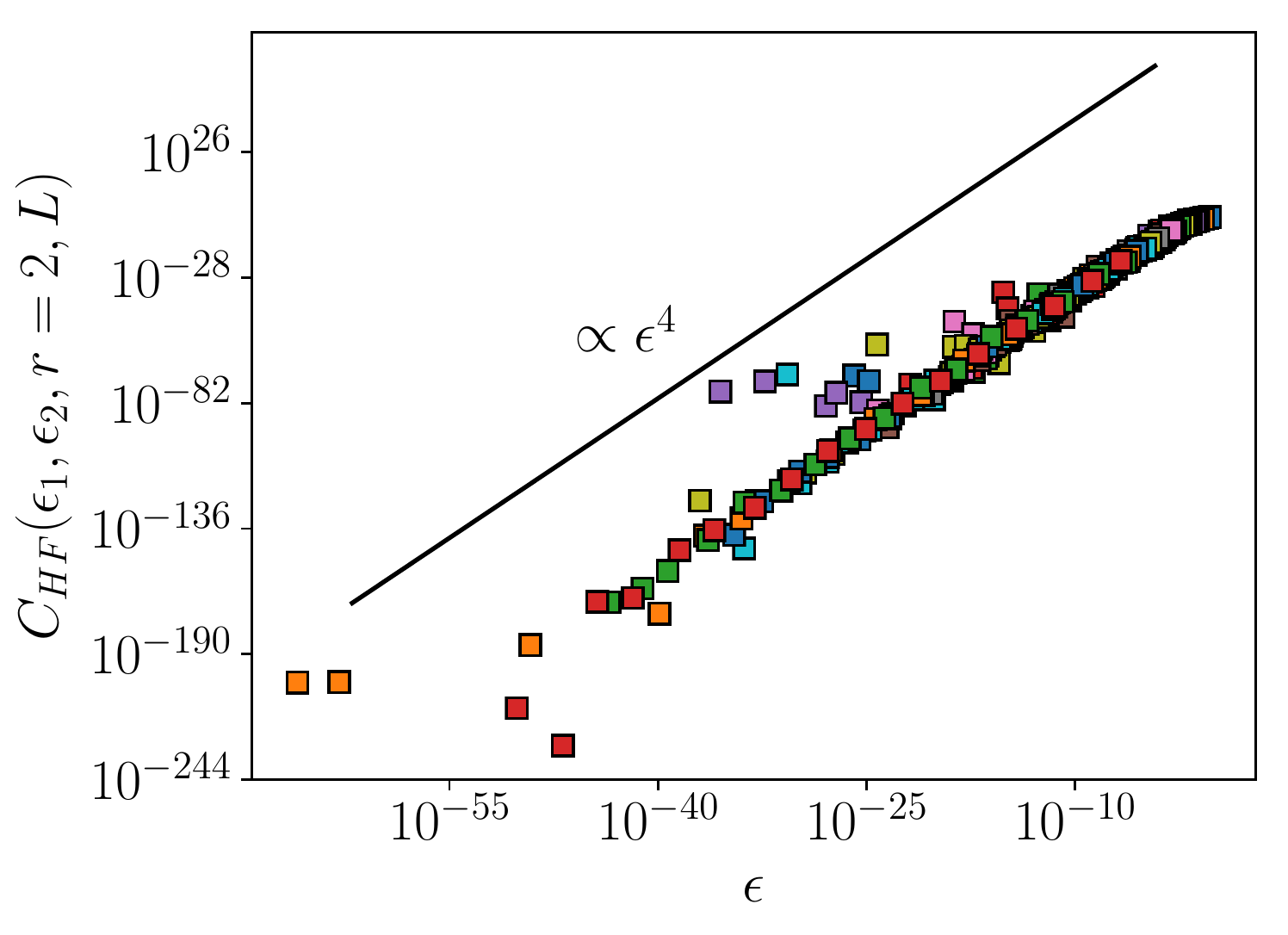}
	\caption{Hartree, Fock, and Hartree-Fock correlation functions
	$C_H(\epsilon_1, \epsilon, r=2, L)$, $C_F(\epsilon_1, \epsilon, r=2, L)$, and
	$C_{HF}(\epsilon_1, \epsilon, r=2, L)$ for a small even distance ($r=2$). 
	{\it
	Upper left:} Scaling of $C_{H}, C_{F}$ with $L$ at $\epsilon = \epsilon_2$. The slope  yields the
	power-law scaling $\sim L^{-3/2}$, implying a relation $\gamma -
	\alpha/2 = 3/2$ for the exponents in Eq.~(\ref{scaling-alpha-beta-gamma}).
	{\it Upper right:} Scaling with energy at fixed $L=4000$. The slope implies the
	scaling $\sim |\ln \epsilon|$ for $C_H$, implying the exponent $\alpha=1$  in
	Eq.~(\ref{scaling-alpha-beta-gamma}). In both panels, the Fock correlation
	function is nearly equal to the Hartree one, which is a characteristic feature
	of the critical regime for even $r$. As a result, $C_{HF}$ shown in lower panels is strongly
	suppressed with respect to $C_H$ and $C_F$.	
	{\it Lower left:} Scaling of $C_{HF}$ with $L$ at $\epsilon = \epsilon_2$.
	{\it Lower right:} Scaling of $C_{HF}$ with energy for $L$ from 100 to 10000. The slope agrees with the analytical prediction
	$C_{HF} \propto \epsilon^4$.
	}
	\label{fig:htfo_even_scal}
\end{figure}

In the left panel of Fig.~\ref{fig:htfo}, the numerically obtained dependence of the correlation functions on $r$ is shown for even $r$.  We see that $C_H$ at not too large $r$ scales $r^{-\beta}$ with $\beta = 3/2$. This scaling is the same as for the single-eigenfunction correlation function $C_2$, see Sec.~\ref{sec:single-wafefunc-corr}. To find the exponent $\alpha$ in the critical scaling of $C_H$, Eq.~(\ref{scaling-alpha-beta-gamma}), we show in the right panel of Fig. \ref{fig:htfo_even_scal} the dependence of correlation functions at small even distance ($r=2$) and fixed $L$ on the energy. The slope yields $\alpha = 1$. To determine $\gamma$, we plot in the left panel of the same figure the dependence on the system size $L$. Here the correlation functions are evaluated for two lowest eigenstates, so that the energy $\epsilon$ is equal to $\epsilon_2 = \exp(-c\sqrt{L/2})$. The obtained scaling of $C_H$ is $L^{-2}$; taking into account the $|\ln \epsilon_2| \sim L^{1/2}$ factor originating from the energy dependence of $C_H$, we find that $\gamma = 2$.
The scaling of $C_H$ in the critical regime  is thus given by
\begin{align}
C_H(0,\epsilon,r,L) &\sim \frac{|\ln \epsilon|}{L^2r^{\frac32}}, \qquad r \text{ -- even}.
\label{eq:ht_lowr}
\end{align}
The Fock correlation function $C_F$ for even $r$ is found to behave in exactly the same way. This is what should be expected: indeed, a particular case of a small even $r$ is $r=0$, for which $C_H$ and $C_F$ are identically the same. The $|\ln \epsilon| L^{-2}$ scaling of $C_H$ and $C_F$ for even $r$ is confirmed also by an analytical calculation of the averaged square of the Green function, see Appendix \ref{app:wave-func-corr-analytics} for details.

As was discussed above, the effect of the interaction is controlled by the scaling of the Hartree-Fock correlation function $C_{HF} = C_H - C_F$.
As the data in Fig.~\ref{fig:htfo_even_scal} clearly demonstrate, this function is strongly suppressed (for small even $r$) as compared to $C_H$  and $C_F$.
This is also what is expected analytically:  as shown in Appendix \ref{AppendixMF}, the suppression factor is $\sim \epsilon^4$. If the correlation function is evaluated for two lowest eigenstates, the suppression factor becomes $\sim \epsilon_2^4 \sim \exp(-4c\sqrt{L/2})$. These analytical predictions are fully confirmed by the numerical results, see Fig.~\ref{fig:htfo_even_scal}.

We turn now to the critical behavior of the correlation functions at odd $r$.  We expect that odd-distance correlation functions $C_H$ and $C_F$ are suppressed with respect to their even-$r$ counterparts.
The reason for this is the same as for the the single-eigenfunction correlation function $C_2$, Sec.~\ref{sec:single-wafefunc-corr}:  odd-$r$ correlations necessarily involve wavefunctions on different sublattices. As shown in Appendix \ref{AppendixMF}, the suppression factor for $C_H$ and $C_F$  with odd $r$ is the same ($\sim \epsilon^2$) as for $C_2$.  Again, this translates into an exponential suppression with respect to $L$.

\begin{figure}
	\centering
	\includegraphics[width=.49\columnwidth]{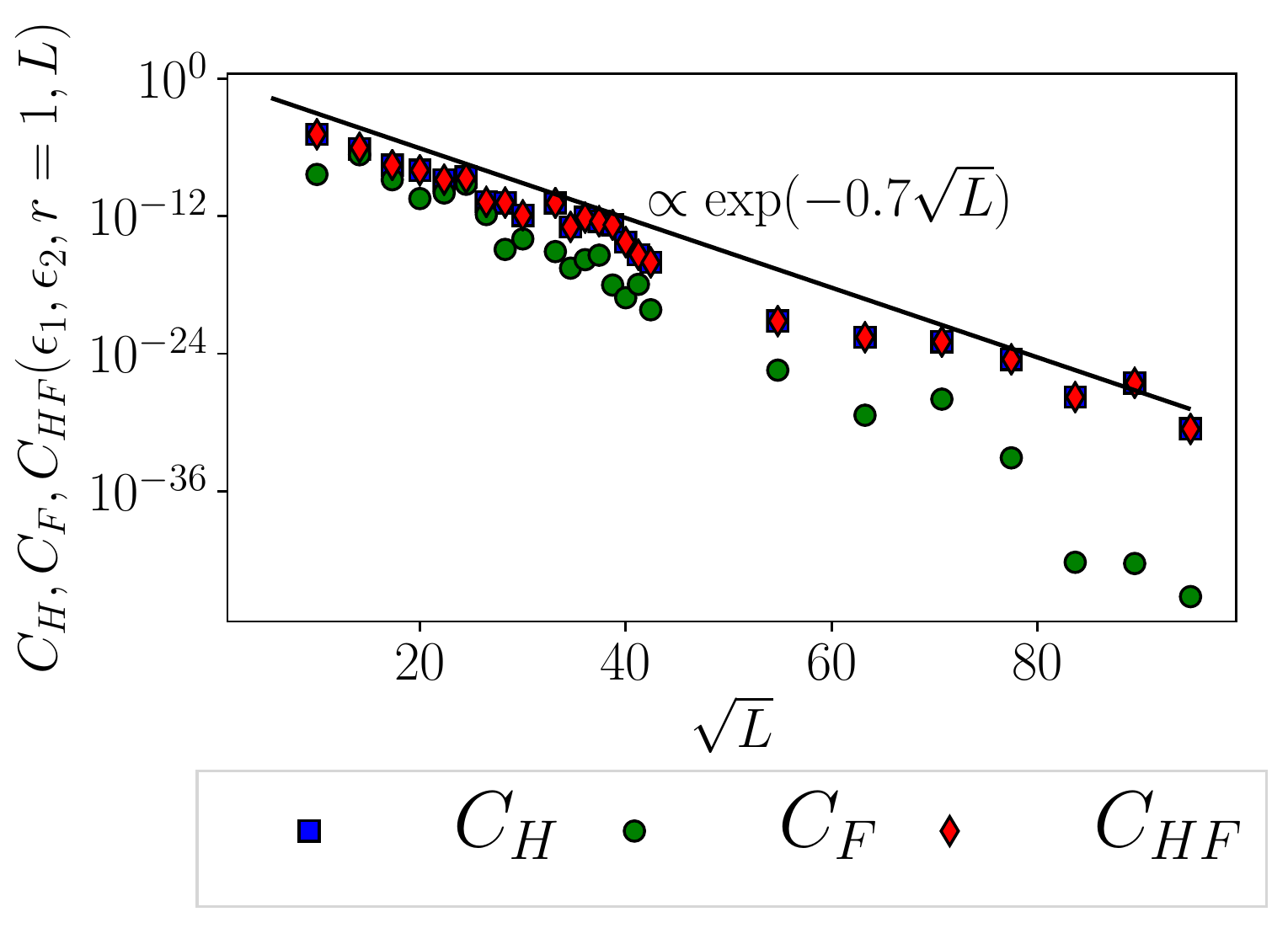}
	\includegraphics[width=.49\columnwidth]{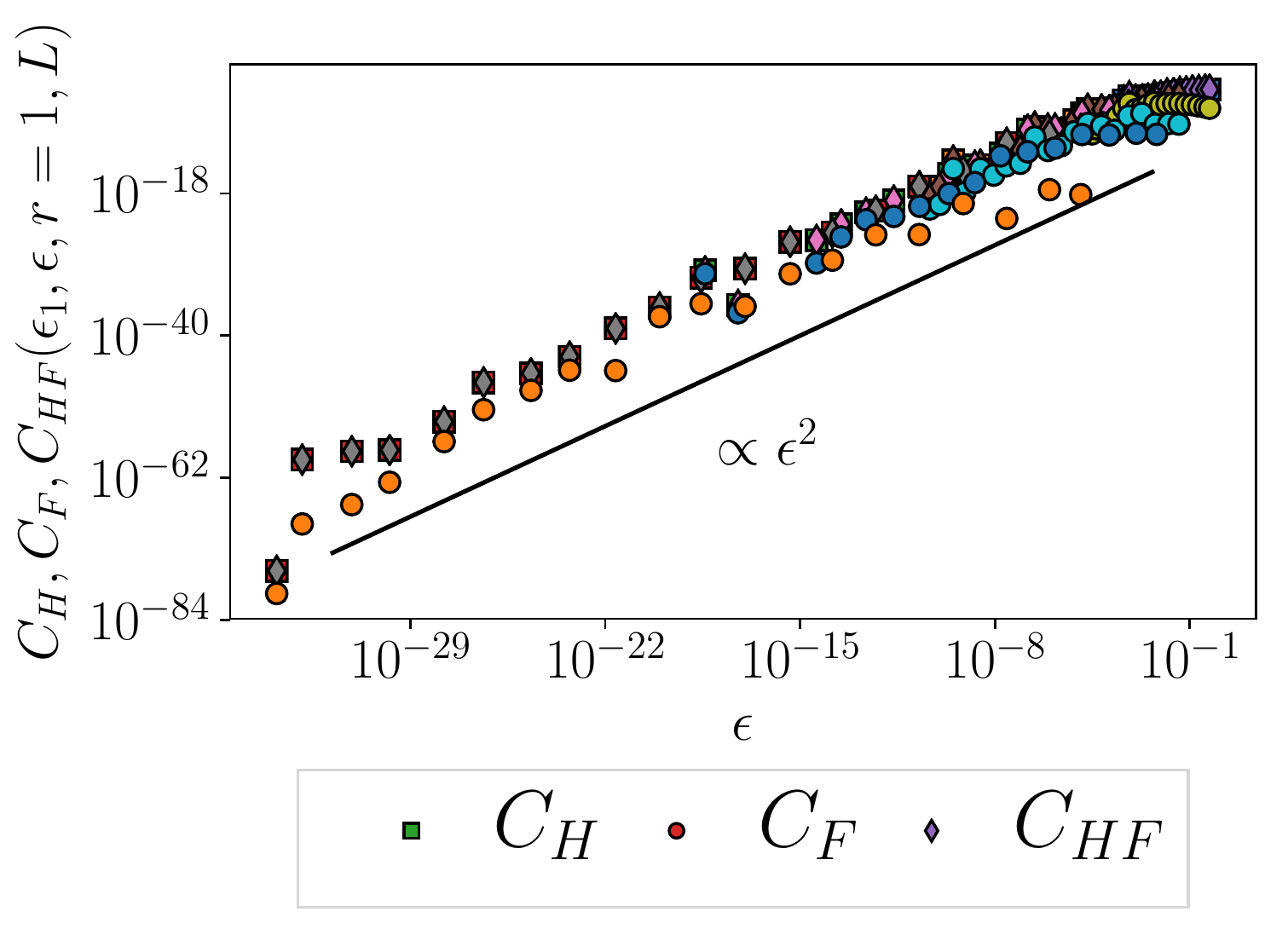}
	\caption{Hartree, Fock, and Hartree-Fock correlation functions $C_H(\epsilon_1, \epsilon, r=1, L)$,
	$C_F(\epsilon_1, \epsilon, r=1, L)$, and $C_{HF}(\epsilon_1, \epsilon, r=1, L)$ for a small odd distance ($r=1$).
	{\it Left:} Scaling with $L$ at $\epsilon = \epsilon_2$. {\it Right:} Scaling with energy. Different colors represent different lengths from 100 to 10000. 
	In both panels, the Fock correlation function is much smaller than the Hartree one, so that $C_{HF} \simeq C_H$.  The dominant scaling for both $C_H$ and $C_F$ is $\sim \epsilon^2$ (which translates into an exponential length dependence in the left panel). The data for the Fock term suggest an additional power-law dependence on length. 
	}
	\label{fig:htfo_odd_scal}
\end{figure}

This expectation is fully supported by the numerical results shown in Fig.~\ref{fig:htfo_odd_scal}. Note that, in the case of odd $r$, the Fock term is considerably smaller than the Hartree one (even though the dominant scaling factor is the same). This, the Hartree-Fock cancellation is not operative and $C_{HF} \simeq C_H$.

We have thus found that the Hartree-Fock correlation function $C_{HF}$ is strongly suppressed at criticality (i.e., at short distances $r$ and low energies, so that $r \ll \xi_\epsilon$). This is valid both for even distances (due to cancellation between Hartree and Fock terms) and for odd
distances (due to different sublattices entering). The suppression factor is $\sim \epsilon^4$ for even $r$ and $\sim \epsilon^2$ for odd $r$.

\begin{figure}
	\includegraphics[width=.66\columnwidth]{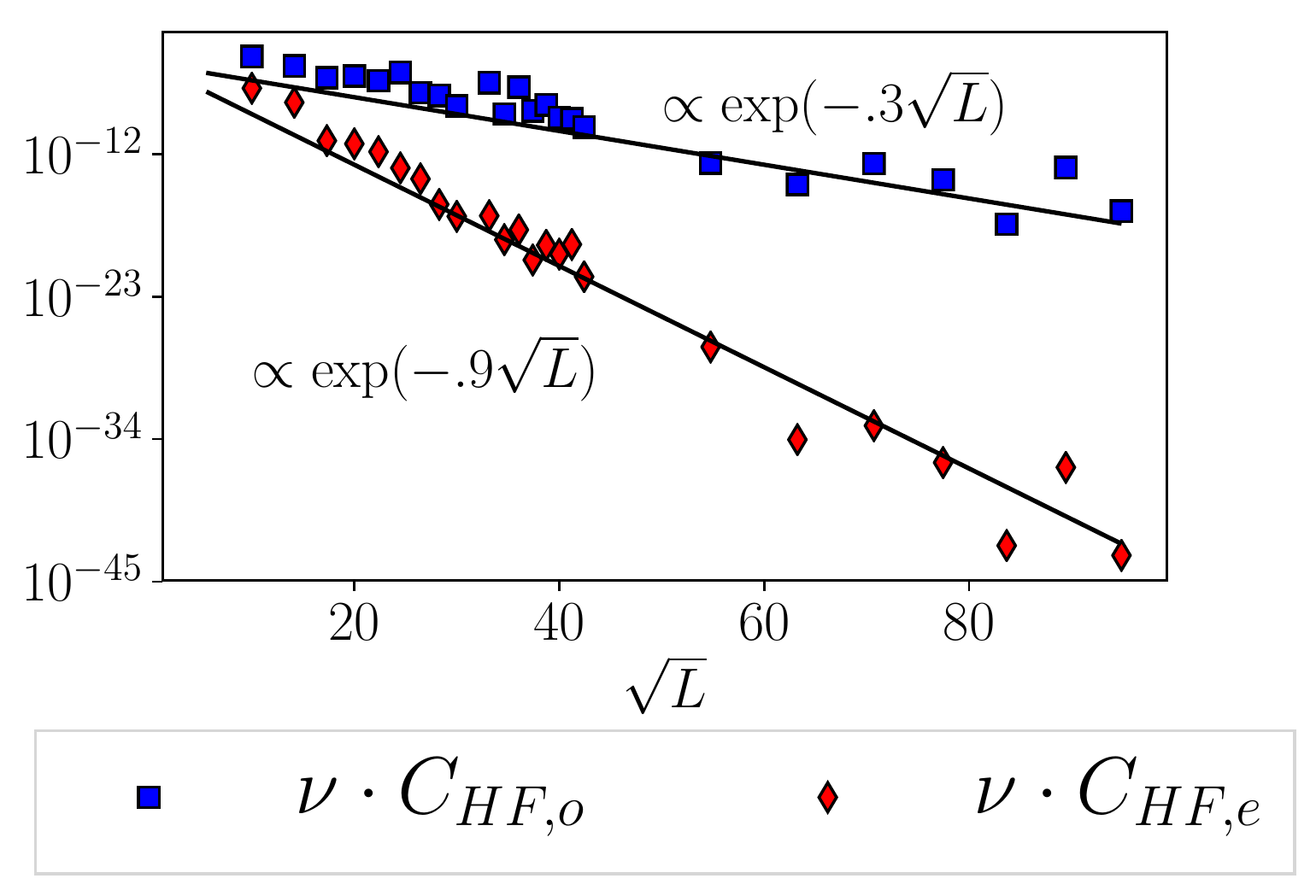}
	\caption{RG irrelevance of interaction at the infinite-randomness fixed point
	of the complex-fermion chain. Product $\nu(L)
	C_{HF}(\epsilon_1,\epsilon_2,r,L)$ of the Hartree-Fock matrix element at
	criticality multiplied by the density of states plotted versus the system size
	$L$, for odd ($r=1$, blue symbols) and even ($r=2$, red symbols). Both for
	even and odd distances, the product decreases quickly with $L$ (as an exponential of $\sqrt{L}$), implying that
	the interaction is irrelevant. }
	\label{fig:fermi_rg}
\end{figure}

We can return now to the question of RG relevance of the interaction which is  determined by Eq.
(\ref{eq:ir_rg}). The right-hand-side of this equation characterizes the scaling of the product of the  interaction matrix element and the density of states with the system size $L$. The matrix element to be used here is the  Hartree-Fock correlation function, see Eqs.~(\ref{correlator-complex-fermions}) and (\ref{corr-wave-func-HF}).
If this product increases (decreases) with $L$, the interaction is relevant (respectively, irrelevant).
The density of states increases exponentially with $\sqrt{L}$ according to Eq.~(\ref{nu-L}) or, equivalently, as $1/\epsilon$ with energy (up to logarithmic correction), see
Eq.~(\ref{nu-epsilon}). On the other hand, the Hartree-Fock correlation function decreases as $\epsilon^2$ (odd $r$) or $\epsilon^4$ (even $r$). Thus, the suppression of the Hartree-Fock correlation function is stronger than the increase of the density of states, and the product decays as a power of $\epsilon$ (and thus exponentially with respect to $\sqrt{L}$).  To illustrate this, we plot in  Fig.~\ref{fig:fermi_rg} the product
$\nu(L) C_{HF}(\epsilon_1,\epsilon_2,r,L)$ for small even and odd distances  ($r=2$ and $r=1$, respectively) as a function of $L$.
We see that both functions decrease exponentially  with  $\sqrt{L}$ as expected.  This implies
that the  interaction in Eq. (\ref{eq:ham_fer}) is
irrelevant  in the presence of disorder, and the system stays critical (at the infinite-randomness fixed point), at
least for sufficiently weak interaction. This is in agreement with our DMRG results in Sec.
\ref{SectionDMRG} and with real-space-RG findings of Refs.~\onlinecite{fisher_random_1994,fisher_critical_1995}.

We have focussed above on the behavior of two-eigenstate correlation functions at criticality ($r \ll \xi_\epsilon$), since such functions emerge when one explores the effect of short-range interaction ($r \sim 1$).  On the other hand, the behavior of the correlation functions at $r \gtrsim \xi_\epsilon$ may be of interest in other contexts.
We briefly discuss this behavior in Appendix \ref{corr-func-away-crit}.

\subsection{Majorana chain}
\label{sec:majorana-correlations}

We turn now to the Majorana model. The simplest interaction term in this model was presented in Eq.~\eqref{eq:ham_disint}. 
However, as was already mentioned before, any fourth order interaction term containing an even number of Majoranas on even sites (and an even number of those on on odd sites) is consistent with the symmetries of the Hamiltonian. In fact, such terms will be generated by RG even if one starts from the simplest term only as in Eq.~\eqref{eq:ham_disint}. 

We generalize first the interaction in Eq.~\eqref{eq:ham_disint} by introducing a distance $r$ separating two nearest-neighbor pairs of Majoranas:
\begin{align}
	H_{\rm int} &= \sum _{j=1}^L \gamma_j\gamma_{j+1}\gamma_{j+r}\gamma_{j+r+1}. 
	\label{eq:ham_majint}
\end{align}
(We will assume $r \ge 2$ to be even but it is not particularly important here.) Such a term is analogous to the odd-$r$ interaction term in the case of complex fermions, see Eq.~(\ref{eq:pj}), since it involves two operators on even sites and two on odd sites. 

We express the Majorana operators $\gamma_i$ in terms of the Bogoliubov
operators $d_\alpha$ using the definitions $c_A=\gamma_A=c^\dagger_A$ and
$c_B=i\gamma_B=-c^\dagger_B$,  and then diagonalizing the Hamiltonian matrix, see Sec.~\ref{scaling-interaction}. 
 At variance with the complex fermion case, these $2L$
Bogolyubov operators are not independent: each operator is related to its
chiral conjugate with inverse sign of the energy, $d_\alpha^\dagger = d_{\bar{\alpha}}$. Thus, we can express the
Majorana operators by using only wavefunctions and Bogolyubov operators
associated with positive energies:
\begin{align}
	\gamma_j&= \sum _{\epsilon_\alpha>0} U_{\alpha,j} (d_{\alpha} + d_{\alpha}^\dagger)&\text{($j$ even),}\\
	\gamma_j&= \sum _{\epsilon_\alpha>0} iU_{\alpha,j} (d_{\alpha} - d_{\alpha}^\dagger)&\text{($j$ odd)}.
	\label{majorana-operators}
\end{align}
Via the same token, the whole Hilbert space of the problem is obtained by acting with operators $d_\alpha^\dagger$ with $\epsilon_\alpha>0$ on the vacuum state. 

Substituting Eq.~\eqref{majorana-operators} into an interaction term in \eqref{eq:ham_majint}, one can evaluate the expectation value 
of the interaction term over any basis state of the non-interacting Fock space. For example, averaging over the non-interacting vacuum (that is annihilated by all $d_\alpha$ with positive energies), we get 
\begin{eqnarray}
&& \langle \gamma_k \gamma_{k+1} \gamma_{k+r} \gamma_{k+r+1} \rangle \nonumber \\
&& =-\sum_{\alpha>0;\beta>0}
\left(U_{k,\alpha}U_{k+1,\alpha}U_{k+r,\beta}U_{k+r+1,\beta}\right.\nonumber \\
&& \left.+ U_{k,\alpha} U_{k+1,\beta}U_{k+r,\alpha}U_{k+r+1,\beta}\right. \nonumber\\
&& \left.-U_{k\alpha}U_{k+1\beta}U_{k+r,\beta}U_{k+r+1,\alpha}\right).
\label{majorana-interaction}
\end{eqnarray}
Three terms here correspond to the expansion of a Pfaffian that is a general form of the Majorana Wick's theorem\cite{bravyi_disorder-assisted_2012}. 

The matrix element in Eq.~\eqref{majorana-interaction} consists of three terms. The first of them is similar to a
Hartree term in the sense that amplitudes of each eigenstates enter at spatial points separated by a minimal distance (one site).
The other two terms are similar to Fock terms. In full  analogy with the case of complex fermions, we define correlation functions depending 
on two energies, distance $r$, and the system size $L$:
$\epsilon_\alpha,\epsilon_\beta$, system size $L$ and distance $r$:
\begin{align}
C_{H,o}(\epsilon_\alpha,\epsilon_\beta,r,L)&= \langle U_{k,\alpha}U_{k+1,\alpha}U_{k+r,\beta}U_{k+r+1,\beta}\rangle _{\rm dis},
\label{CH-majorana-odd} 
\\
C_{F,1,o}(\epsilon_\alpha,\epsilon_\beta,r,L)&= \langle U_{k,\alpha} U_{k+1,\beta}U_{k+r,\alpha}U_{k+r+1,\beta}\rangle_{\rm dis},
\label{CF1-majorana-odd}
\\
C_{F,2,o}(\epsilon_\alpha,\epsilon_\beta,r,L)&= \langle U_{k,\alpha}U_{k+1,\beta}U_{k+r,\beta}U_{k+r+1,\alpha}\rangle_{\rm dis},
\label{CF2-majorana-odd}
\\
C_{HF,o}(\epsilon_\alpha,\epsilon_\beta,r,L)&= \langle U_{k,\alpha}U_{k+1,\alpha}U_{k+r,\beta}U_{k+r+1,\beta} \nonumber\\
&+ U_{k,\alpha} U_{k+1,\beta}U_{k+r,\alpha}U_{k+r+1,\beta}\nonumber\\
&-U_{k,\alpha}U_{k+1,\beta}U_{k+r,\beta}U_{k+r+1,\alpha}\rangle_{\rm dis}.
\label{CHF-majorana-odd}
\end{align}
The subscript ``o'' serves to indicate that, as was explained above, these correlation functions bear analogy with odd-$r$ correlations introduced for the model of complex fermions.

\begin{figure}
	\centering
	\includegraphics[width=.9\columnwidth]{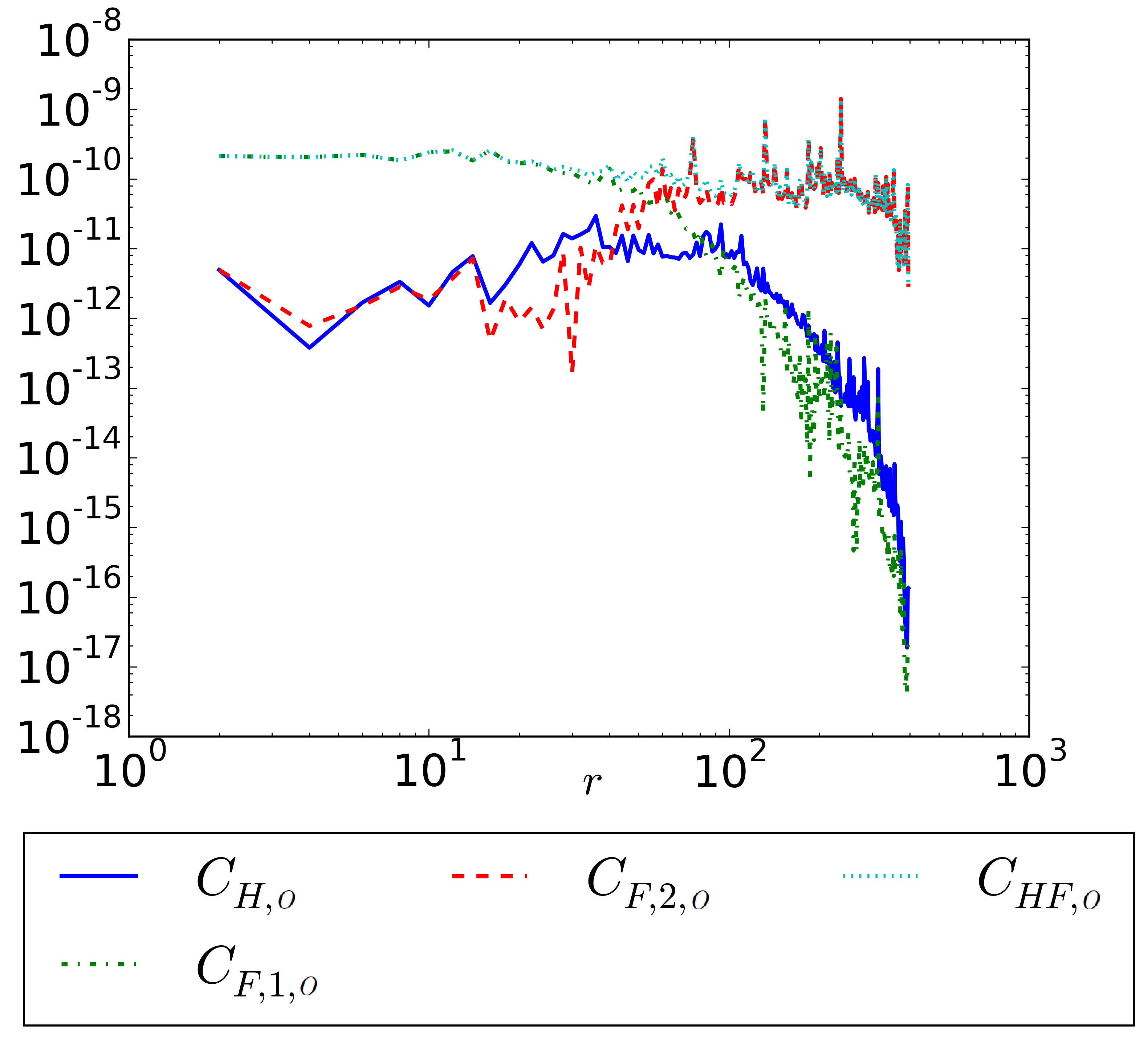}
	\caption{Eigenstate correlation functions (\ref{CH-majorana-odd}) - (\ref{CHF-majorana-odd})
	corresponding to the four-point Majorana interaction for two lowest-energy eigenstates with 
		even $r \ge 2$ in the system of size $L=400$. For sufficiently short distances, $r < \xi_\epsilon$ (critical regime), the term $C_{F,1,o}$ is dominant and $r$-independent. The magnitude of all terms is rather small in view of the $\epsilon^2$ suppression that is demonstrated in Fig.~\ref{fig:majo_e_scal}. }
	\label{fig:majo_e}
\end{figure}

\begin{figure}
	\centering
	\includegraphics[width=.49\columnwidth]{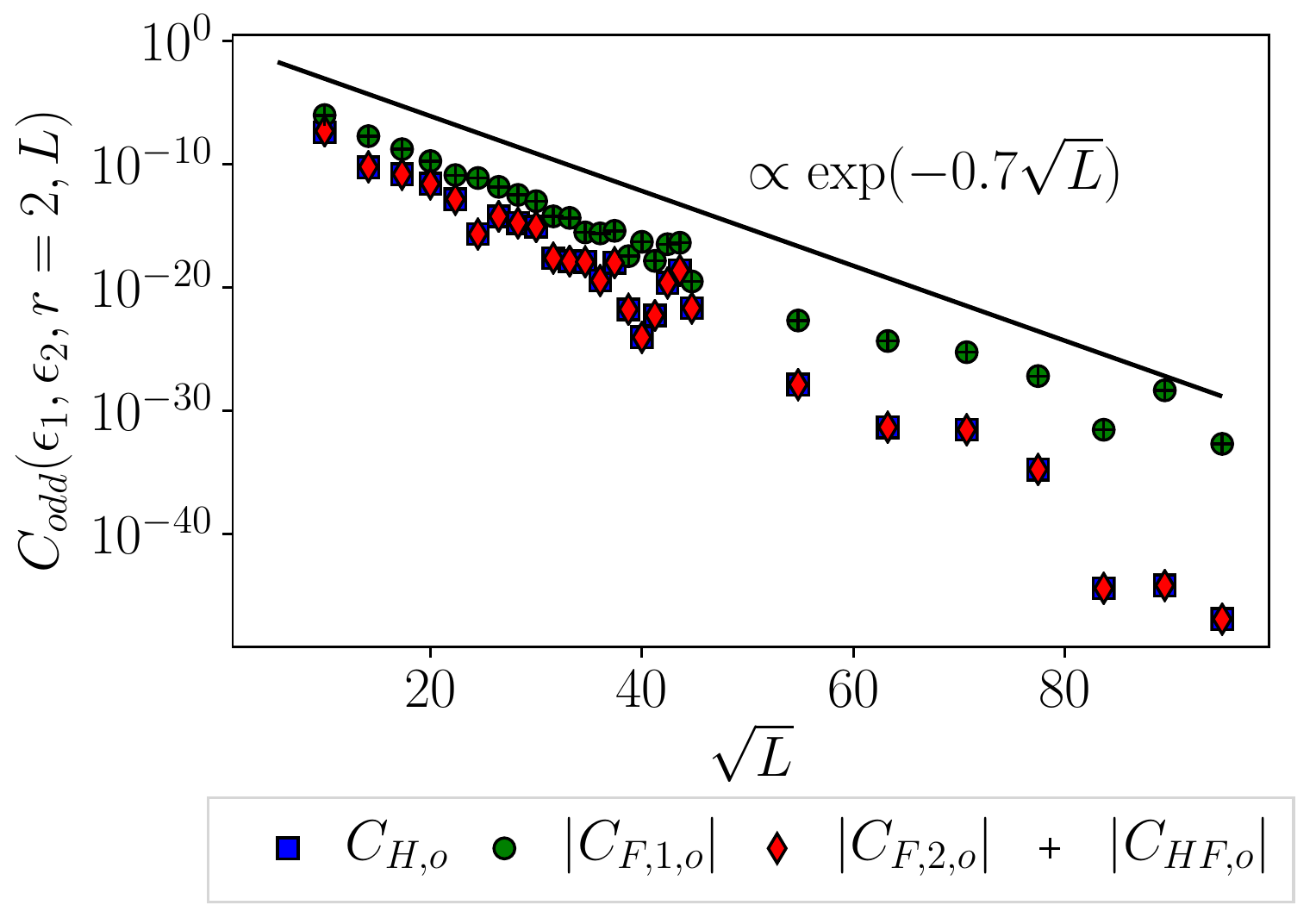}
	\includegraphics[width=.49\columnwidth]{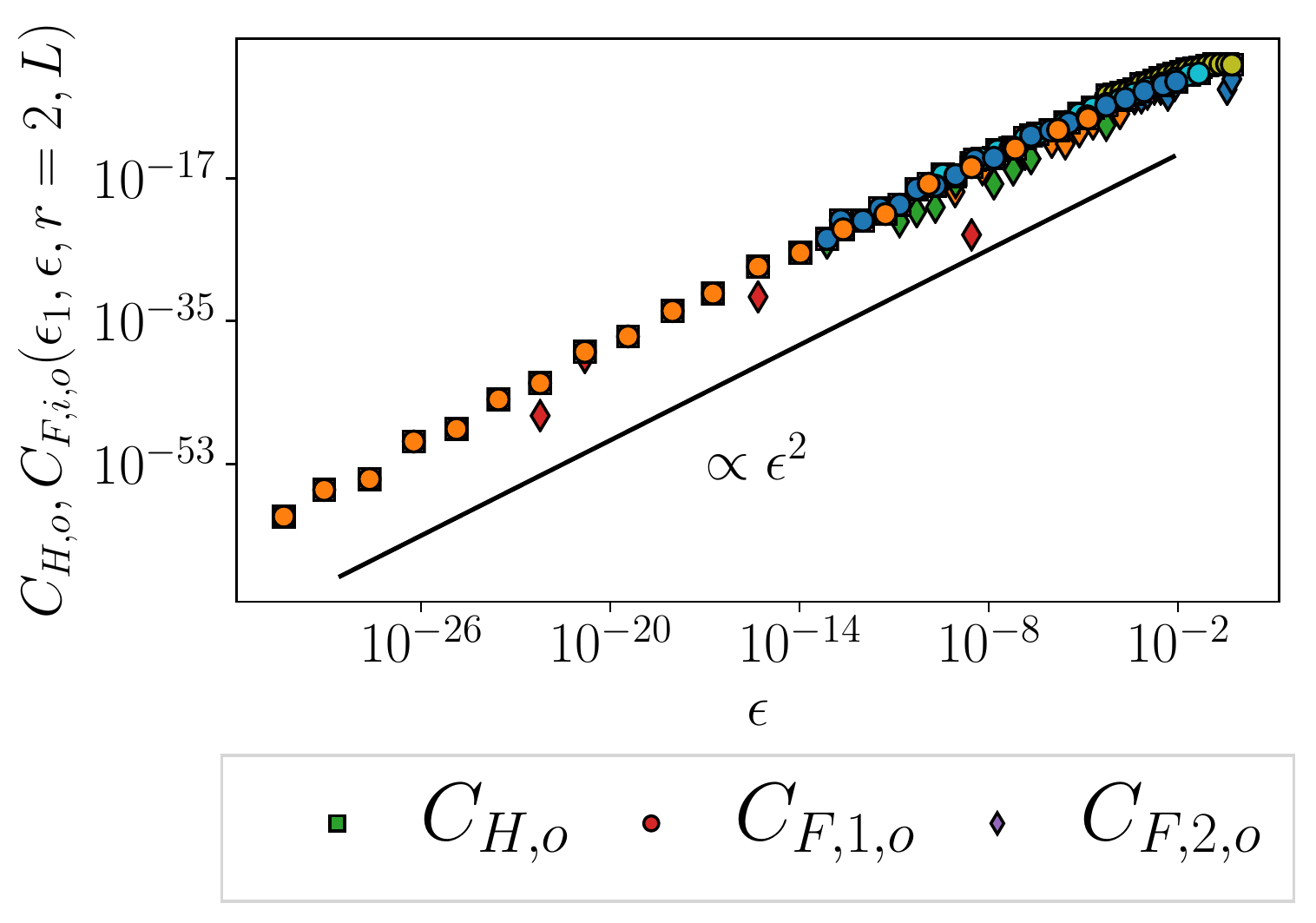}
	\caption{{\it Left:} Scaling of the correlation functions
		$C_{H,o}(\epsilon_1,\epsilon_2,r=2,L)$,
		$C_{F,1,o}(\epsilon_1,\epsilon_2,r=2,L)$, $C_{F,2,o}(\epsilon_1,\epsilon_2,r=2,L)$,
		$C_{HF,o}(\epsilon_1,\epsilon_2,r=2,L)$, Eqs.~(\ref{CH-majorana-odd}) - (\ref{CHF-majorana-odd}) with respect to system size $L$. 	
		{\it Right:} Scaling of the same correlation functions with energy. Different colors represent $L$ from 100 to 10000. The data clearly demonstrated the $\epsilon^2$ scaling that is also expected analytically.}
	\label{fig:majo_e_scal}
\end{figure}

The same analytical consideration as were used in the case of correlation functions \eqref{corr-wave-func-H} - \eqref{corr-wave-func-HF} with odd $r$ suggest that all the correlation functions  (\ref{CH-majorana-odd}) - (\ref{CHF-majorana-odd}) should be suppressed by the factor $\sim \epsilon^2$. We show now by numerical analysis that the correlation functions (\ref{CH-majorana-odd}) - (\ref{CHF-majorana-odd})  indeed behave in a way very similar to the correlation functions \eqref{corr-wave-func-H} - \eqref{corr-wave-func-HF} with odd $r$.  In Fig. \ref{fig:majo_e} we show the $r$-dependence of the correlation functions  (\ref{CH-majorana-odd}) - (\ref{CHF-majorana-odd}) in a system of length $L=400$.
We observe that in the critical regime of not too large $r$  (the condition is $r \ll \xi_\epsilon$) the function $C_{F,1,o}$ dominates. It is also seen that the magnitude of this term is quite small.  To understand the source of this smallness and its parametric dependence,  we show in Fig.~\ref{fig:majo_e_scal} the dependence of the correlation functions on system size $L$ and on the energy $\epsilon$.  The right panel clearly shows the $\epsilon^2$ scaling  that is expected from the analytical argument and is fully analogous to the scaling in Fig.~\ref{fig:htfo_odd_scal}. This is translated into an exponential scaling with respect to $\sqrt{L}$ of correlation functions evaluated on two lowest- energy states, as is seen in the left panel of Fig.~\ref{fig:majo_e_scal} and is again in full analogy with the corresponding behavior in Fig.~\ref{fig:htfo_odd_scal}.

\begin{figure}
	\includegraphics[width=.66\columnwidth]{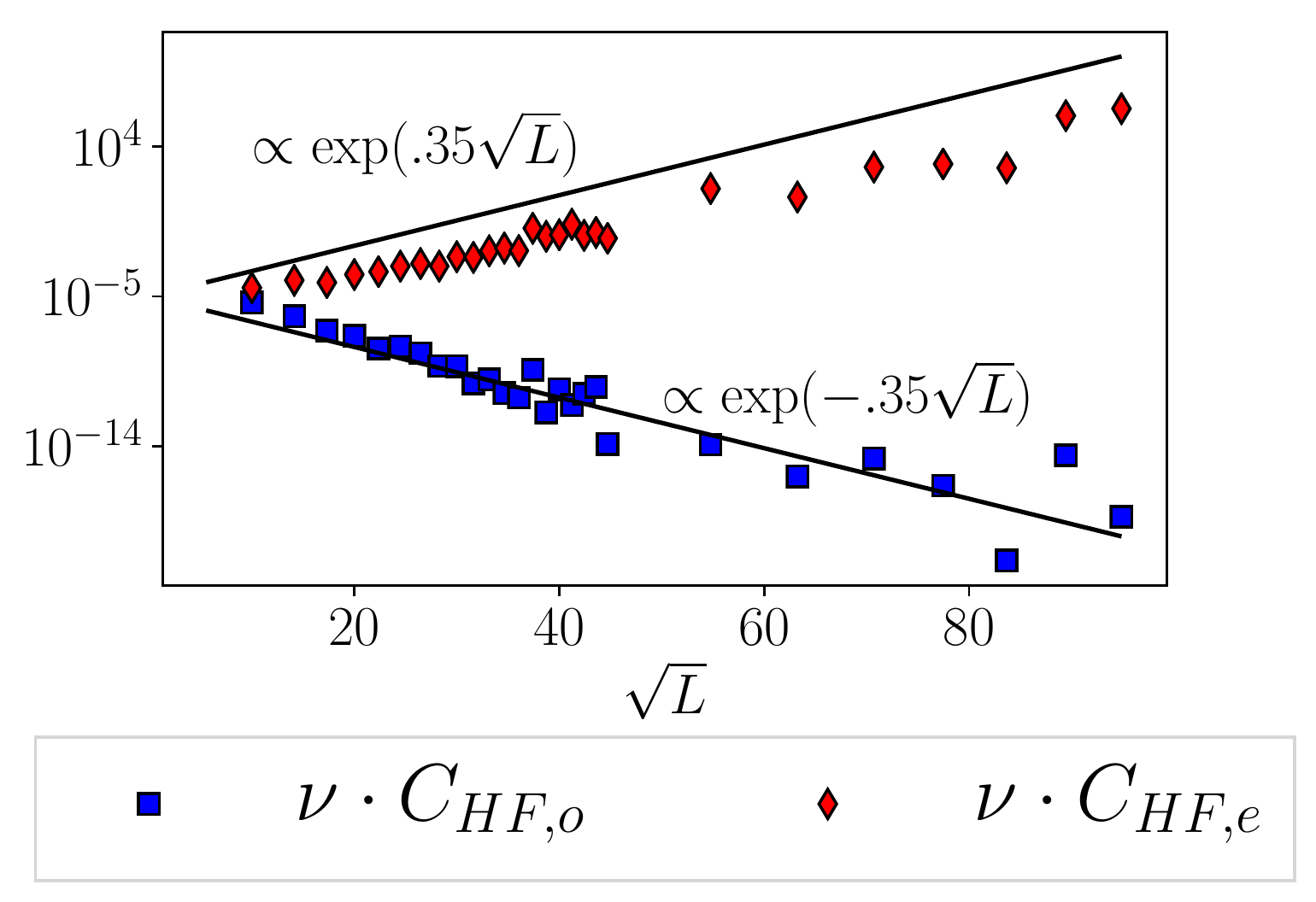}
	\caption{RG (ir)relevance of interaction at the infinite-randomness fixed point
		of the majorana chain. Product $\nu(L)
		C_{HF,\{e,o\}}(\epsilon_1,\epsilon_2,r,L)$ of the Hartree-Fock correlation function and the density of states is plotted versus the system size
		$L$.  Blue symbols: $\nu C_{HF,o}$ for $r=2$ quickly decreases with $L$, implying RG irrelevance of the corresponding interaction terms. Red symbols: $\nu C_{HF,e}$ for $r=4$ quickly increases with $L$, indicating RG relevance of the corresponding interaction term.}
	\label{fig:majorana_rg}
\end{figure}

The $\epsilon^2$ scaling of the correlation functions (\ref{CH-majorana-odd}) - (\ref{CHF-majorana-odd}) implies the RG irrelevance of the corresponding interaction term. Indeed, the density of states increases only as $1/\epsilon$ with logarithmic correction, see Eq.~(\ref{nu-epsilon}), and thus the suppression of the interaction wins over the increase of the density of states. We will verify this numerically below (Fig.~\ref{fig:majorana_rg}). 
As explained above, the reason behind the $\epsilon^2$ suppression of the matrix elements is the fact that both even and odd sites are involved. This tells us which correlation functions may escape such a suppression: those that involve sites of one sublattice only, i.e., with all distances between the sites being even. We thus consider such a generalized interaction term:
\begin{align}
\hat{O}=\gamma_k\gamma_{k+2}\gamma_{k+r}\gamma_{k+r+2},
\label{interaction-majorana-even}
\end{align}
with an even $r\ge 4$.  Such a term is allowed by symmetries and will be generalized by RG from the original interaction. 
This leads us to introduce the corresponding 
generalization  of the correlation functions  (\ref{CH-majorana-odd}) - (\ref{CHF-majorana-odd}):
\begin{align}
C_{H,e}(\epsilon_\alpha,\epsilon_\beta,r,L)&= \langle U_{k,\alpha}U_{k+2,\alpha}U_{k+r,\beta}U_{k+r+2,\beta}\rangle _{\rm dis},
\label{CHe}
\\
C_{F,1,e}(\epsilon_\alpha,\epsilon_\beta,r,L)&= \langle U_{k,\alpha} U_{k+2,\beta}U_{k+r,\alpha}U_{k+r+2,\beta}\rangle_{\rm dis},
\label{CF1e}
\\
C_{F,2,e}(\epsilon_\alpha,\epsilon_\beta,r,L)&= \langle U_{k,\alpha}U_{k+2,\beta}U_{k+r,\beta}U_{k+r+2,\alpha}\rangle_{\rm dis},
\label{CF2e}
\\
C_{HF,e}(\epsilon_\alpha,\epsilon_\beta,r,L)&= \langle U_{k,\alpha}U_{k+2,\alpha}U_{k+r,\beta}U_{k+r+2,\beta} \nonumber\\
&+ U_{k,\alpha} U_{k+2,\beta}U_{k+r,\alpha}U_{k+r+2,\beta}\nonumber\\
&-U_{k,\alpha}U_{k+2,\beta}U_{k+r,\beta}U_{k+r+2,\alpha}\rangle_{\rm dis}.
\label{CHFe}
\end{align}
The subscript ``e'' indicates that all distances between the sites involved are even, in analogy with correlation functions \eqref{corr-wave-func-H} - \eqref{corr-wave-func-HF} at even $r$. 

\begin{figure}
	\centering
	\includegraphics[width=.49\columnwidth]{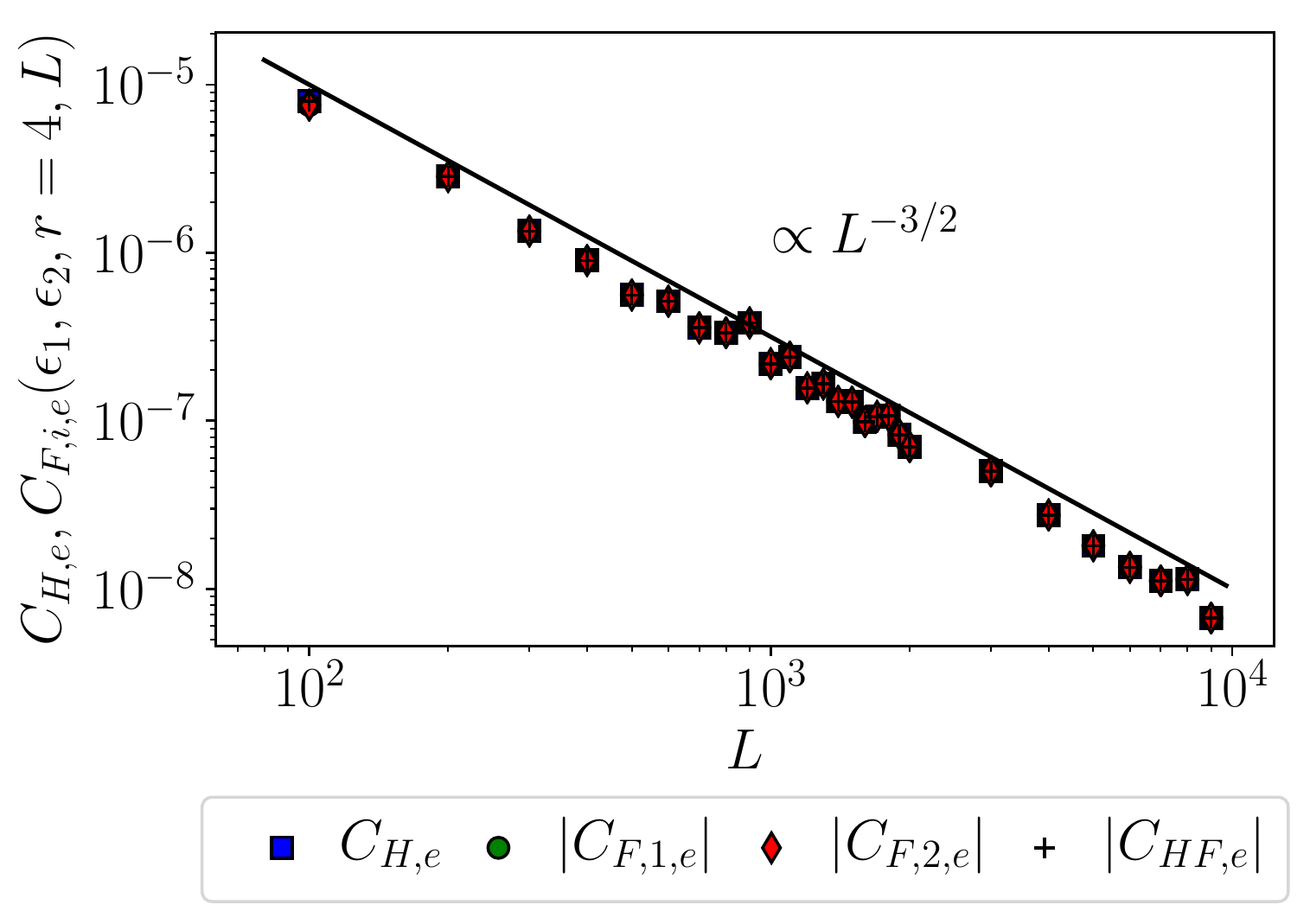}
	\includegraphics[width=.49\columnwidth]{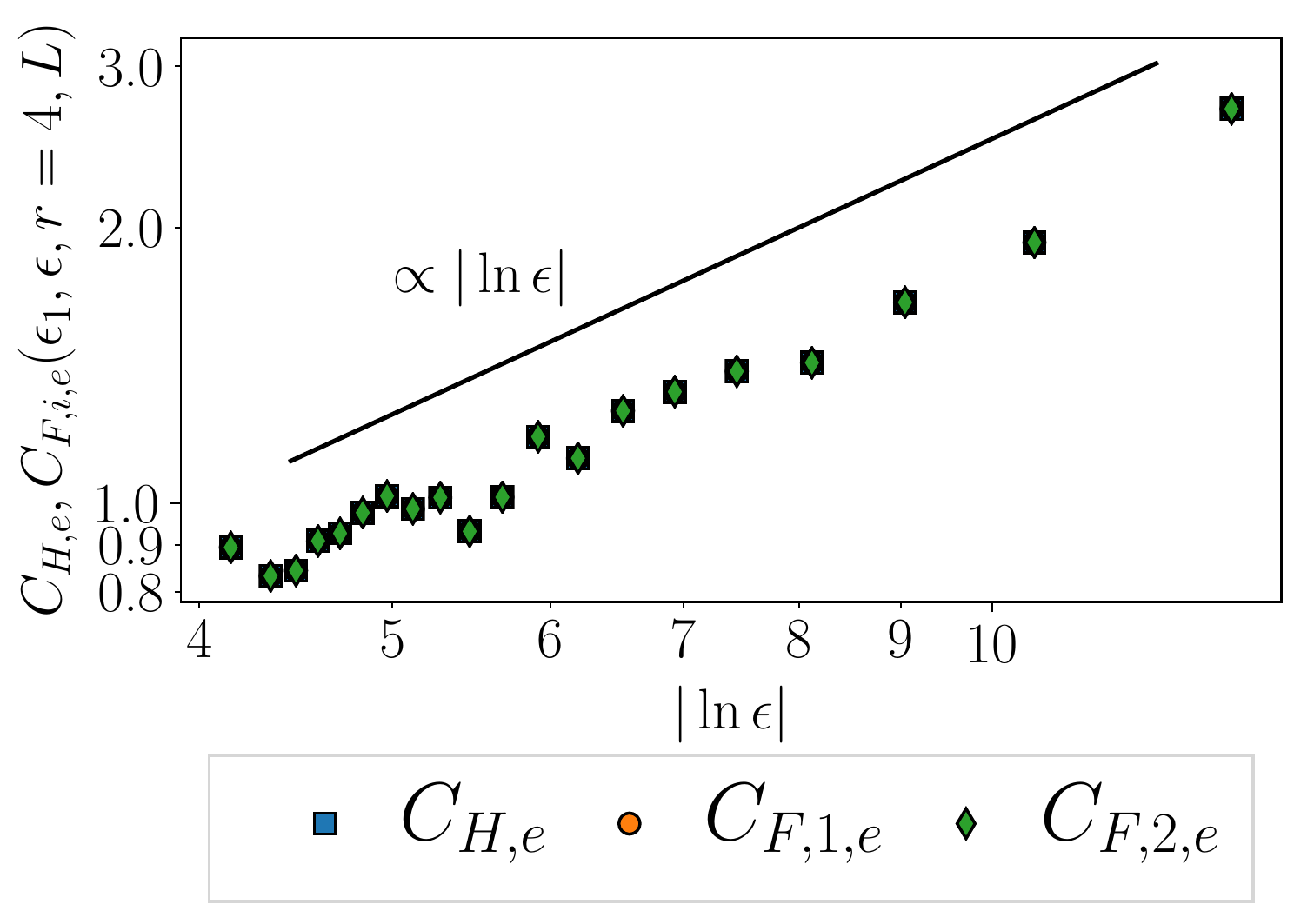}
	\caption{Correlation functions (\ref{CHe})-(\ref{CHFe}) with $r=4$. 
	{\it Left:} Scaling with the system size $L$ of the correlation functions evaluated on two lowest-energy eigenstates.
 The slope corresponds to a power law with
		an exponent $3/2$. 
	{\it Right:} 	Dependence on energy  at fixed $L=4000$. The slope corresponds to the $|\ln \epsilon|$ scaling. 
		 The total scaling with $L$ and $\epsilon$ is therefore the same as for the complex-fermion correlation function $C_H$ with even $r$, Eq.~(\ref{eq:ht_lowr}). }
	\label{fig:ec8_scale}
\end{figure}

In view of the analogy that we have just emphasized, we can expect that (i) the correlation function $C_{H,e}$ scales similarly to $C_H$, \eqref{corr-wave-func-H}, and (ii) the correlation functions $C_{F,1,e}$ and $C_{F,2,e}$ scale in the same way and, moreover, are equal in the leading order to $C_{H,e}$, in analogy with the corresponding behavior of $C_F$, \eqref{corr-wave-func-F}. However, since we now have three terms rather than two, the strong Hartree-Fock compensation should not happen, leaving us with 
$C_{HF,e} \simeq C_{H,e}$. These expectation are fully supported by the numerical simulations. In Fig.~\ref{fig:ec8_scale} we show the $r$ dependence of the correlation functions \eqref{CHe} - \eqref{CHFe} evaluated on two lowest-energy eigenstates in a system of size $L=400$.  All four  correlations functions $C_{H,e}$, $C_{F,1,e}$, $C_{F,2,e}$, and $C_{HF,e}$ are nearly equal in the critical regime (not too large $r$) and show the $r^{-3/2}$ scaling in analogy with $C_H$ and $C_F$. In fact, the overall behavior 
of the correlation function $C_{H,e}$  ($C_{F,1,e}$ and $C_{F,2,e}$) in Fig.~\ref{fig:ec8_scale} is remarkably similar to that of $C_H$ (respectively, $C_F$)  in Fig.~\ref{fig:htfo_even_scal}. 
We turn now to the scaling  of the correlation functions \eqref{CHe} - \eqref{CHFe} with energy $\epsilon$ and length $L$, see Fig.~\ref{fig:ec8_scale}. The figure is very similar to the upper two panels of Fig.~\ref{fig:htfo_even_scal} and confirms that $C_{H,e}$, $C_{F,1,e}$, and $C_{F,2,e}$  scale exactly in the same as $C_H$ with even $r$, \eqref{eq:ht_lowr}. Since the Hartree-Fock compensation is not operative now, the correlation function $C_{HF,e}$ scales in the same way. 

\begin{figure}
	\centering
	\includegraphics[width=.9\columnwidth]{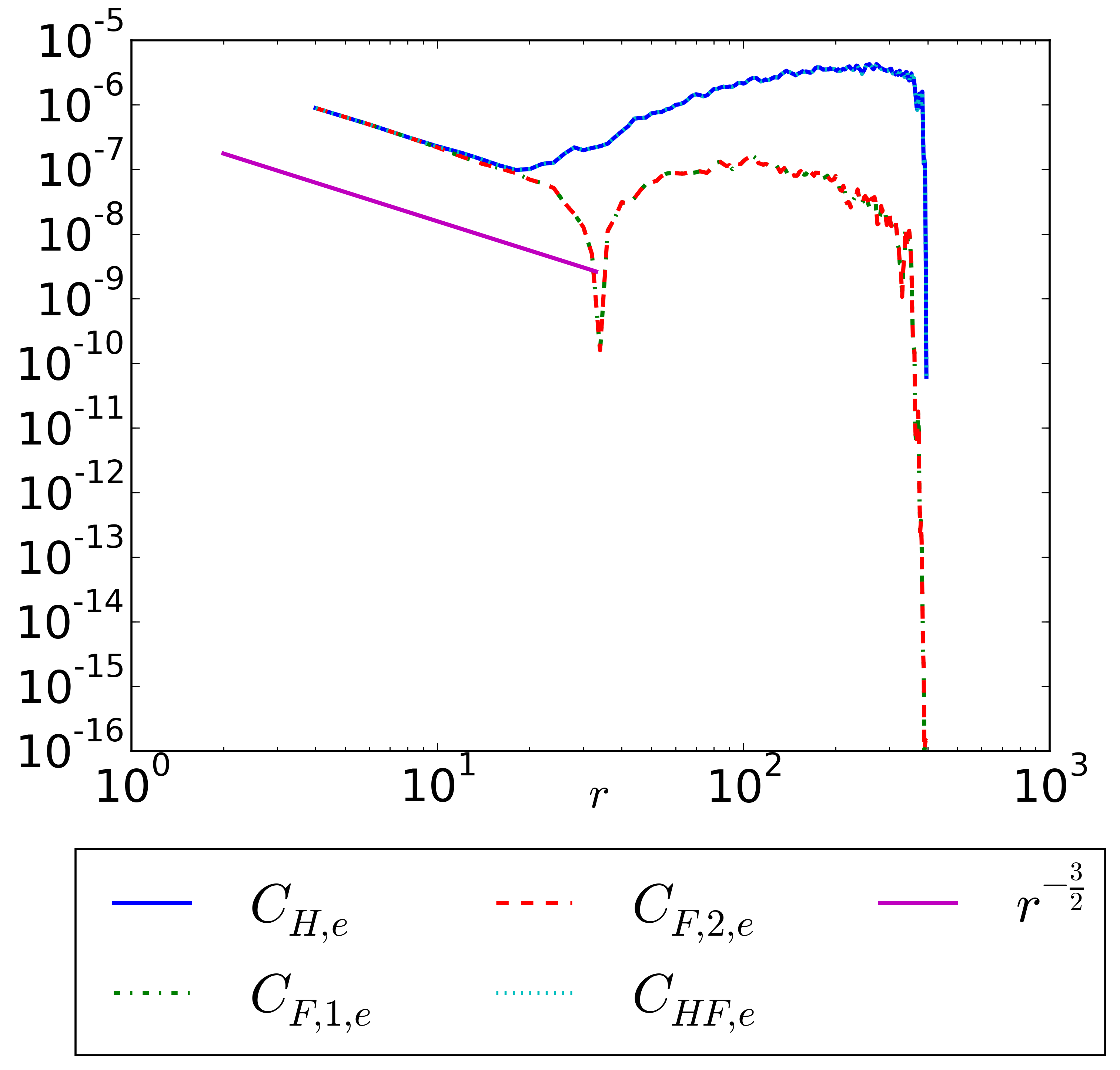}
	\caption{Correlation functions
		$C_{H,e}(\epsilon_1,\epsilon_2,r,L=400)$,
		$C_{F,1,e}(\epsilon_1,\epsilon_2,r,L=400)$,
		$C_{F,2,e}(\epsilon_1,\epsilon_2,r,L=400)$, and 
		$C_{HF,e}(\epsilon_1,\epsilon_2,r,L=400)$ evaluated on two lowest-energy eigenstates, as functions of even $r$. 
		The functions $C_{H,e}$, $C_{F,1,e}$, and $C_{F,2,e}$ are nearly equal to each other and scale as $r^{-3/2}$. 
		In $C_{HF,e}$ two out of three terms approximately cancel, leaving $C_{HF,e}\simeq C_{H,e}$.
		}
	\label{fig:ec8}
\end{figure}

Since the correlation function $C_{HF,e}$ decreases with $L$ in a power-law fashion only, and the density of states increases in an exponential way, they product should clearly increase exponentially. This is explicitly demonstrated in Fig.~\ref{fig:majorana_rg}. For comparison, we also show there the product $\nu C_{HF,e}$ that decreases with increasing $L$ as discussed above. The exponential increase of $\nu C_{HF,e}$ indicates the RG relevance of the corresponding interaction term. This explains why the interaction drives the system away from the  the infinite-randomness fixed point and establishes the spontaneous symmetry breaking and localization,  as exhibited by the DMRG results,  Sec.~\ref{sec:majorana-int-dis-dmrg}. 

At this point, the following comment is in order. The completeness of eigenstates in combination with the chiral symmetry implies that $\sum_{\epsilon_\alpha > 0} U_{k,\alpha} U_{k+r,\alpha}$ is equal to zero for any even $r \ne 0$. As a result, the correlation functions \eqref{CHe} - \eqref{CHFe} are zero when summed over all states with positive energies. Exactly such sums will arise if we calculate the expectation of the interaction (\ref{interaction-majorana-even}) over the vacuum state (or, more generally, over any  Fock-space basis state). However, what we are actually interested in is not this expectation value but rather the effect of non-diagonal matrix elements of the interaction. In more conventional problems, it turns out that it is sufficient to study the scaling of the expectation value to understand the effect of the interaction. It turns out that the situation with the term of the type (\ref{interaction-majorana-even})  in the present problem is more delicate. The full analysis of the effect of non-diagonal matrix elements of such an interaction at the infinite-randomness fixed point is a very challenging task that we leave to future work. We expect that two properties of the correlation functions \eqref{CHe} - \eqref{CHFe} that we have identified above---namely, (i) the contributions that, when multiplied with the density of states, strongly increase with $L$ and (ii) the absence of Hartree-Fock cancellation of such contributions---will be also key ingredients of such a more sophisticated analysis, thus governing the RG relevance of the interaction for the disordered Majorana chain.

\section{Summary and Outlook}
\label{summary}

The main goal of this work was the investigation of the low-energy physics of a
chain of Majorana fermions  in the presence of interaction and disorder. One of
intriguing questions was a difference between this interacting Majorana problem
and the 1D model of interacting complex fermions with chiral symmetry that
belongs to the same symmetry class BDI.  In the absence of interaction, both
models are equivalent (apart from halving the number of states in the Majorana
case), and flow into the same infinite-randomness fixed point. It turns out that
the interaction makes them drastically different. To explore and understand the
physics of these models, we have used a combination of several computational and
analytical approaches, including DMRG, mean-field analysis, and two different
types of RG (around the clean interacting fixed point and around the
non-interacting disordered fixed point). The latter type of RG required
investigation of statistical properties of eigenfunction correlations at
infinite-randomness fixed point, which has turned out to be a very interesting
and non-trivial problem by itself. Our key results can be summarized as follows:

(1)  We have carried out the DMRG analysis of the models (in their spin
representations), by calculating the entanglement entropy as well as the
spin-spin correlation functions. This has allowed us to determine the
corresponding phase diagrams and to understand some physical properties of the
emerging phases. More specifically:

(i) We have first considered an interacting Majorana chain with staggering, see
Figs.~\ref{fig:phase_cc} and \ref{fig:phase_sz} for the color-code
representation of the entanglement entropy and the spin correlations in the
interaction-staggering plane. The obtained phase diagram is shown in
Fig.~\ref{fig:phase_skizze}. On the no-staggering (self-dual) line we observe
the Ising (central charge $c=\frac12$) and Ising+LL ($c=\frac32$) phases, in agreement
with  Ref.~\onlinecite{rahmani_phase_2015}. Away from the self-dual line (i.e.,
in the presence of staggering), we find gapped phases as well as a LL critical
phase with $c=1$.  The distinct character of phases manifests itself in the
spatial dependence of the spin-spin correlation functions,
Fig.~\ref{fig:cleanweak}. The $c=1$ critical phase can be understood as the
result of gapping the Ising sector of the LL+Ising phase, with LL sector
remaining gapless. The gapped phases on both side of the self-dual line are
topologically distinct. We have also found interesting parts of the gapped
phases with entanglement entropy showing relatively sharp maxima at points where
the antiferromagnetic ordering of spins experience certain ``phase slips''.

(ii) We have then applied the  DMRG analysis to interacting disordered Majorana
chains. Here we focussed on the systems without staggering, which are critical
in the absence of disorder.  In the case of attractive interaction, our DMRG
results on entanglement show that the system remains critical also in the
presence of disorder. Moreover,  as shown in Fig.~\ref{fig:att_ent}, we find
(within the numerical accuracy) the same value of the central charge, $c=\frac12$,
as for the clean system. The situation is radically different for the repulsive
interaction, where we find that the system gets localized. This happens already
for weak repulsion (for which the clean system hat $c=\frac12$ central charge), as
is seen from the behavior of the entanglement entropy, Fig.~\ref{fig:weak_ent}.
The behavior of the spin correlation function, Fig.~\ref{fig:disweak},
demonstrates that the system finds itself spontaneously in one of two
topological phases. A similar behavior is observed for the intermediate strength
of the interaction, Figs.~\ref{fig:med_ent} and \ref{fig:disfloating}. Thus, an
interplay of repulsive disorder and interaction leads to a spontaneous symmetry
breaking that results in localization and topological ordering.

(iii) In the case of disordered interacting complex fermions, the DMRG shows
(both in the cases of attraction and repulsion) the same behavior as for the
non-interacting model. Specifically, the found value of the central charge is $c
= \ln 2$, Fig.~\ref{fig:fermi_dmrg}, which is a hallmark of the
infinite-randomness fixed point.

(2)   As a first attempt to add analytical understanding to the numerical
results, we have developed a weak-disorder RG  in spirit of Giamarchi-Schulz.
This was done in the vicinity  of all three clean critical theories: $c=\frac12$ and
$c=\frac32$ for Majorana chain and $c=1$ for complex fermions. In all the cases, the
disorder is RG-relevant and drive the system away from the corresponding clean
fixed point, towards the strong-disorder regime. Therefore, this approach is not
sufficient for exploring  the infrared behavior of the models.

(3)  The flow of disorder to strong coupling has motivated an alternative RG
analysis, in which the starting point is the non-interacting disordered theory
that is at the strong-randomness fixed point. Investigation of the effect of
interaction requires understanding of the scaling of eigenfunction correlations
at this fixed point. This theory is a remarkable strong-disorder
Anderson-localization critical theory, and the corresponding eigenfunction
statistics is also highly interesting on its own, so that we have studied it in
some detail. For the Hartree-type correlations of two eigenfunctions at even
distance $r$, we have determined, by combination of numerical and analytical
means, the critical scaling (\ref{eq:ht_lowr}). This formula shows that, in
analogy with Anderson-transition critical points in higher dimensions,
correlations are strongly enhanced at criticality (small $\epsilon$) and at
small $r$. On the other hand, for odd $r$ the correlations turn out to be
strongly suppressed at criticality, in view of the chiral symmetry. Furthermore,
we show that a strong cancellation between the Hartree and Fock terms leads to a
strong suppression of Hartree-Fock correlation function also for even $r$. We
have shown that this suppression overweights the divergence of the density of
states at criticality, Fig.~\ref{fig:fermi_rg}.  As a result, the interaction
turns out to be RG-irrelevant at the strong-disorder fixed point for the
complex-fermion model, in full consistency with the corresponding DMRG results.

For Majorana problem, the interaction matrix elements involves four sites. For
even separation between the sites, the critical scaling of the corresponding
eigenstate correlation functions, Fig.~\ref{fig:ec8}, is analogous to that of
two-point correlation function $C_H$, Eq.~(\ref{eq:ht_lowr}). The crucial
difference is that in the Majorana case, for given two eigenstates and a give
set of spatial points there are three contributing correlations functions
instead of two (Hartree and Fock) in the complex-fermion case. As a result, the
Hartree-Fock cancellation is not operative in the Majorana problem, and the
interaction is relevant at the infinite-randomness fixed point. This is again
consistent with the DMRG results and explains a dramatic difference between the
behavior of interacting disordered Majorana chains and that of its
complex-fermion counterpart.

Before closing the paper, we make several comments on possible extensions of our
work that represent prospective directions for future research.

(i) It would be interesting to extend our analysis of disordered interacting
Majorana systems to higher-dimensional systems including quasi-1D (ladders) and
2D geometry. Clean version of such models was studied in
Ref.~\onlinecite{rahmani_Majorana-hubbard_2019}.

(ii) Another potential extension concerns the symmetry class. We recall that the Majorana model that we have considered in this paper belongs to the symmetry class BDI. If the sublattice symmetry is violated, the system will be in the symmetry class D. It would be interesting to study the interacting Majorana models of this symmetry class in 1D, quasi-1D, and 2D geometry. In particular, an intriguing question is how generic is the difference between interacting Majorana and complex-fermion models from the same symmetry class.

(iii) \ADDED{Our numerics show that the behavior of the disordered Majorana chain differs strongly for attractive and repulsive interaction. Specifically, we find localization in the repulsive case, whereas the system remains critical for attractive $g$, see right panel of Fig.~\ref{fig:att_ent}. Analytical understanding of the impact of the sign of the interaction would be desirable.}
Further, the physics of the disordered attractive interaction case itself deserves a more detailed study. The numerical data suggest the value $c=\frac12$ of the central charge, different from the value $c= (\ln 2)/2$ characterizing the non-interacting system. This difference is consistent with our finding that the interaction in Majorana chain is relevant at the infinite-randomness fixed point of the non-interacting system.  On the other hand, we also know that the disorder is relevant at the clean fixed point, so that the coincidence of the found central charge with that of the clean system appears surprising. Further investigation of other physical observables should help to clarify the precise physical nature of this phase.

(iv)  The spontaneous symmetry breaking in disordered interacting Majorana
chains, which leads to localization and topological order, calls to exploring
the physics of these systems at high temperatures. It is expected that they will
undergo a many-body (de-)localization transition accompanied by restoration of
symmetry. Transitions between many-body localized and ergodic phases have attracted a
great deal of attention in recent years.\cite{gornyi_interacting_2005,
pal_many-body_2010,nandkishore_many-body_2015,abanin_many-body_2018}

(v) A complete analysis of statistical properties of various eigenfunction
correlations (also those including a larger number of eigenstates and/or spatial
points) at the infinite-randomness fixed point represents a very interesting (and also very challenging) problem. This fixed point represents an intriguing example
of a strong-disorder Anderson-localization critical theory. In fact, it was argued in Ref. \onlinecite{gruzberg_localization_2005} that a ``superuniversality'' holds in the sense that the same fixed point describes critical theories of all five symmetry classes (BDI, AIII, CII, D, DIII) that can host 1D topological insulators according to the ``periodic table''. 
 This fixed point exhibits
criticality in various observables, but at the same time many properties are
similar to those in the localized phase. In this respect, this non-interacting
1D critical point bears a certain similarity with the transition between the
localized and ergodic phases on random regular graphs
\cite{tikhonov_statistics_2019} that serves as a toy-model for the many-body
localization transition.

\ADDED{(vi) Another interesting generalization of our models is including disorder in the interaction terms. For relatively weak randomness of the interaction, the results derived here are expected to retain validity, since such terms are generated anyway during RG flow.
On the other hand, if the random interaction is a dominant part of the Hamiltonian, the models will resemble those of Sachdev-Ye-Kitaev (SYK) type \cite{sachdev_gapless_1993}. 
The scenario of not fully quenched kinetic energy is considered in Refs. \cite{altland_syk_2019, altland_quantum_2019}, where coupled quantum dots are studied. It would be interesting to see whether the SYK-like physics may emerge in our model in the case of strong random interaction. In this case, one could study a crossover between the SYK and the infinite-randomness fixed point. }

\section{Acknowledgements}

We gratefully acknowledge collaboration with N. Kainaris at the early stage of this work. We also thank E. Doggen and K. S. Tikhonov for help with numerical simulations and M. Foster for discussions and for sharing his unpublished notes on 1D chiral-class models. The work was supported by the Deutsche Forschungsgemeinschaft via the grant MI 658/7-2 (Priority Programme 1666 ``Topological Insulators'').

\appendix

\section{Weak-disorder RG around the Ising + LL fixed point of the interacting Majorana chain}
\label{AppendixRG}

In this Appendix, we provide details of the weak-disorder RG treatment of the interacting Majorana chain in the Ising+LL fixed point, Sec.~\ref{sec:ising-luttinger-weak-disorder-rg}. The starting point is the effective mean-field Hamiltonian \eqref{eq:rg:H0} including the third-nearest-neighbor hopping as well as a weak randomness
in the nearest-neighbor hopping $t+\delta t_j$, supplemented with the interaction term $g\gamma_j\gamma_{j+1}\gamma_{j+2}\gamma_{j+3}$.

Using  the low energy expansion \eqref{eq:lowdec} for the nearest-neighbor hopping operator $\gamma_j\gamma_{j+1}$ yields oscillatory contributions
with wave vectors $k_i =0$, $k_0$, $k_0+\pi$, $2k_0$, $2k_0+\pi$, and $\pi$ that can be dropped in the clean case. In the presence of randomness, they couple, however, to the corresponding Fourier harmonics of disorder  $\delta t_j$.  We employ the replica trick  to average over disorder. As a result, the following terms in the action representing effective ``interactions'' between different replica species $a, b$ are generated:
\begin{widetext}

\begin{eqnarray}
S_{k_0} &=& -\dfrac{8(1-\cos k_0)}{\pi a}D_{k_0}\int \mathrm{d}x\mathrm{d}\tau\mathrm{d}\tau'\sum_{a,b}\left[ i \gamma_L^a\gamma_L^b\sin(\phi_a+\theta_a-\phi_b-\theta_b)\right.\nonumber\\
& + & \left. i\gamma_R^a\gamma_R^b\sin(\phi_a-\theta_a-\phi_b+\theta_b)
-i\gamma_L^a\gamma_R^b\sin(\phi_a+\theta_a-\phi_b+\theta_b)\right],  \nonumber \\
S_{k_0+\pi} &=& -\dfrac{8(1+\cos k_0)}{\pi a}D_{k_0+\pi}\int \mathrm{d}x\mathrm{d}\tau\mathrm{d}\tau'\sum_{a,b}\left[ i \gamma_L^a\gamma_L^b\sin(\phi_a-\theta_a-\phi_b+\theta_b)\right.\nonumber\\
&+ &\left.  i\gamma_R^a\gamma_R^b\sin(\phi_a+\theta_a-\phi_b-\theta_b)
-i\gamma_L^a\gamma_R^b\sin(\phi_a-\theta_a-\phi_b-\theta_b)\right],    \nonumber  \\
S_{2k_0} &=&  -\dfrac{1}{\pi^2 a^2}D_{2k_0}\int \mathrm{d}x\mathrm{d}\tau\mathrm{d}\tau'\sum_{a,b}\cos(2\phi_a-2\phi_b)\sin(2\theta_a)\sin(2\theta_b), \nonumber \\
S_{2k_0+\pi} &=&  -\dfrac{1}{(\pi a)^2}D_{2k_0+\pi}\int \mathrm{d}x\mathrm{d}\tau\mathrm{d}\tau'\sum_{a,b}\cos(2\phi_a-2\phi_b), \nonumber \\
S_{\pi} &=& -\dfrac{8}{\pi a}D_{\pi}\int \mathrm{d}x\mathrm{d}\tau\mathrm{d}\tau'\sum_{a,b}\left[ 4 \gamma_L^a\gamma_R^a\gamma_R^b\gamma_L^b+\cos^2k_0 \cos(2\theta_a)\cos(2\theta_b)\right]
\label{eq:rg:terms}.
\end{eqnarray}
\end{widetext}
Each term $S_{k_i}$ is labeled by the corresponding momentum component $k_i$.
Some of the terms allow for a simple physical explanation. In particular, the action term
$S_{2k_0+\pi}$ represents the backscattering between the right and left Fermi-point of the emergent Luttinger-liquid sector, while
$S_{\pi}$ corresponds to backscattering processes commensurate with the lattice.
The RG equations summarized in Table \ref{tab:couplings} and Eq. \eqref{eq:rg:end0} are then inferred in analogy with Ref. \onlinecite{giamarchi_anderson_1988}. The most relevant terms are $S_{2k_0+\pi}$ and $S_{\pi}$. The contribution of the term $S_{2k_0+\pi}$  to the renormalization of $K$, Eq.~\eqref{eq:rg:end0}, is analogous to backscattering in Giamarchi-Schulz RG. For the other term, $S_{\pi}$, the duality exchanging $\phi \leftrightarrow \theta$ and $K \leftrightarrow K^{-1}$ may be used to find the contribution to $K$. 

While the forward scattering can be completely gauged away in the standard Giamarchi-Schulz RG, here the transformation gauging it out generated additional terms. However, a direct inspection shows that they are irrelevant in the RG sense.

The interaction generates a replica-diagonal term that couples the Luttinger-liquid and Majorana sectors:
\begin{align}
S_{int} &= - g' \int \mathrm{d}x\mathrm{d}\tau\sum_a \gamma_L^a \gamma_R^a (\Psi_L \Psi_R + \Psi_L^{\dagger} \Psi_R^{\dagger})\nonumber\\
&=- 2 g' \int \mathrm{d}x\mathrm{d}\tau\sum_a \gamma_L^a \gamma_R^a \cos(2\theta_a) \label{eq:rg:int}.
\end{align}
This term is RG-irrelevant in the range of interest, $K<1$;  the corresponding dimensional coupling is denoted $y'$ in Table  \ref{tab:couplings}.  Higher terms respecting the symmetry are, of course, also generated. It can be checked by dimension counting that all terms arising due to interaction remain irrelevant in the range $1/4 < K < 1$.

\section{Origin of low-energy suppression of wave function correlations in disordered complex-fermion chain}
\label{AppendixMF}

In this Appendix, we present analytical arguments explaining the origin of the suppression of eigenstate correlations in a complex-fermion chain at low energies found numerically in Sec.  \ref{sec:complex-fermion-correlations}. An eigenvector $U_{i+1,\epsilon}$ of Hamiltonian \eqref{eq:bdimat} fulfills the following transfer matrix equation:
\begin{align}
\begin{pmatrix}
U_{i+1,\epsilon}\\
U_{i,\epsilon}
\end{pmatrix}
&=
\begin{pmatrix}
\epsilon /t_{i+1} & -t_i/t_{i+1}\\
0 & 1
\end{pmatrix}
\begin{pmatrix}
U_{i,\epsilon} \\
U_{i-1,\epsilon}
\end{pmatrix} \label{eq:transfer}
\end{align}
For zero energy, $\epsilon=0$, two sublattices are decoupled, so that the wave function lives on one sublattice. For finite (but small) $\epsilon$ the wave function on the second sublattice is suppressed by $\epsilon$. This implies the suppression of the correlation functions $C_2(\epsilon,r,L)$,
$C_H(\epsilon_\alpha,\epsilon_\beta,r,L)$, and $C_F(\epsilon_\alpha,\epsilon_\beta,r,L)$  for odd $r$ by a factor $\sim \epsilon_>^2$, where $\epsilon_>$ is the larger of two energies $\epsilon_\alpha,\epsilon_\beta$. This suppression is indeed numerically observed, see Fig.~\ref{fig:htself_e} and the right panel of Fig.~\ref{fig:htfo_odd_scal} which make evident the $ \epsilon_>^2$ scaling of the odd-$r$ correlation functions. As is seen in this figure, for odd $r$ the Fock term is substantially smaller than the Hartree one, so that there is no cancellation between them and $C_{HF} \simeq C_H$.

\begin{figure}
	\centering
	\includegraphics[width=.85\columnwidth]{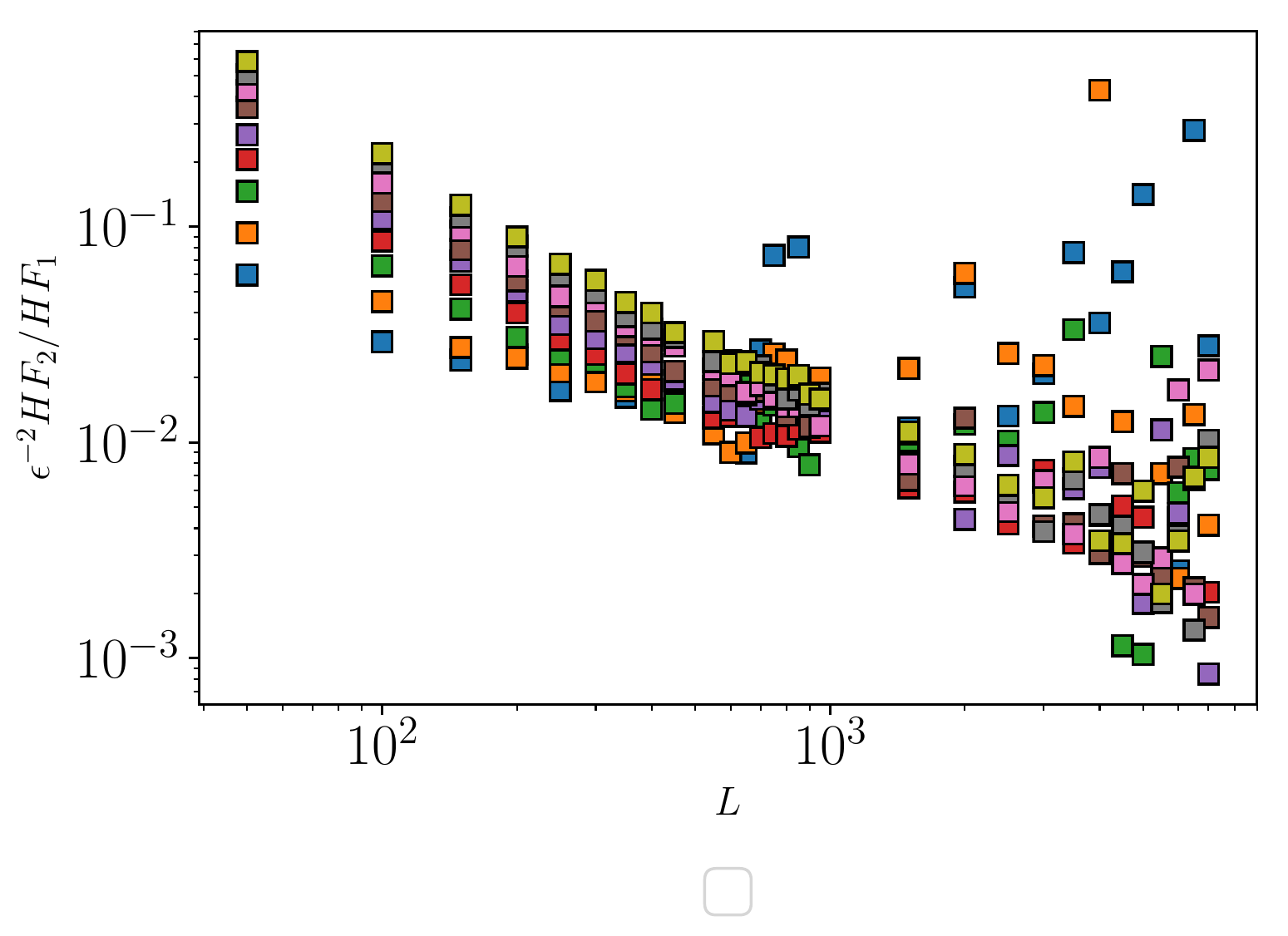}
	\caption{Ratio $C_ {HF}(\epsilon_1,\epsilon_n,2,L) / C_ {HF}(\epsilon_1,\epsilon_n,1,L)$  multiplied by $\epsilon_n^{-2}$ for the first twenty levels $n$ (in distinct colors) as a function of $L$. The data show only a weak (at most linear) dependence on $L$ (that corresponds to a logarithmic energy dependence), which should be contrasted to the exponential $L$ dependence of both entering correlation functions $C_ {HF}$ and of the energy $\epsilon_n$.  This confirms the analytic prediction in Eq. (\ref{even-odd-ratio}), with possible logarithmic-in-energy corrections. }
	\label{fig:hf_even_e2}
\end{figure}

For even $r$,  an even stronger suppression holds  for the Hartree-Fock correlation function. As an example, consider $r=2$.
Using the transfer-matrix equation \eqref{eq:transfer}, we get the relation
\begin{align}
&|U_{i,\epsilon_\alpha}|^2
|U_{i+2,\epsilon_\beta}|^2 +
|U_{i+2,\epsilon_\alpha}|^2
|U_{i,\epsilon_\beta}|^2 \nonumber \\
& - 2U_{i,\epsilon_\alpha}U_{i+2,\epsilon_\alpha}U_{i+2,\epsilon_\beta}U_{i,\epsilon_\beta}\nonumber\\
&=
\dfrac{1}{t_{2+i}^2}\left(\epsilon_\alpha^2
|U_{i+1,\epsilon_\alpha}|^2
|U_{i,\epsilon_\beta}|^2 \right.\nonumber\\
&\left.- 2\epsilon_\alpha\epsilon_\beta
U_{i,\epsilon_\alpha}
U_{i+1,\epsilon_\alpha}
U_{i+1,\epsilon_\beta}
U_{i,\epsilon_\beta}
+\epsilon_\beta^2
|U_{i,\epsilon_\alpha}|^2
|U_{i+1,\epsilon_\beta}|^2\right).
\label{HF-epsilon}
\end{align}
The left-hand side of Eq.~(\ref{HF-epsilon}) is the difference between the Hartree and Fock terms that enters the correlation function $C_{HF}$ for $r=2$.
On the other hand, the right-hand-side is the linear combination of $C_H$ and $C_F$ terms for $r=1$, each of them multiplied by a factor quadratic in energies.
We have thus proven that $C_{HF}$ for $r=2$ is suppressed by an additional factor $ \sim \epsilon_>^2$ in comparison with the $r=1$ correlation function $C_{HF} \simeq C_H$,
\begin{equation}
C_ {HF}(\epsilon_\alpha,\epsilon_\beta,2,L) \sim \epsilon_>^2 C_ {HF}(\epsilon_\alpha,\epsilon_\beta,1,L).
\label{even-odd-ratio}
\end{equation}
 The same argument holds for other even $r$.
This is fully supported by the numerical data, as shown in Fig. \ref{fig:hf_even_e2} where we plot the ratio $C_ {HF}(\epsilon_1,\epsilon_n,2,L) / C_ {HF}(\epsilon_1,\epsilon_n,1,L)$ multiplied by $\epsilon_n^{-2}$ for different $n$, as a function of $L$. We remind the reader that $\epsilon_n$ scales exponentially as a function of $L$ and $n$, see
Eq.~(\ref{eq:epsilon_scale}). Each of the factors $C_ {HF}(\epsilon_1,\epsilon_n,2,L)$, $C_ {HF}(\epsilon_1,\epsilon_n,1,L)$, and $\epsilon_n^{-2}$, when taken separately, changes within an enormous range of many dozens of decades, see, e.g. Figs.~\ref{fig:htfo_even_scal} and \ref{fig:htfo_odd_scal}. On the other hand, the product plotted in Fig. \ref{fig:hf_even_e2} changes only weakly (at most linearly in $L$, which means logarithmically in $\epsilon$), in full agreement with the analytical argument.

Since we have shown above that the odd-$r$ correlation function  in the right-hand side of Eq.~(\ref{even-odd-ratio}) scales as $\epsilon_>^2$, the even-$r$ Hartree-Fock correlator should scale as $\epsilon_>^4$ according to this equation. The  $\epsilon_>^4$ scaling of $C_ {HF}$ for even $r$ is indeed observed numerically,  see Fig.~\ref{fig:htfo_even_scal}.

\section{Disordered Majorana chain with mean-field treatment of interaction in the Ising + LL phase}
\label{sec:majorana-disorder-mean-field}

In this Appendix, we present an analysis of the disordered Majorana chain that treats disorder exactly and the interaction on the mean-field level. This approach is in a sense complementary to those in the main text of the paper. In the weak-disorder RG of Sec.~\ref{SectionClean} the interaction was treated exactly and the disorder was considered as a perturbation. Contrary to this, the analysis of Sec.~\ref{sec:SectionIRFP} considered disorder exactly and the interaction perturbatively. Here, we treat the disorder by using the field-theoretical $\sigma$ model approach.  This treatment is essentially exact, in analogy with Sec.~\ref{sec:SectionIRFP}. The key differences with Sec.~\ref{sec:SectionIRFP} are that (i) we consider a sufficiently strong repulsive interaction for which the clean system is in the Ising+LL phase, and (ii) we include the interaction on the mean-field level only. This allows us to obtain the phase diagram of the system in the plane spanned by the disorder strength and the staggering. The phase diagram contains four distinct topological phases. Of course, we know from Sec.~\ref{sec:SectionIRFP} and from the numerical study in Sec.~\ref{sec:majorana-int-dis-dmrg}
that including effects of interaction beyond the mean-field level destabilizes the system on the critical line.
This means that the transitions between the topological phases are in fact not of second order (as found in the mean-field treatment below) but rather of first order.
On the other hand, the phase diagram is expected to remain applicable also beyond the mean-field level.

At mean-field level with respect to the interaction, the third nearest neighbor hoppings are generated and the nearest-neighbor hopping is renormalized. The full mean-field Hamiltonian, including the randomness $\delta t_j$ in the nearest neighbor hopping, reads
\begin{align}
H_{I+LL}^{MF}&=\frac{i}{2}\sum_j \left[(t_1+t_2 + (-1)^j(t_1-t_2) +2\delta t_j) \gamma_j\gamma_{j+1} \right. \nonumber \\
& \left. + ((t_1'+t_2')+(-1)^j(t_1'-t_2')) \gamma_j\gamma_{j+3}\right].
\label{eq:HILLMF}
\end{align}
By choosing $t_1\neq t_2$ or $t_1'\neq t_2'$, the system can be staggered. The random component $\delta t_j$ of the hopping is assumed to have Gaussian statistics, with zero average.

The formalism presented in Refs. \onlinecite{altland_quantum_2014,altland_topology_2015} for a
particular model can be extended to the case of generic banded Hamiltonians. For
convenience, we have performed computations in class AIII instead of BDI (i.e.,
allowing for complex $\delta t_j$). The results for AIII shown here remain essentially the same for
the class BDI as can be checked numerically using transfer matrices.

The calculations proceed by integrating out the disorder using the supersymmetry
formalism. After Hubbard-Stratonovich decomposition and saddle-point expansion
(which yields the self-consistent Born approximation), one arrives at a
non-linear sigma model describing the disordered wire.
The action describes the soft modes $T\in
\mathrm{GL}(1|1)$: \begin{align} S[T] &= \tilde{\chi} \mathrm{str}(T\partial
T^{-1}) -\frac{\tilde{\xi}^2}{4} \mathrm{str}(T\partial^2 T^{-1})
\label{eq:ind:th}. \end{align} There are two coupling constants here:
$\tilde{\xi}$ has a meaning of the bare conductance, and $\tilde{\chi}$ of the
bare topological index. Under RG, these coupling constants get renormalized. The
theory thus exhibits a two-parameter RG flow, which is largely analogous to the
Khmelnitskii-Pruisken flow for the 2D theory describing the quantum Hall effect.

Except for the case of half-integer bare values, $\tilde{\chi}$ flows to the
nearest integer value, which is the actual topological index $\chi$.
Half-integer values of $\tilde{\chi}$ are stable under RG-flow and correspond to
critical theories  at the boundary of two topologically distinct phases. To
determine the phase diagram, one thus should compute the dependence of the bare
index $\tilde{\chi}$ on parameters of the chain. These dependences are obtained
when one derives the $\sigma$ model from the microscopic model, as sketched
above. We skip details of this calculation, since it is analogous to that
carried out for a different microscopic model in Ref.
\onlinecite{altland_topology_2015}.
A general 1D non-interacting Hamiltonian $H$ with chiral symmetry and with translational invariance in average
 can be written as:
\begin{equation}
H = h_n \sum_i a_{i+n}^\dagger b_i + \sum_i r_{n,i} a_{i+n}^\dagger b_i + h.c.
\label{H-general}
\end{equation}
Here $a_i$ and $b_i$ are operators on two sublattices, $h_n$ are the average hopping matrix elements, and $r_{n,i}$ are random contributions to hopping that are characterized by zero mean and by the variance
\begin{equation}
\left\langle r_{n,i} r_{m,j}^* \right\rangle = w_n \delta_{i,j}\delta_{n,m}.
\end{equation}
We find the following result for the bare index $\tilde{\chi}$ in terms of the parameters of $H$:
\begin{equation}
\tilde{\chi} = \sum_q \dfrac{h^-(q)v^+(q)}{\Sigma_0^2+h^+(q)h^-(q)} + \sum_n n u_n,
\end{equation}
where
\begin{eqnarray}
h^-(q) &=& \sum_n h_n e^{-i n q}, \\ h^+(q) &=& \sum_n h_n e^{i n q},\\
v^+(q) &=& \sum_{m,n} (n-m)u_n h_m e^{i m q}, \\
u_n &=& \frac{w_n^2}{\sum_m w_m^2},
\end{eqnarray}
and the self-energy $\Sigma_0$ is a solution of the equation
\begin{equation}
\left(\sum_n w_n^2\right) \sum_q\dfrac{1}{\Sigma_0^2 - h^+(q)h^-(q)} = 1
\label{sigma-0}\\
\end{equation}
representing the self-consistent Born approximation.

\begin{figure}
\centering
\includegraphics[width=\columnwidth]{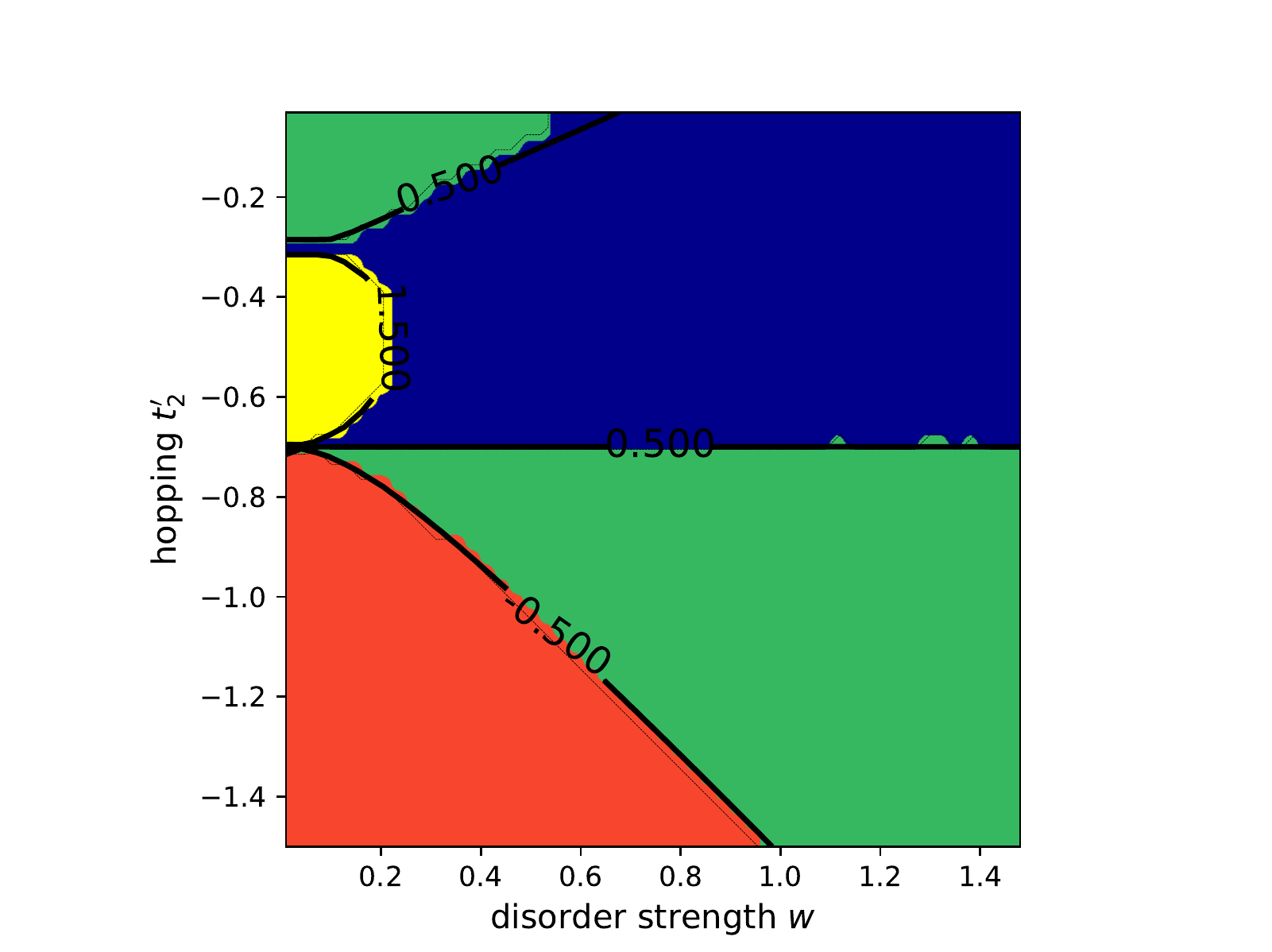}
\caption{Phase diagram of the mean-field Hamiltonian (\ref{eq:HILLMF}) describing the Ising+LL phase of the disordered Majorana chain. The parameters
$t_1'=-0.7$ and $t_2=t_1=t=1$ are fixed. The phase diagram is shown in the plane spanned by disorder $w$ and the hopping $t_2'$. The zero staggering corresponds to $t_2' = t_1' = -0.7$. Black lines are phase boundaries as obtained analytically via mapping on the $\sigma$ model from the condition that the bare index $\tilde{\chi}$ is half-integer.
Colored regions are four distinct topological phases with the values of the topological index $\chi$ equal to -1 (red), 0 (green), 1 (blue), and 2 (yellow), as obtained from the transfer-matrix numerics.  A perfect agreement between numerical and analytical results is observed. At zero disorder, $w=0$, and zero staggering, $t_1'-t_2'=0$, three critical lines meet, yielding a critical theory with central charge $c=3/2$.
}
\label{fig:MeanFieldPhases}
\end{figure}

Our Hamiltonian (\ref{eq:HILLMF}) is a particular case of Eq.~(\ref{H-general}).  The nearest and third nearest neighbor hopping   of Eq.~(\ref{eq:HILLMF}) are encoded in
terms of Eq.~(\ref{H-general}) in $h_1 = t_1$, $h_2 = t_1'$, $h_0 = t_2$, and $h_{-1} = t_2'$. Further, the randomness in the nearest neighbor hopping of Eq.~(\ref{eq:HILLMF}) translates into $u_0 = 1/2$ and $u_1 = 1/2$.
The resulting phase diagram in the parameter plane
spanned by disorder strength $w$ and staggering $t_1'-t_2'$ is shown in Fig.
\ref{fig:MeanFieldPhases}.

We have compared the analytical results (black lines show the corresponding phase boundaries in Fig. \ref{fig:MeanFieldPhases}) with those of direct transfer matrix numerics.  Four topological phases ( with $\chi=-1$, 0, 1, and 2) as obtained by the latter approach are shown by different colors in Fig. \ref{fig:MeanFieldPhases}. An excellent agreement between the analytical and numerical data is observed. This is quite non-trivial since (i) the $\sigma$ model derivation holds in the limit of large number of channels, $N\gg 1$, whereas our model corresponds to $N=3$, (ii) the analytical calculation of parameters of the $\sigma$ model is controlled fir weak disorder, $w/t \ll 1$, whereas we find a very good agreement also for $w/t \sim 1$.

The self-duality transformation ensures that the zero-staggering line ($t_2'=-0.7$ in Fig.\ref{fig:MeanFieldPhases}) is critical within this mean-field analysis. An important observation is that the critical line is adjacent only to 0 (green) and 1 (blue) topological phases for finite disorder.

In the clean DMRG analysis, Sec.~\ref{sec:majorana-clean-dmrg}, only two  distinct topological phases were observed, which correspond to the green and blue phases of Fig.~\ref{fig:MeanFieldPhases}. The other two phases (red and yellow) can only be reached by adding the third nearest neighbor hopping explicitly \cite{rahmani_phase_2015}
since otherwise the Hamiltonian (\ref{eq:HILLMF}) with the corresponding parameters can not be obtained as a mean-field Hamiltonian of an interacting Majorana chain.
When disorder is added to the mean-field model, we observe that the parameter space for the red and yellow phases shrinks.

\section{Analytical approach to wave function correlations}
\label{app:wave-func-corr-analytics}

In this Appendix, we provide analytical results for the scaling of eigenfunction correlation functions at the infinite-randomness fixed point. These results complement, support, and explain the corresponding numerical results in Sec. \ref{sec:SectionIRFP}.

In Ref. \onlinecite{balents_delocalization_1997} the average of one Green's function in a non-interacting 1D model of class BDI was computed by means of supersymmetry formalism that allowed to map the problem onto quantum mechanics of a $SU(1\vert 1)$ spin. In order to obtain directly the correlation functions of two eigenstates, one would need to average products of two Green's functions with the corresponding energy and spatial arguments. While the mapping on a supersymmetric quantum mechanics can be generalized to this situation, the solution of the corresponding problem becomes extremely difficult.  For this reason, we choose below a slightly different approach and calculate, by using the supersymmetry technique, the averaged square of the Green's function at an imaginary frequency. This average is related, by virtue of a spectral decomposition, to the two-eigenstates correlation functions. The resulting conclusions on the scaling of the two-eigenstates correlations are in agreement with our numerical fundings in Sec. \ref{sec:SectionIRFP}.

\begin{figure}
	\centering
	\includegraphics[width=\columnwidth]{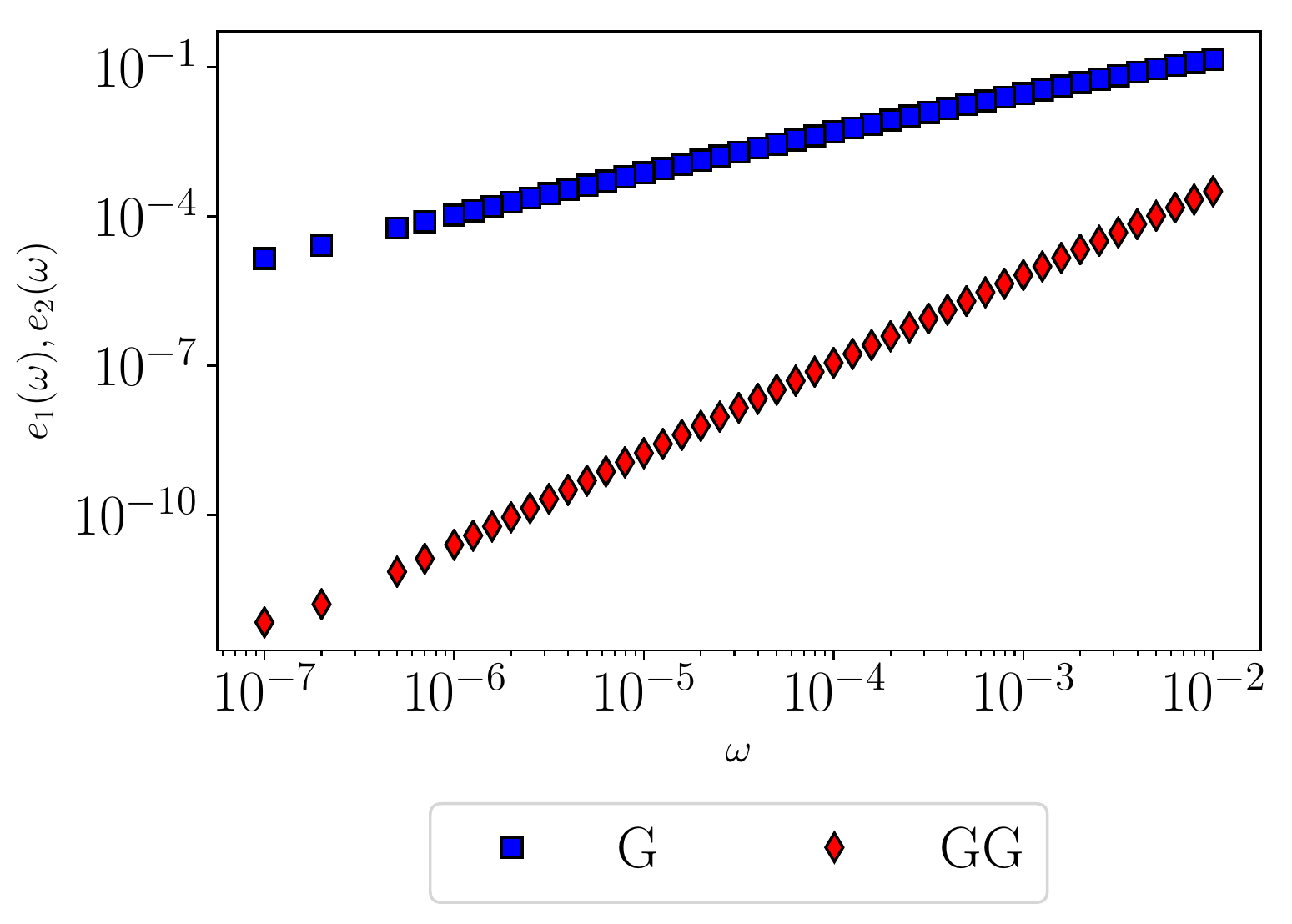}
	\caption{In this plot, the validity of Eqs. \eqref{averaged-GF} and \eqref{averaged-GF-squared} for $\langle G(i\omega)\rangle_{\rm dis}$  and $\langle G(i\omega) G(i\omega)\rangle_{\rm dis}$ derived in a continuum-limit approximation to the effective Schr\"odinger equation is verified numerically. For this purpose, we plot
		$e_1(\omega)=i\omega \langle G(i\omega)\rangle_{\rm dis} - |\ln(\omega/a_0)|^{-2}$ and $e_2(\omega) = 3\omega^2\langle G(i\omega) G(i\omega)\rangle_{\rm dis}-|\ln(\omega/a_0)|^{-2}$ computed numerically. The disorder strength is set $W=1$. The constant $a_0 \simeq 0.8$ is determined to minimize the errors $e_i$. It can be seen that $e_1(\omega) \propto \omega$ and $e_2(\omega) \propto \omega^2$.
	}
	\label{fig:Omega}
\end{figure}

We follow the formalism of Ref. \onlinecite{balents_delocalization_1997} and map the original lattice model with random hopping onto a continuous model of a Dirac fermion with random mass, cf. Sec.~\ref{majorana-12-fixed-point}.  The latter is considered to be delta-correlated and gaussian-distributed disorder, with the strength $W$ (which sets the ultraviolet cutoff for the critical theory and can be set to unity).  Within the mapping onto the supersymmetric quantum mechanics, the averaged Green's function at an imaginary frequency $i\omega$ and with coinciding spatial arguments is obtained from the ground state of the corresponding effective Schr\"odinger equation.
We obtain, in agreement with Ref. \onlinecite{balents_delocalization_1997},
\begin{equation}
\langle G(i\omega)\rangle_{\rm dis} = \dfrac{a_1W}{i\omega|\ln(\omega/a_0 W)|^2}.
\label{averaged-GF}
\end{equation}
We have found the constants $a_1$ and $a_0$ by a numerical solution of the effective Schr\"odinger equation of the supersymmetric quantum mechanics; the results are $a_1 = 1$ (which holds with a very high accuracy and is apparently exact) and $a_0 \simeq 0.8$. 
Extending this analysis to the averaged square of the Green's function, we obtain
\begin{equation}
\langle G(i\omega)G(i\omega)\rangle_{\rm dis} = \dfrac{a_2 W}{\omega^2|\ln(\omega/a_0 W)|^2},
\label{averaged-GF-squared}
\end{equation}
where $a_2 = 1/3$ (which again holds numerically with a very high accuracy and should thus be exact).  Equations (\ref{averaged-GF}) and (\ref{averaged-GF-squared}) are derived in the continuum-limit approximation to the effective Schr\"odinger equation. We have verified, however, by a numerical solution of the exact (discrete) equation that they hold with an outstanding accuracy. Specifically, as shown in Fig.~\ref{fig:Omega}, the relative correction to Eq.~(\ref{averaged-GF}) is of the order $\omega$ and that to Eq.~(\ref{averaged-GF-squared}) is of the order $\omega^2$.  This means, in particular, that all orders of expansion of Eqs. (\ref{averaged-GF}) and (\ref{averaged-GF-squared}) in $1/|\ln \omega|$ are fully reliable.

Now we connect these results to the correlation functions of eigenstates $\psi_\alpha(r)$ (which are continuum limit counterparts of the states $U_{i\alpha}$ studied numerically in Sec. \ref{sec:SectionIRFP}. Since all arguments of Green's functions that we consider are equal (we set them $r=0$), only eigenstates at this point will enter. Using the spectral decomposition of the single-particle Green's function, we get
\begin{align}
\langle G(i\omega)\rangle_{\rm dis} &= \sum_\alpha \left \langle \dfrac{\psi_\alpha^2(0)}{i\omega-\epsilon_\alpha} \right \rangle_{\rm dis} \nonumber\\
&= \int \dd \epsilon\: L\: \nu(\epsilon) \dfrac{\langle \psi_\alpha^2(0)\rangle_{\rm dis} }{i\omega-\epsilon}.
\label{G-spectral}
\end{align}
The average entering here is $\langle \psi_\alpha^2(0)\rangle_{\rm dis}= L^{-1}$ due to eigenfunction normalization. 
Further, the density of states is
\begin{equation}
\nu(\epsilon) \simeq \frac{c^2}{\epsilon |\ln (\epsilon / \Lambda)|^3 },
\label{nu-epsilon-with-coeff}
\end{equation}
see Eq.~(\ref{nu-epsilon}), where $c \sim 1$ is the constant defined in Eq.~(\ref{eq:epsilon_scale}) and we have introduced the ultraviolet cutoff  $\Lambda \sim 1$. 
Substituting this in Eq.~(\ref{G-spectral}), we get 
\begin{eqnarray}
\langle G(i\omega)\rangle_{\rm dis} &=& c^2\left[\dfrac{1}{i\omega |\ln \omega/\Lambda |^2} + \dfrac{\ln 2}{i\omega |\ln \omega/\Lambda |^3} \right. \nonumber \\
&+& \left. \mathcal{O} (\omega^{-1} |\ln \omega |^{-4})\right].
\label{G-omega-from-spectral}
\end{eqnarray}
We see that Eq.~(\ref{G-omega-from-spectral}) is in full agreement with the result (\ref{averaged-GF}) of the supersymmetric calculation. 
Indeed, not only the leading behavior agrees but also Eq.~(\ref{averaged-GF}) can be expanded to bring it to the form (\ref{G-omega-from-spectral}). 
This confirms that the formula (\ref{nu-epsilon-with-coeff}) for the density of states that we have used when deriving Eq.~(\ref{G-omega-from-spectral}) from the spectral decomposition (\ref{G-spectral}) is correct. One can, of course, also obtain (\ref{nu-epsilon-with-coeff}) by performing an analytical continuation of Eq.~(\ref{averaged-GF}). Note, however, that we used different models of disorder in the numerical and analytical calculations, so that numerical value of the coefficient $c^2$ in Eq.~(\ref{nu-epsilon-with-coeff}) cannot be directly obtained from the analytical result. 

Having satisfied ourselves that the spectral decomposition works properly for $\langle G(i\omega)\rangle_{\rm dis}$, we turn to $\langle G(i\omega) G(i\omega)\rangle_{\rm dis}$ that provides information about correlations of different eigenfunctions. The spectral decomposition now yields
\begin{align}
\langle G(i\omega) G(i\omega)\rangle_{\rm dis} &= \sum_\alpha \left\langle \dfrac{\psi_\alpha^4(0)}{(\epsilon_\alpha-i\omega)^2}  \right \rangle_{\rm dis}\nonumber\\
&+\sum_{\alpha\neq \beta} \left \langle  \dfrac{\psi_\alpha^2(0)\psi_\beta^2(0)}{(\epsilon_\alpha-i\omega)(\epsilon_\beta-i\omega)}  \right \rangle_{\rm dis} .
\label{GG-spectral}
\end{align}
In Sec. \ref{sec:SectionIRFP}, we have found numerically the following scaling of the eigenstates correlation functions entering Eq.~(\ref{GG-spectral}):   $\langle \psi^4_\alpha(0) \rangle_{\rm dis} = aL^{-1}$, Eq.~(\ref{eq:htself_balents}), and $\langle \psi^2_\alpha(0)\psi^2_\beta(0)\rangle_{\rm dis} = bL^{-2}\ln\epsilon_>$, Eq.~(\ref{eq:ht_lowr}), where $a$ and $b$ are numerical coefficients, and $\epsilon_>$ is the larger of the two energies $\epsilon_\alpha$ and $\epsilon_\beta$. 
Substituting them into Eq.~(\ref{GG-spectral}) and rewriting the sum over energies as integrals with the density of states (\ref{nu-epsilon-with-coeff}), we obtain
\begin{eqnarray}
\langle G(i\omega) G(i\omega)\rangle_{\rm dis} &=& c^2\left[\dfrac{a}{i\omega^2 |\ln \omega/\Lambda |^2} + \dfrac{a-(2/3)bc^2}{i\omega^2 |\ln \omega/\Lambda |^3} \right.
\nonumber \\
&+& \left. \mathcal{O} (\omega^{-1} |\ln \omega |^{-4})\right].
\label{eq:av_g2n}
\end{eqnarray}
We observe now that two leading terms of Eq.(\ref{eq:av_g2n}) fully correspond to the expansion of the result (\ref{averaged-GF-squared}) of the supersymmetry-formalism calculation. This proves that the numerically  found values of the exponents, $\alpha =1$ and $\gamma=2$, in the scaling of eigenstate correlations,  
Eq.~(\ref{eq:ht_lowr}), are indeed exact.

\section{Entanglement entropy in gapped regime of the Majorana chain with repulsive interaction and staggering}
\label{ap:red_patch}

\begin{figure}
  \centering
  \includegraphics[width=.9\columnwidth]{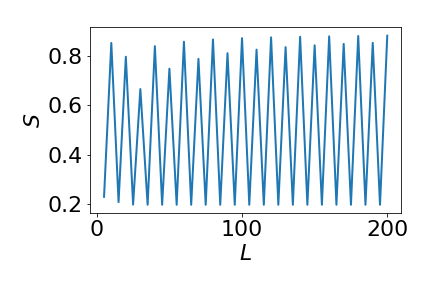}
  \caption{Entanglement entropy $S$ of the central bond vs system size $L$ of
  the interacting Majorana chain with staggering, Eq. (\ref{eq:spins}). The
  parameters $t^{(1)}=1.00$, $t^{(2)}=0.72$, $g^{(1)}=1.5$, and $g^{(2)}=1.08$
  are chosen in such a way that the system belong to the red region in Fig.
  \ref{fig:phase_cc}. Apart from even-odd oscillations, the entanglement entropy
  stays constant with system size. Thus the system is gapped for these
  parameters.}
  \label{fig:red_phase_size}
\end{figure}

In the phase diagram of the clean interacting Majorana chain with staggering
(Fig. \ref{fig:phase_cc} in Sec.~\ref{sec:clean-repulsive}) we observe a region
(plotted in red) where application of formula (\ref{eq:scaling}) yields a very
high apparent central charge. This is in contrast to the dual region (obtained
by reflection with respect to the self-dual line) where the formula yields a
central charge of zero consistent with the expectation of a gapped phase. Thus
the region above the critical line should be gapped as well. To check this, we
have calculated the entanglement entropy $S$ at the central bond for different
system sizes. The result shown in Fig. \ref{fig:red_phase_size} unambiguously
exhibits the area law for $S$ (i.e. no increase with $L$), so that the region is
gapped. This is not in contradiction with the high apparent central charge
observed in Fig. \ref{fig:phase_cc}, since the formula (\ref{eq:scaling}) is
guaranteed to be valid only in conformal theories. On the other hand, in other
gapped regions the entanglement entropy did not show any anomalies of this type.
It is thus interesting to look more closely at this region in order to
understand the reasons for the anomalous behavior of $S$ there.

\begin{figure}
  \centering
  \includegraphics[width=.45\columnwidth]{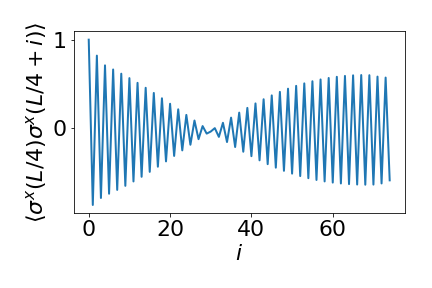}
  \includegraphics[width=.45\columnwidth]{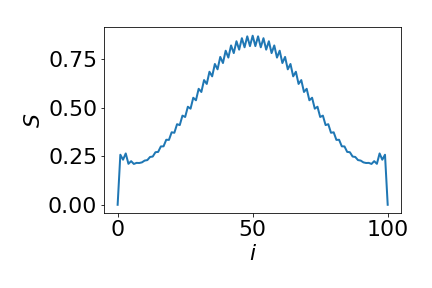}
  \caption{Spin-spin correlator $\langle \sigma^x(L/4) \sigma^x(L/4+i) \rangle$
  versus distance $i$	(left) and entanglement entropy at the bond $i$ (right) of
  a system with Hamiltonian Eq. (\ref{eq:spins}) and parameters $t^{(1)}=1.00$,
  $t^{(2)}=0.72$, $g^{(1)}=1.5$, $g^{(2)}=1.08$, and $L=100$. These parameters
  belong to the red region in Fig. \ref{fig:phase_cc}. The $\sigma^x$ spin
  component shows the antiferromagnetic order but the $\pi$ phase shifts occurs
  at the central bond. In view of this, the spin correlator in the left panel
  takes there a zero value. The entanglement entropy in the right panel has a
  peak around the same spatial point.}
  \label{fig:red_phase_sx}
\end{figure}

To shed light on the behavior of the entanglement entropy, we compare in Fig.
\ref{fig:red_phase_sx} the $\sigma^x$ correlator with the entanglement entropy
as function of bond position. The entanglement entropy increases sharply around
the central bond leading to a spurious high value of the central charge if it is
calculated by formula Eq. (\ref{eq:scaling}). The correlator $\langle
\sigma^x_{L/4}\sigma^x_{L/4+i}\rangle$ shows two regions of antiferromagnetic order
with a phase shift at $i=L/2$.

\begin{figure}
  \centering
  \includegraphics[width=.45\columnwidth]{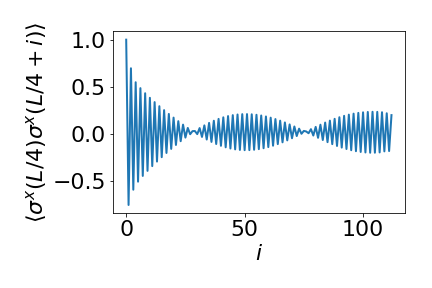}
  \includegraphics[width=.45\columnwidth]{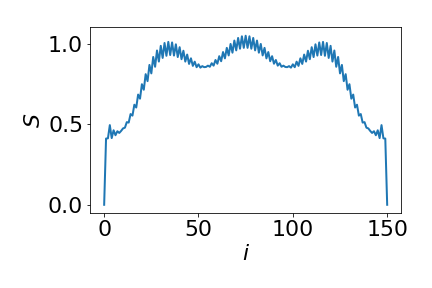}
  \caption{Spin-spin correlator $\langle \sigma^x(L/4) \sigma^x(L/4+i) \rangle$
  versus distance $i$	(left) and entanglement entropy at bond $i$ (right)
  of a system with Hamiltonian Eq. (\ref{eq:spins}) and parameters
  $t^{(1)}=1.00$, $t^{(2)}=0.90$, $g^{(1)}=1.5$, $g^{(2)}=1.35$, and $L=150$. 
  \ADDED{In the indices, $L/4$ denotes the integer part $[150/4] = 37$.}  In the phase
  diagram of Fig. \ref{fig:phase_cc}, these parameters put the system just below the red patch, but still
  outside the LL region. For these parameters and length, the antiferromagnetic
  ordering of the $\sigma^x$ spin component changes phase several times, as seen in the left panel. The entanglement entropy in the right panel exhibit peaks at the corresponding bonds.}
  \label{fig:red_phase_node}
\end{figure}

Considering points in the phase diagram of Fig. \ref{fig:phase_cc} in a narrow
region between the red patch and the extended critical region with $c=1$. Here
the $\sigma^x$ correlator looks very similar, except that there is more then one
node where the phase of the antiferromagnetic ordering shifts. A characteristic
example is shown in the left panel of Fig. \ref{fig:red_phase_node}. By
comparing this plot with the entanglement entropy of the same system (right
panel of Fig. \ref{fig:red_phase_node}), we see that each of these nodes is
associated with a maximum in the entanglement entropy. We find that the number
of such nodes depends on  parameters of the Hamiltonian and on the system size.
Furthermore, it also depends on whether the system size is even or odd. This
explains the difference between even and odd system sizes in Fig.
\ref{fig:red_phase_size}. We leave a more detailed analysis of the physics in
this regime to future work.

We have verified that the peculiarity of this type does not arise in other regions of the phase diagram in Fig. \ref{fig:phase_cc}. In those regions the central charge obtained by fitting the $x$-dependence of the entanglement entropy $\mathcal{S}$ at fixed $L$ according to Eq. \eqref{eq:scaling} is consistent with that found from the $L$-dependence of $\mathcal{S}$.

\section{Correlation functions away from criticality}
\label{corr-func-away-crit}

\begin{figure}
	\centering
	\includegraphics[width=.66\columnwidth]{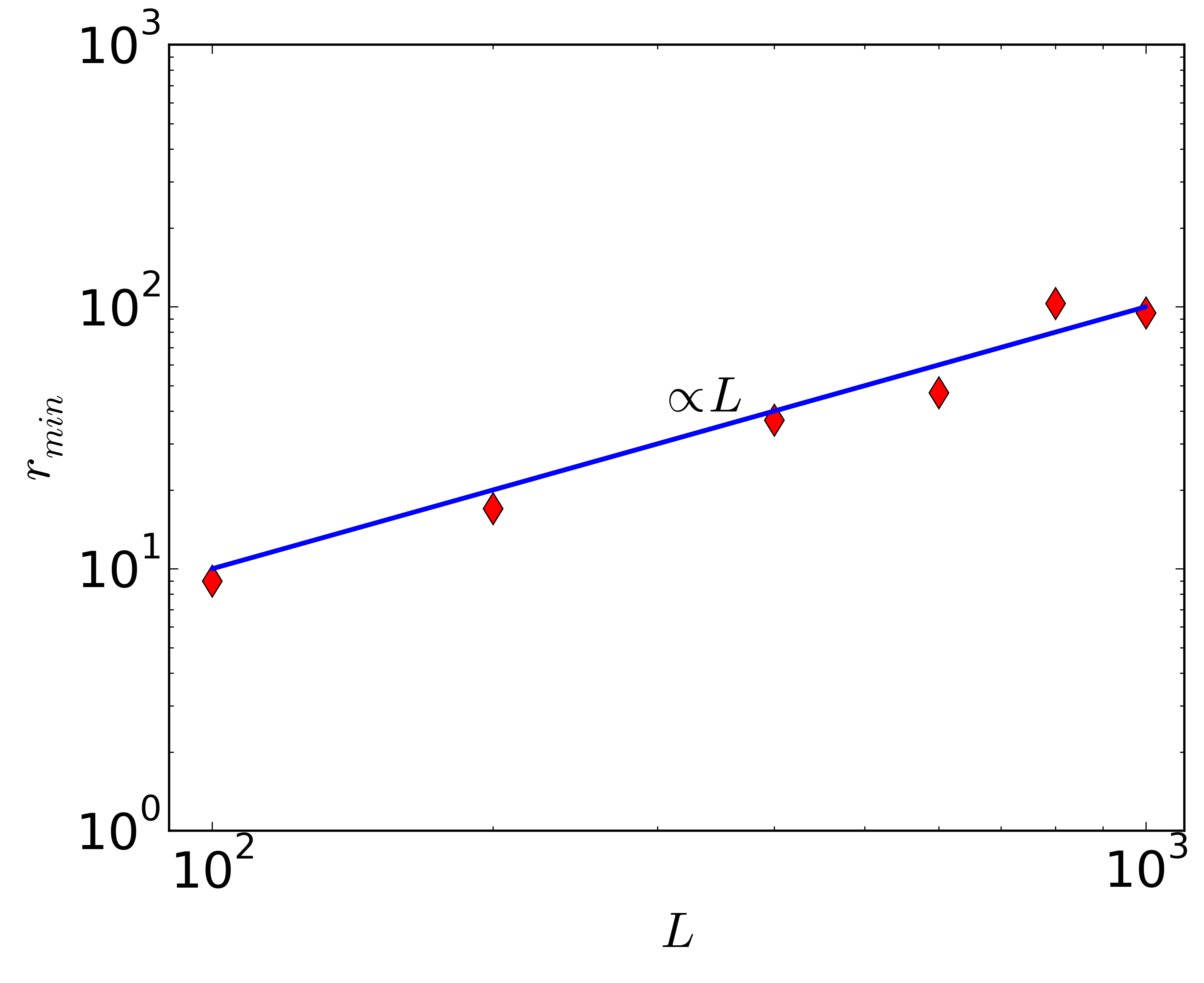}
	\caption{Numerically found position $r_{\rm min}$ of the minimum of $C_H(\epsilon_1,\epsilon_2,r,L)$ with respect to $r$, as a function of length $L$.
This minimum determines the upper border of the critical regime, see left panel of Fig.~\ref{fig:htfo}.
	The scaling $r_{\rm min} \propto L$ is found, confirming the expectation that the critical regime extends up to $\xi_\epsilon$ with $\epsilon = \epsilon_2$.
	}
	\label{fig:minimum}
\end{figure}

In Sec.~\ref{sec:two-wavefunc-corr}, we studied Hartree, Fock, and Hartree-Fock
correlations of two eigenfunctions. We focussed there on the critical regime of
sufficiently small $r$, which is of particular physical interest and also the
one needed to describe the effect of a finite-range interaction. For
completeness, we discuss here the range of large $r$, such that the system is
away from criticality. The scaling (\ref{eq:ht_lowr}) of the correlation
function $C_H$  is expected to hold as long as the system is at criticality,
i.e., at $r < \xi_\epsilon$. (The same applies to $C_F$, which is nearly equal
to $C_H$ in the critical regime.)  According to Eqs.~(\ref{xi-epsilon}) and
(\ref{eq:epsilon_scale}), the localization length $\xi_\epsilon$ is equal to the
system size $L$ times some numerical coefficient, if we choose the second level
$\epsilon_2$ as the larger of two energies, as is done, e.g., in the left panel
of Fig.~\ref{fig:htfo}. In this figure $L=400$, and the critical regime extends
up to $L \simeq 20$. In order to check that the upper border of the critical
regime is indeed equal to $L$ times a numerical coefficient, we plot in
Fig.~\ref{fig:minimum} the position $r_{\rm min}$ of the minimum of
$C_H(\epsilon_1,\epsilon_2,r,L)$ with respect to $r$, as a function of $L$.  As
is clear from the left panel of Fig.~\ref{fig:htfo}, this minimum essentially
marks the upper border (with respect to $r$) of the critical regime, which is
expected to be $ \sim \xi_\epsilon$.  We see that the expectation that $r_{\rm
min}$ scales as $L$ is confirmed, i.e., the critical regime extends up to
$\xi_\epsilon$, as expected.

\begin{figure}
	\centering
	\includegraphics[width=.49\columnwidth]{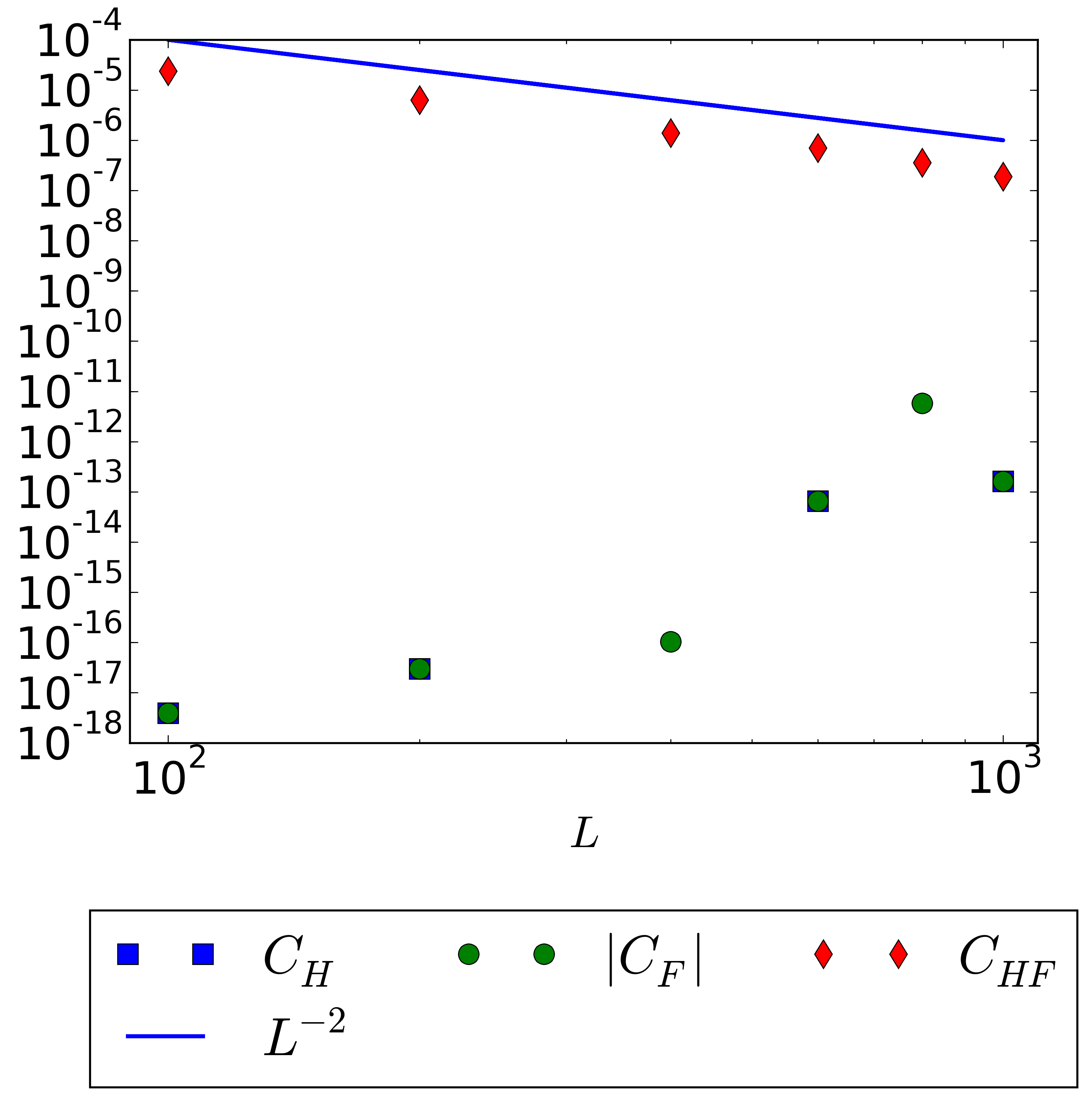}
	\includegraphics[width=.49\columnwidth]{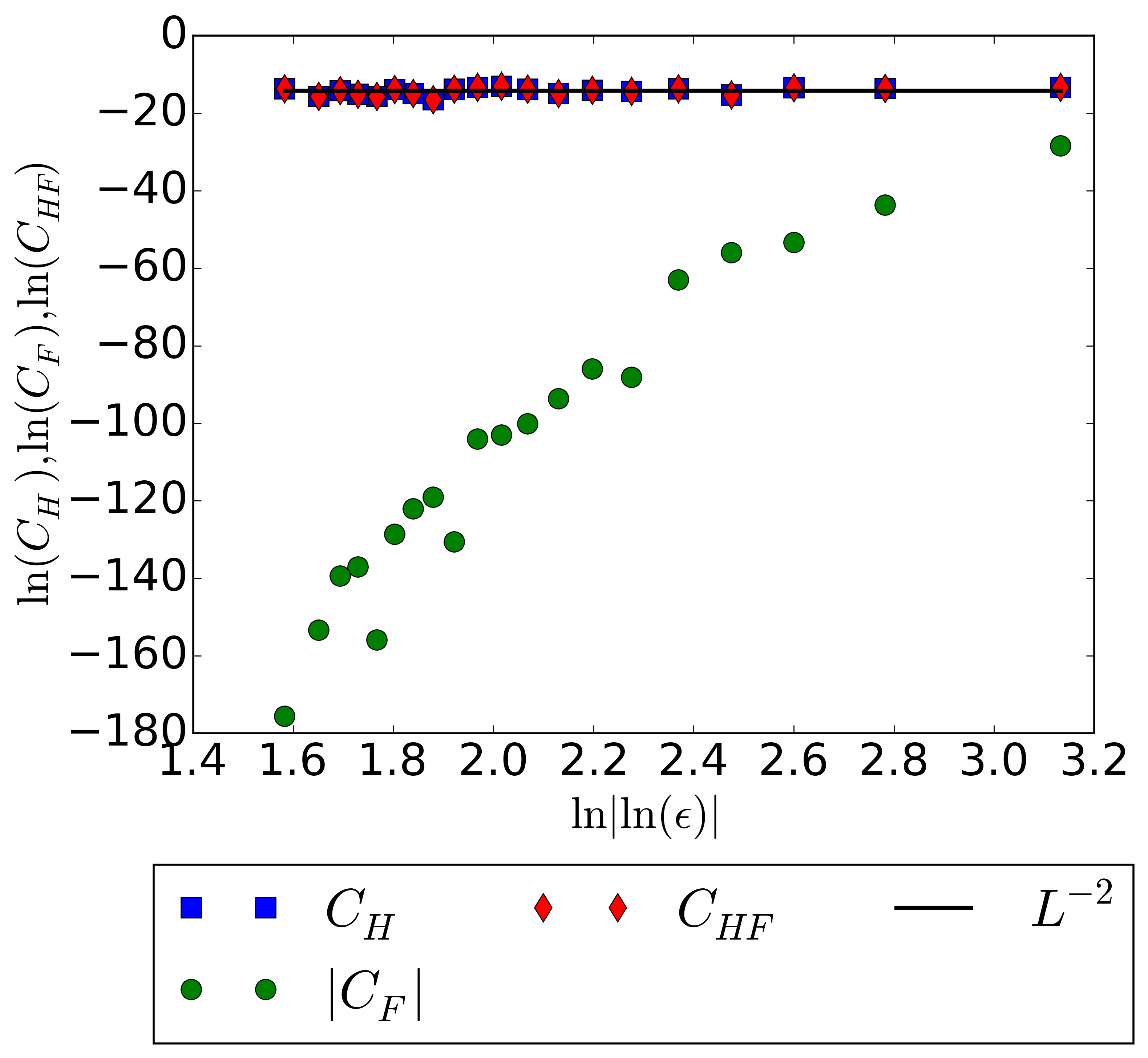}
	\caption{Correlation at large distances. {\it Left:} matrix elements
		$C_H(\epsilon_1,\epsilon_{20},r=150,L)$,
		$C_F(\epsilon_1,\epsilon_{20},r=150,L)$, and
		$C_{HF}(\epsilon_1,\epsilon_{20},r=150,L)$ as functions of the system size $L$.
		{\it Right:}  $C_H(\epsilon_1,\epsilon,r=1000,L=1200)$,
		$C_F(\epsilon_1,\epsilon,r=1000,L=1200)$,,
		$C_{HF}(\epsilon_1,\epsilon,r=1000,L=1200)$, as functions of energy $\epsilon$.
		In both panels, $C_H$ shows the behavior (\ref{eq:ht_highr}) corresponding to the loss of correlations. The Fock correlation functions is much smaller in this regime, $C_F \ll C_H$.
		}
	\label{fig:htfo_scal}
\end{figure}

Now we turn to the behavior of the correlation functions for $r > \xi_\epsilon$,
i.e., outside of the critical regime.  For separation $r \sim L$ the states are
expected to lose all correlations, which implies that
\begin{equation}
C_{H}(\epsilon, r \sim L ,L) \sim \frac{1}{L^2}.
\label{eq:ht_highr}
\end{equation}
The saturation of the correlation function $C_H$ at a value $\sim 1/L^2$ at
large $r$ is evident in the left panel of Fig.~\ref{fig:htfo}. As a further
check, we show in  Fig.\ref{fig:htfo_scal} the $L$-dependence (left panel) and
$\epsilon$ dependence (right panel) of $C_H$ for $r \sim L$.  The figure
confirms that, in this regime, $C_H \sim L^{-2}$ and is essentially
$\epsilon$-independent.

It is interesting to notice that the critical behavior (\ref{eq:ht_lowr}) at its
upper border $r \sim \xi_\epsilon \sim \ln^2\epsilon$  yields $C_H \sim
1/L^2\ln^2\epsilon$, which does not match Eq.~(\ref{eq:ht_highr}) due to an
additional factor $1/\ln^2\epsilon \ll 1$. Thus, there should be an intermediate
regime for $ \xi_\epsilon < r < L$ located between the critical regime
(\ref{eq:ht_lowr}) and the uncorrelated regime (\ref{eq:ht_highr}). This regime,
where $C_H$ rapidly increases with $r$, is clearly observed in the left panel of
Fig.~\ref{fig:htfo}. We leave an analysis of this regime to a future work.

Finally, we note that for large distances $r > \xi_\epsilon$ (i.e., outside of
the critical regime), the Fock term becomes much smaller than the Hartree one,
$C_H \gg C_F$, see Figs. \ref{fig:htfo} and \ref{fig:htfo_scal}. Therefore, the
strong Hartree-Fock cancellation (which occurs for even $r$) is only a property
of the critical regime. Another interesting observation is that the Fock term 
changes sign around $r\sim \xi_\epsilon$. This explains the dips in the curves 
for $|C_F|$, see Figs. \ref{fig:htfo} and \ref{fig:ec8}.

\bibliography{interacting_bdi}

\end{document}